\documentclass[twocolumn,showpacs,showkeys,preprintnumbers,aps,floatfix, pre]{revtex4-1}
\usepackage{graphicx} 
\usepackage{amsmath}
\usepackage{color}
\usepackage{multirow}
\usepackage{amssymb}
\usepackage{dcolumn}%
\usepackage{bm}%
\usepackage{ctable}
\usepackage{tabularx}
\usepackage{graphicx}

\begin{document}

\title[]{\Large \bf Memory formation in cyclically deformed  amorphous solids and sphere assemblies}
\author{Monoj Adhikari}
\author{Srikanth Sastry}
\affiliation{Theoretical Sciences Unit, Jawaharlal Nehru Centre for Advanced Scientific Research,\\Jakkur, Bengaluru 560064, India }

\begin{abstract}
We study a model amorphous solid that is subjected to repeated athermal cyclic shear deformation. It has previously been demonstrated that the memory of the amplitudes of shear deformation the system is subjected to (or trained at) is encoded, and can be retrieved by subsequent deformation cycles that serve as {\it read} operations. Here we consider different read protocols and measurements and show that single and multiple memories can be robustly retrieved through these different protocols. We also show that shear deformation by a larger amplitude always erases the stored memories. These observations are similar to those in experiments with non-Brownian colloidal suspensions and corresponding models, but differ in the possibility of storing multiple memories non-transiently. Such a possibility has been associated with the presence of cycles of transitions that take place in the model amorphous solids, between local energy minima. Here, we also study low density sphere assemblies which serve as models for non-Brownian colloidal suspensions, under athermal deformation, and identify a regime where the signatures of memory encoding are similar to the model glass, even when transition between local energy minima are absent.  We show that such a regime corresponds to the presence of {\it loop reversibility}, rather than {\it point reversibility} of configurations under cyclic deformation. 
\end{abstract} 

\maketitle

\section{Introduction}
Retention of memory of past history arises ubiquitously in describing the properties of condensed matter, ranging from near equilibrium conditions to far from equilibrium conditions, including in the presence of external driving. Simple examples may merely involve a dependence on history that breaks a symmetry or leads a system to reside in a metastable state. Indeed such history dependence forms the basis of conventional memory devices, such as magnetic or phase change memory devices.
Other popular memory devices, such as shape memory materials  \cite{Kbhattacharya}, rely on the presence of phase transformations, but also on the ability of a material to reside in one among a large number of possible structures, in order to accommodate externally applied deformation. The presence of multiple distinct internal structures or states in which a material can exist for long times is a generic condition for the presence of memory effects, seen particularly in systems that exhibit some form of disorder. The range of examples is vast, and includes structural glasses \cite{Kovacs,kovacsfs,kovacsbertin,kovacsjim} and spin glasses \cite{agingrejuvenation}, magnetic systems with disorder 
 \cite{SethnaPRL,pierceetal,PhysRevLettPierce, PhysRevBPierce,PhysRevLetDhar} that exhibit return point memory, and 
charge density waves systems that exhibit return point memory and pulse duration memory \cite{PhysRevLetMiddleton,coppersmith1987,tang1987,CDWLittlewood,CDWNagel,Kaspar2013,Iseri2016a,Mungan2018}, crumpled thin sheets and elastic foams \cite{PhysRevLettLahini}, systems exhibiting echoes \cite{echoes}, sheared  colloidal suspensions \cite{PhysRevLett.107.010603,PhysRevE.88.032306,PhysRevLettPaulsen}, glasses and related model systems \cite{PREFiocco,MemFiocco,JPCMFiocco}, and shaken granular systems \cite{shakengrains,Bandi2017d,Royer2015g}, to name a few examples. This list of largely condensed matter examples does not include the large array of biological contexts in which memory formation is important and interesting, such as neuronal, genetic, epigenetic, immunological {\it etc.} memories, but some approaches to modeling such memories \cite{hopfield,amit} have been developed with input from theories of disordered spin models, and in turn, such approaches inform some recent work on self assembly and design of functional materials \cite{murugan2015multifarious,MuruganJSP,Rocks,yan2017architecture,tlusty}. 

In this paper, we address memory effects in two broadly related systems, namely athermally sheared glasses and non-Brownian colloidal suspensions. Non-Brownian suspensions, when subjected to large amplitude oscillatory shear \cite{naturePine} show a transition from an {\it absorbing state} at low amplitudes of shear (wherein, particles cease to move when observed stroboscopically, {\it i. e.} at the end of each cycle,  after a transient) to a diffusing state at large amplitudes. The threshold or critical amplitude $\gamma_c$ displays features akin to a continuous phase transition, with diverging time scales to reach steady states, and a continuous rise of the fraction of active particles (defined as particles that move during a cycle, which serves as an order parameter). These features are realised through a simple model \cite{natphysCorte} in which pairs of particles that overlap when subjected to shear deformation are given random displacements or {\it kicks} after they are returned to their undeformed positions, and the process is repeated for each cycle. Memory effects were studied in this model by Keim {\it et al}  \cite{PhysRevLett.107.010603,PhysRevE.88.032306}. After a number of cycles of shear deformation with a fixed amplitude $\gamma_1 < \gamma_c$, the system reaches an  absorbing state. When this  {\it training} process of repeatedly shearing by $\gamma_1$ is complete ({\it i. e.} when particles cease to move), a shear cycle  with any $\gamma<\gamma_1$ results in no rearrangements of particles since a larger amplitude shear includes smaller amplitudes within its cycle. On the other hand, a deformation by an amplitude bigger than $\gamma_1$ will result in particle rearrangements.  As a result, this procedure encodes a memory, which can be read by performing shear deformation cycles with increasing amplitude and measuring the fraction of particles which are displaced as a function of amplitude. The fraction of particles that move is zero below the training amplitude, and becomes finite for amplitudes beyond $\gamma_1$. When 
the training phase involves cycles of more than one amplitude, the system can encode multiple memories transiently, but when
the number of cycles increases and the system reaches a steady state, memory of all but the highest amplitude are lost. Interestingly, addition of noise during the training cycles induces the memories of multiple training amplitudes to be retained. Subjecting the system to deformation by an amplitude larger than the largest training erases the memories, but gradually. 
These features of memory have also been realised experimentally in sheared non-Brownian suspensions \cite{PhysRevLettPaulsen}. 

Cyclically sheared amorphous solids (glasses) under athermal conditions \cite{PREFiocco,PKetal,Regev2013d,Priezjev2013c,Kawasaki2016a} reveal a transition, associated with yielding behaviour, that bears resemblance to the absorbing to diffusive transition in athermal suspensions. At amplitudes of shearing below a threshold value, the model amorphous solids studied computationally reach stroboscopically invariant states, whereas they reach diffusive states above the threshold. The threshold strain amplitude is characterised by diverging times to reach the steady state. Despite such similarity with athermal suspensions, there are significant differences, since a sheared amorphous solid never reaches a state where the particles do not interact with each other. It is thus interesting to consider the nature of memory effects in amorphous solids, which was addressed in Ref. \cite{MemFiocco}. It was found that cyclically deformed amorphous solids also show memory similar to athermal suspensions, but with key differences. In deformed glasses, a read cycle leaves the system unperturbed only at the training amplitude in the case of single memory, and the lowest amplitude of shear in the case of multiple memories. The origin of this behaviour was analysed in the case of single memories in \cite{MemFiocco}. It was shown that in the steady state reached after training, the system reaches the same configuration at the end of each cycle, but it does so at the end of a sequence of transitions between local energy minima, or {\it inherent structures}. A read cycle of any amplitude other than the training amplitude will disrupt this cycle of transitions, and will lead to a measurable signature. In the work described, the mean squared displacement with respect to the trained configuration was used as the measurement. 

The work in Ref. \cite{MemFiocco} raises a number of obvious questions which we address in the present paper.  Fiocco \emph{et. al.} \cite{MemFiocco}  studied memory effects in the absorbing state at a single training amplitude (or a single pair of amplitudes). We study memory effects in absorbing state with many different amplitudes and address how the memory effects ({\it e. g.} their strength) depend upon the amplitude of deformation below $\gamma_c$, and we study whether memory effects are possible above  $\gamma_c$. Fiocco \emph{et. al.} used the simplest possible way to read off the memory -- since the investigation was {\it in silico},  copies of the trained system were made, and each copy was independently subjected to a different, single, read cycle with a different strain amplitude. We refer to this protocol as a {\it parallel read}. Such a procedure is, of course, not available for experimental investigation, wherein the read cycles must be applied sequentially. We thus address whether the memory effects seen earlier are reproduced also with a {\it sequential read} protocol. We consider different measurements, namely, measuring mean squared displacements with respect to the final configuration from the previous read cycle instead of the trained configuration, and also the computation of the fraction of active particles. We consider whether the previous results concerning multiple memories can be extended beyond two memories, and investigate further whether such memories are persistent or transient. We also consider the conditions under which memories are erased. Finally, we consider structural signatures of training by considering x, z dependent pair correlations, following previous work on athermal suspensions \cite{PhysRevE.88.032306}.

As mentioned before, the differences in the memory signatures in athermal suspensions and glasses has been rationalised by the presence in the latter case of a non-trivial energy landscape, and transitions of the trained system between energy minima during a cycle of shear, even after reaching a stroboscopically invariant state. In the case of athermal suspensions and models thereof, in the steady state after full training, the system undergoes cyclic shear without any of the particles colliding (or interacting) with the other particles, whereas in a glass, particles always have finite interactions among them, leading to a non-trivial energy landscape that is traversed by the system. In probing this distinction further, we consider a different model of sheared athermal assemblies of particles, namely, soft sphere assemblies at densities below the jamming density, that are subjected to cyclic deformation under athermal conditions. In such systems, in addition to the absorbing state wherein spheres do not interact with each other any longer and the diffusive or active regime wherein they do, a third, intermediate state has been identified \cite{PhysRevESchreck}, where the sphere coordinates  are stroboscopically invariant, but spheres undergo collisions during the strain cycling. This state has been termed {\it loop reversible}. We investigate how the memory effects may be different in the loop reversible state as compared to the absorbing state (also referred to as point reversible) and the diffusive states. We show that in the loop reversible state, memory effects very similar to those in glasses are observed, thereby indicating that the distinction between the earlier studied cases of suspensions and glasses lies in the presence of absence of non-trivial  displacements during cyclic deformation, rather than the presence or absence of a non-trivial landscape. 

The rest of the article is organised as follows: In section \ref{secII}, we describe the two models which we study, and provide the various definitions and descriptions of protocols used. In section \ref{secIII}, we describe our results for the model glass, and in section \ref{secIV} results for the soft sphere system. Finally, in section \ref{secV}, we summarise our results and conclude with a discussion of their implications. 

\section{Models and Definitions}  \label{secII}

We describe below the two model systems we study in this work, and provide details of the investigations we carry out computationally. 

\paragraph{The Kob-Andersen Binary mixture ($A_{80}B_{20}$)  with Lennard-Jones interactions between particles (BMLJ)} \cite{PhysRevKob} is a model   glass former that has been extensively investigated. The interaction potential, with a quadratic cut-off, is given by 
\begin{equation}V_{\alpha \beta}(r) = \begin{cases} 4\epsilon _{\alpha\beta} [(\frac{\sigma_{\alpha \beta}}{r})^{12}-(\frac{\sigma_{\alpha \beta}}{r})^{6}]- \\ 4\epsilon _{\alpha\beta}(c_0+c_2(\frac{r}{\sigma_{\alpha \beta}})^2), r_{\alpha \beta}$ $ \leq  r_{c, \alpha \beta } \\
0 ,     \quad \quad \quad \quad \quad \quad \quad \quad  r_{\alpha \beta } > r_{c, \alpha \beta }
\end{cases} 
\end{equation} 
where $\alpha ,\beta$ $\in$ (A,B),  $\epsilon_{AB}/\epsilon_{AA}  = \epsilon_{BA}/\epsilon_{AA} = 1.5, \epsilon_{BB}/\epsilon_{AA}  = 0.5 $, and 
$\sigma_{AB}/\sigma_{AA} = \sigma_{BA}/\sigma_{AA} = 0.8 $, $\sigma_{BB}/\sigma_{AA} = 0.88 $. The interaction potential has cut off, $r_{c, \alpha \beta }$ = $2.5\sigma_{\alpha \beta}$.  We report results in reduced units, with units of length, energy and time scales being $\sigma_{AA}$, $\epsilon_{AA}$  and  $\sqrt \frac{\sigma^2_{AA}m_{AA}}{\epsilon_{AA}} $ respectively. We simulate BMLJ samples consisting of $N = 4000$ particles.  The system, at fixed number density ($N/V$, $V$ being the volume) $\rho = 1.2$  is equilibrated at reduced temperature $T=0.466$ {\it via} a constant temperature molecular dynamics simulation. All the simulations reported here are performed in LAMMPS \cite{plimpton1995fast}. 

\paragraph{Soft Sphere binary mixture (SS)} is also used as a model glass former and in studies of jamming \cite{PhysRevLettsoft}. The interaction potential is given by:
\begin{equation}
V_{ij} = \epsilon_{ij}\left (1-\frac{r_{ij}}{\sigma_{ij}}\right)^2,
\end{equation}
where $i,j$ $\in$ (A,B), indicate the type of particle. The two types of particle differ in their size, with $\sigma_{BB} =1.4 \sigma_{AA}$, but with the interaction strengths being the same for all pairs. In reporting results for this system, we use reduced units, with units of length, energy and time scales being $\sigma_{AA}$, $\epsilon_{AA}$  and  $\sqrt \frac{\sigma^2_{AA}m_{AA}}{\epsilon_{AA}} $ respectively.
 We simulate 50:50 soft sphere mixtures consisting of 2000 particles, at packing fraction $\phi = 0.61$  (where  $\phi$ for the binary mixture considered is related to the number density, $\rho$ by  $\phi =  \frac{\pi}{6} (x_A \sigma^3_{AA} + x_B\sigma^3_{BB}) \rho$ where $x_A, x_B$ are the fractions of  $A$,$B$ type of particles, each equal to $0.5$ in this case). The initial configurations are obtained from Monte Carlo simulations of  hard sphere mixtures of the same size ratio, equilibrated  at packing fraction $\phi = 0.363$. The higher density configurations are obtained starting from these initial configurations by performing 
a fast initial compression of the hard sphere system using a Monte Carlo simulation till the desired density is reached.
 
\paragraph{Training Protocols:}
Configurations taken from the equilibrated liquid are subjected to energy minimization using the conjugate-gradient algorithm  \cite{sastry1998signatures}, to obtain sets of local energy minimum structures which are termed as {\it inherent structures}. 
The inherent structures are then subjected to cyclic shear deformation using the  Athermal Quasi-Static (AQS) procedure, consisting of two steps: (i) Particles are displaced by applying an affine transformation, $x{\prime}= x+ d\gamma~ z$, where $d\gamma$ is the strain increment in the $xz$ plane, with $y$ and $z$ coordinates unaltered. Shear strain $\gamma$ is incremented by small strain steps ($d\gamma=2 \times 10^{-4}$ for the BMLJ system and $d\gamma= 10^{-3}$ for soft sphere system). (ii) The energy of the deformed configuration is minimized, subject to  Lees- Edwards periodic boundary conditions \cite{LeesEdwards}, which are appropriate for shear deformed simulated systems.   These two steps, which closely approximate quasi-static, athermal, deformation,  are repeated many times to produce configurations with any desired strain $\gamma$. 
Samples are subjected to oscillatory shear deformation  (0 $\rightarrow$ $\gamma_1$ $\rightarrow$  0 $\rightarrow$ $-\gamma_1$ $\rightarrow$ 0) at fixed amplitude $\gamma_1$ repeatedly till they reach a steady state. This procedure is referred to as  {\it training} the samples. The BMLJ system is trained at five different amplitudes, $\gamma_{train}= 0.02, 0.03, 0.06$ (which are below $\gamma_c$, the yielding strain  $\gamma_c$ \cite{PKetal}) and $\gamma_{trained} =0.09, 0.11$  (which are above $\gamma_c$). When a single training amplitude is applied, the results shown are averaged over $30$ independent samples. When multiple training strain amplitudes are applied, the data shown are averaged over 50 independent samples. The soft sphere system is trained at two different amplitudes $\gamma_1= 0.03, 0.12$. For both single memory and multiple memory cases, the data shown here are averaged over 10 independent samples. 
 
\paragraph{Reading Protocols:}  
 After training,  in the read procedure we refer to as ``Parallel read", identical copies of the samples are subjected to a single cycle of shear deformation each, with such amplitudes covering the range of strain amplitudes from $0$ to $0.13$.   We also consider a second read protocol which we refer to as ``Sequential read". In this case, after training, we apply single cycles of shear for an increasing sequence of amplitudes, using the final configuration after a cycle at one amplitude as the starting configuration for the cycle at the next (higher) amplitude. As the measurement that is used to reveal the presence or absence of memory of the training, we use the mean squared displacement  (MSD) of particles in configurations at the end of a full cycle of deformation, with respect to the reference configuration.  We compute the MSD either with respect to the final configuration of the training phase (or the initial configuration for the read protocol), which we denote by $MSD_0$, or compute the MSD with respect to the final configuration of the previous read cycle, in which case (for the $i^{th}$ cycle) we denote it as $MSD(i,i-1)$.  $MSD_0$ is defined as 
 
\begin{eqnarray}
MSD_0 = \frac{1}{N} \sum_k ({\bf r}_k (read) - {\bf r}_k (trained))^2
\end{eqnarray}
where ${\bf r}_k (trained)$ is the position of particle $k$ in the trained configuration, and ${\bf r}_k (read)$ is the position of particle $k$ after the relevant read cycle. $MSD(i,i-1)$  is defined as 
\begin{eqnarray}
MSD(i,i-1) = \frac{1}{N} \sum_k ({\bf r}_k (i)- r_k (i-1))^2\rangle
\end{eqnarray} 
where $i$ and $i-1$ are cycle indices. 

 We also compute the fraction of active particles ($f_{active}$) to characterize the memory. We define $f_{active}$ as the fraction of particles that move larger than $0.1 \sigma_{AA}$ during a read cycle, following \cite{PKetal}. We use the notation $N_{cycles}$ for the number of training cycles.
  
\paragraph{Two dimensional pair correlation function:}
In order to assess the structural change resulting from cyclic deformation, we compute a  two dimensional directional pair correlation function $g(x,z)$ in the shear plane $xz$, which is defined as: 
\begin{equation} 
\begin{split}
g(x,z)  =  \frac{1}{N \rho} \times  ~~~~~~~~~~~~~~~~~~~~~~~~~~~~~~~~~~~~~~~~~~~~~~~~~~~~~~~~ \\  
\left \langle \sum_{i=1}^{N-1}\sum_{j=i+1}^{N}\delta(x-(x_i-x_j)) \delta(z- (z_i-z_j))  \theta(a - |y_i-y_j|) \right \rangle  
\end{split}
\end{equation}
 where $\langle ..\rangle$ implies averaging over independent samples. $x_i,  y_i,  z_i$ are the particle coordinates. 
Since we compute a two dimensional correlation function in a three dimensional system, we consider pairs of particles with are in the same (shear) plane, by demanding that their vertical ($y$) separations do not exceed a specified value, $a = 0.2 \sigma_{AA}$. This is enforced by the Heaviside function $ \theta(a - |y_i-y_j|)$. In practice, we divide the simulation box into slabs of fixed width $a$ along the $y$ direction and compute $g(x,z)$ 
for pairs of particles within each slab, averaging over all the slabs. The data shown are averaged over $30$ independent samples.

\section{Results: The BMLJ system}\label{secIII}
\subsection{Single Memory}
First, we study memory effects in the absorbing states ($\gamma < \gamma_c = 0.08$) prepared with different amplitudes of cyclic shear deformation. The samples are trained at $\gamma_1$ = 0.02, 0.03 and 0.06. After training, parallel reading is performed on the trained samples.
\subsubsection{Parallel read}  

\begin{figure}[h!]
\centering
\includegraphics[width = 0.40\textwidth]{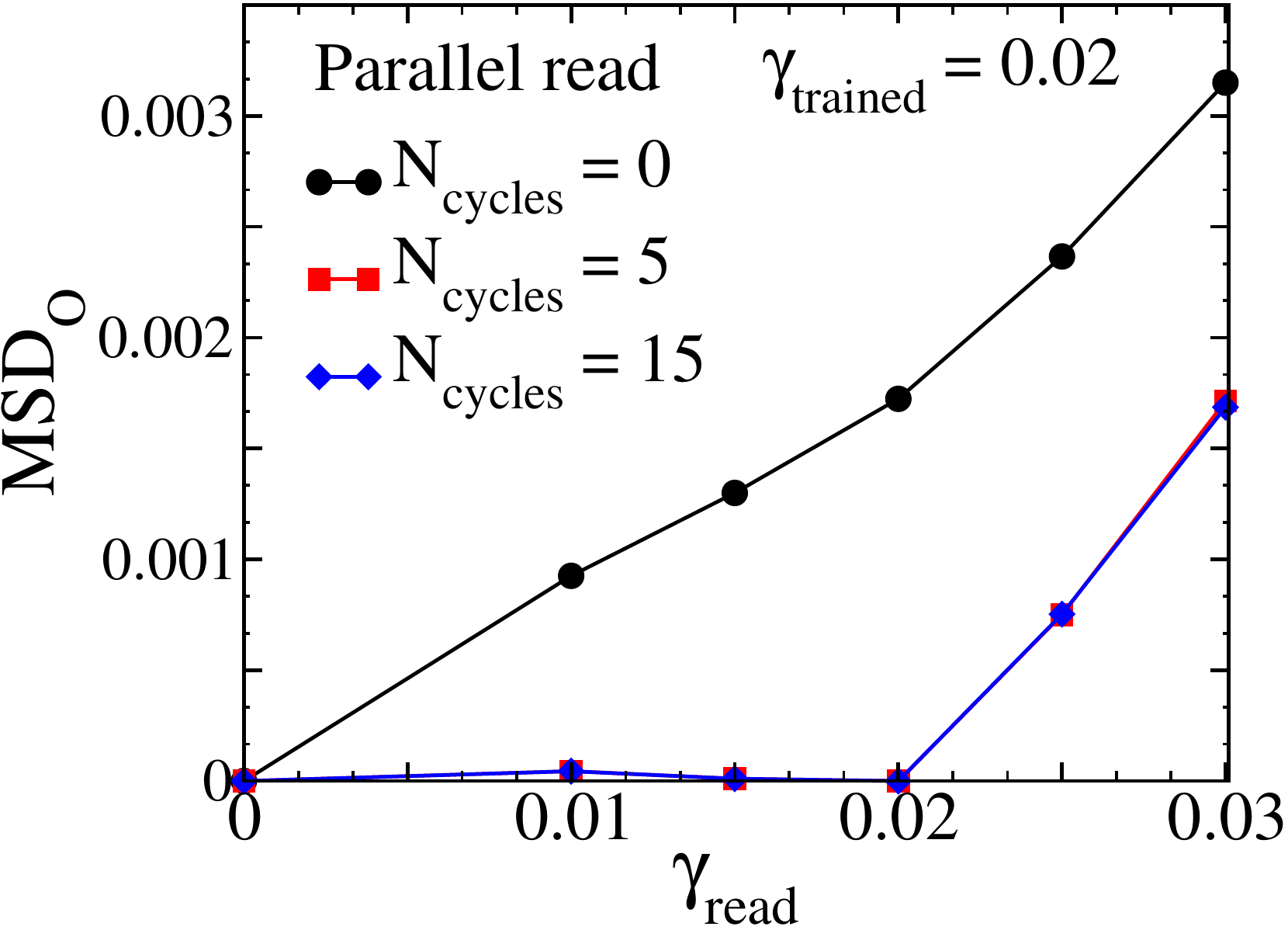}
\includegraphics[width = 0.40\textwidth]{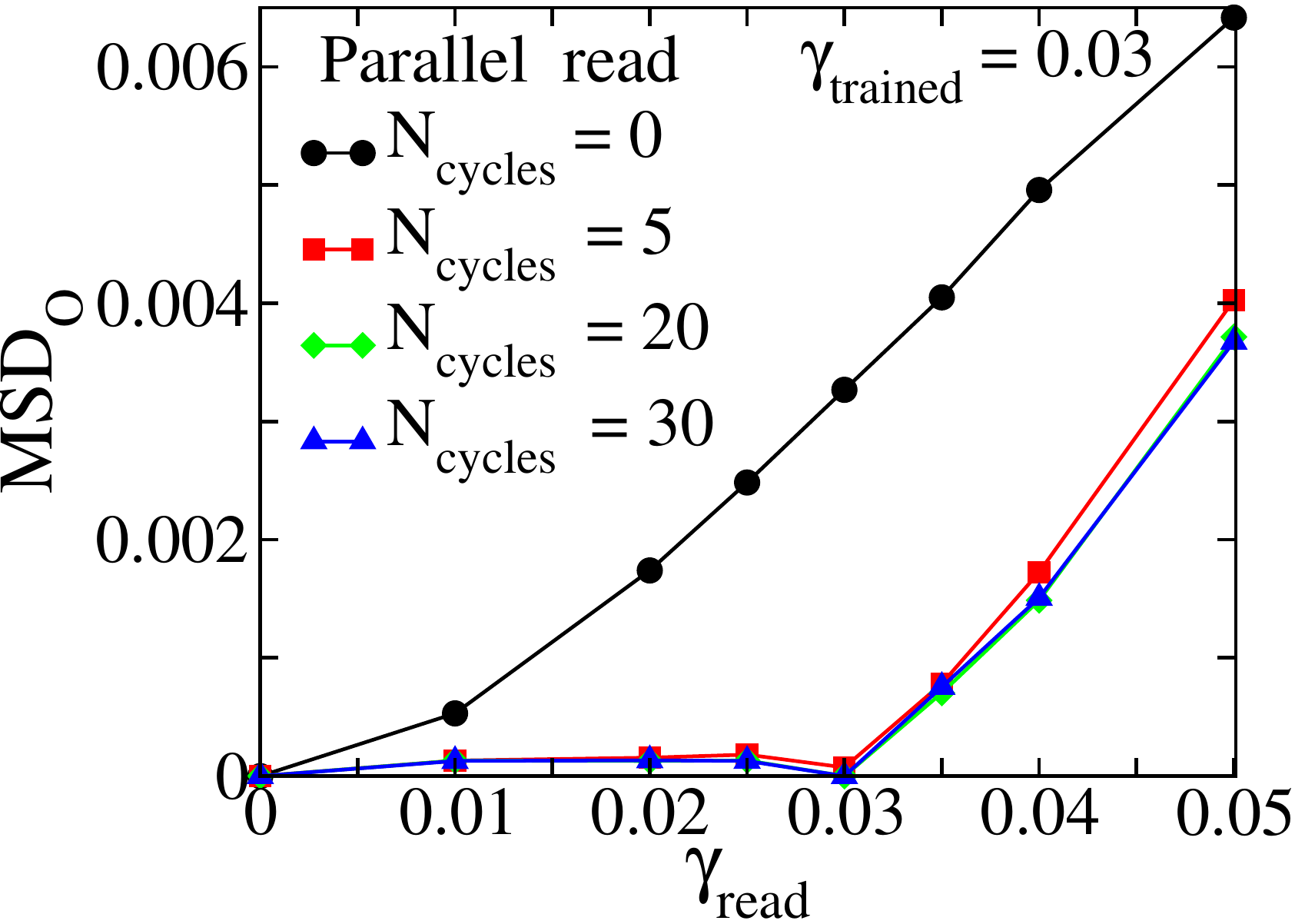}
\includegraphics[width = 0.40\textwidth]{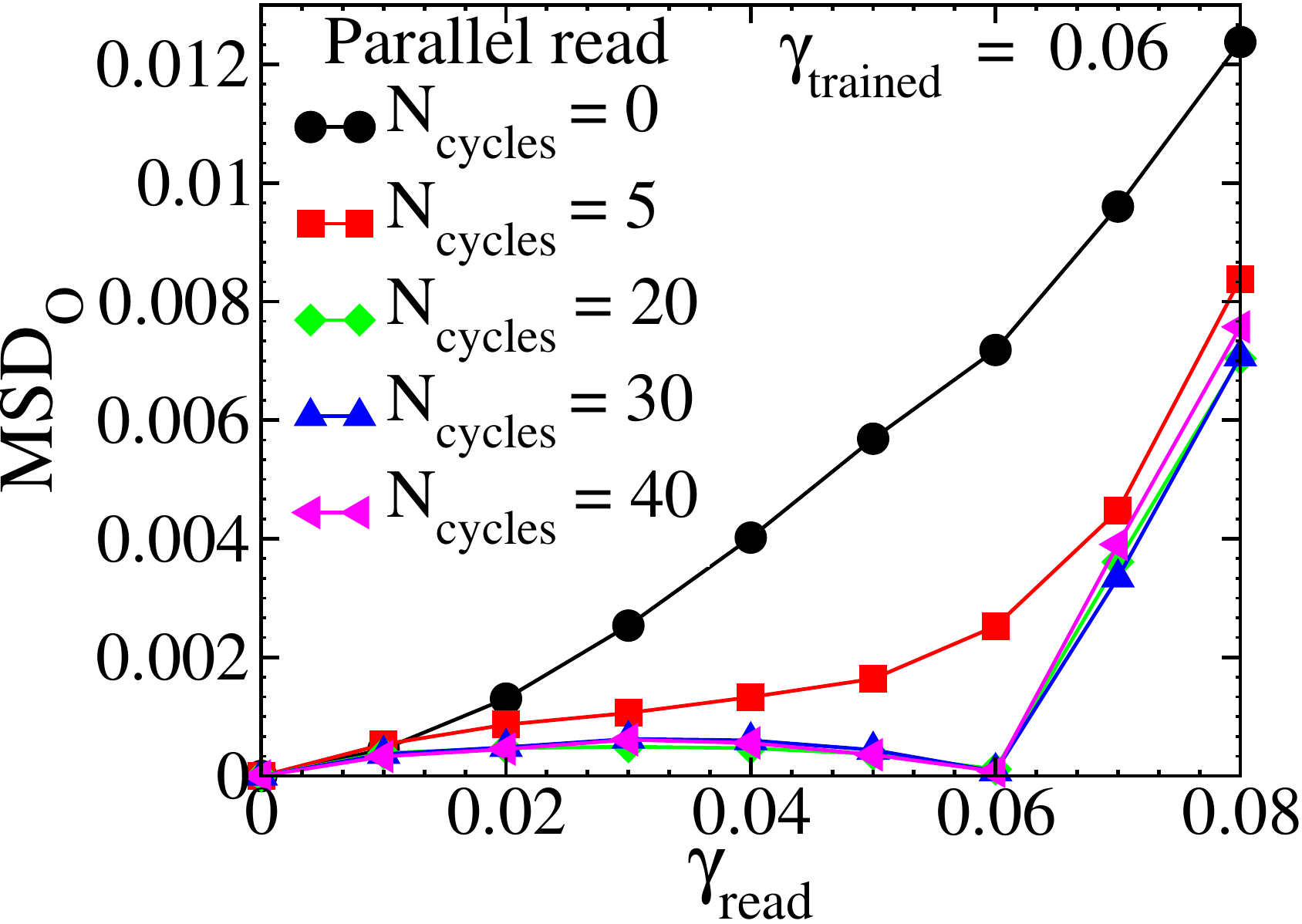}
\caption{The mean squared displacement (MSD) as a function of $\gamma_{\text read}$ for different training cycles. The system is trained at $\gamma_{trained}=0.02$ (top) and  0.03 (middle) and 0.06 (bottom). In each case, the MSD  at $\gamma_{trained}$ is either lower than other $\gamma$ values (partially trained) or zero (fully trained), constituting a memory of the training amplitude. } 
\label{fig-1} 
\end{figure}
In Fig.\ref{fig-1}, the MSD is plotted against $\gamma_{\text read}$ for different numbers of training cycles. In the untrained system ($N_{cycles} =0$), with the application of shear deformation in the read cycles, particles move by larger amounts for larger  strain values. As a result, the MSD increases with $\gamma_{\text read}$  monotonically for the untrained system. However, as the number of the training cycles is increased, the system evolves towards the absorbing state. Correspondingly, the MSD for $\gamma_{\text read}$ close to  $\gamma_{\text trained}$ is seen to decrease.
 After a significant number of training cycles, when the system has reached the  absorbing state, the MSD becomes zero at $\gamma_{\text read} = \gamma_{\text trained}$ since one more cycle of shear deformation with amplitude $\gamma_{\text trained}$ leaves the system unchanged. The MSD vs. $\gamma_{\text read}$ curve thus displays a clear signature or memory of the training deformation amplitude. If we increase the number of training cycles further, the nature of the MSD vs. $\gamma_{\text read}$ curve does not change, which is expected as the system already is in the absorbing state. As previously noted \cite{MemFiocco}, the MSD is finite not only for read amplitudes bigger than the training amplitude, but also for smaller read amplitudes, which is significantly different from the case of sheared suspensions \cite{PhysRevLett.107.010603}. This can be rationalised by the differences in the nature of reversibility in the two cases. 

To illustrate the reasons for the memory signatures observed, we consider the changes in the energy and particle positions during a read cycle. 

\paragraph{Energy changes during a read cycle:}

\begin{figure}[h!]
\centering
\includegraphics[scale=0.36]{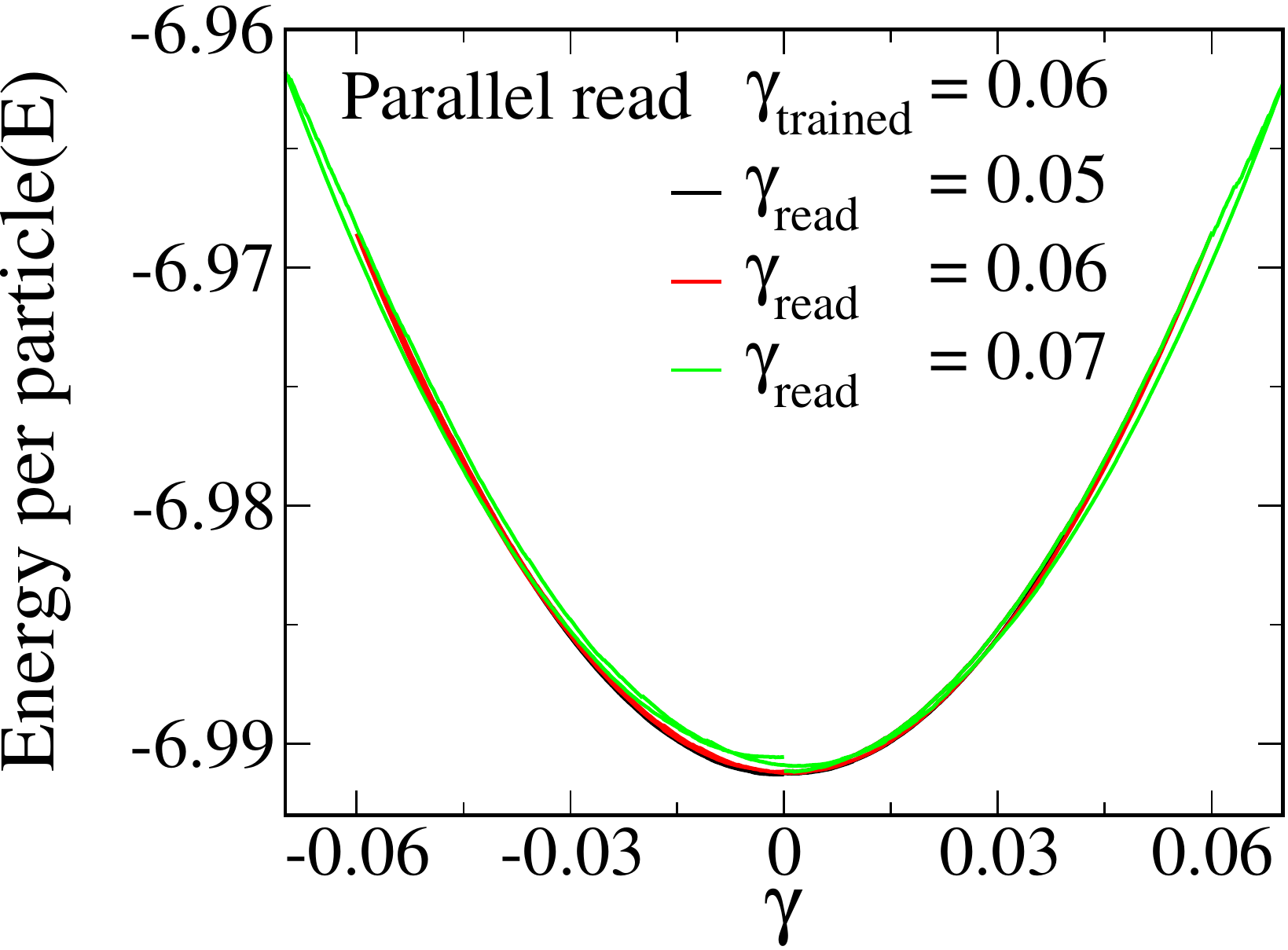}
\includegraphics[scale=0.36]{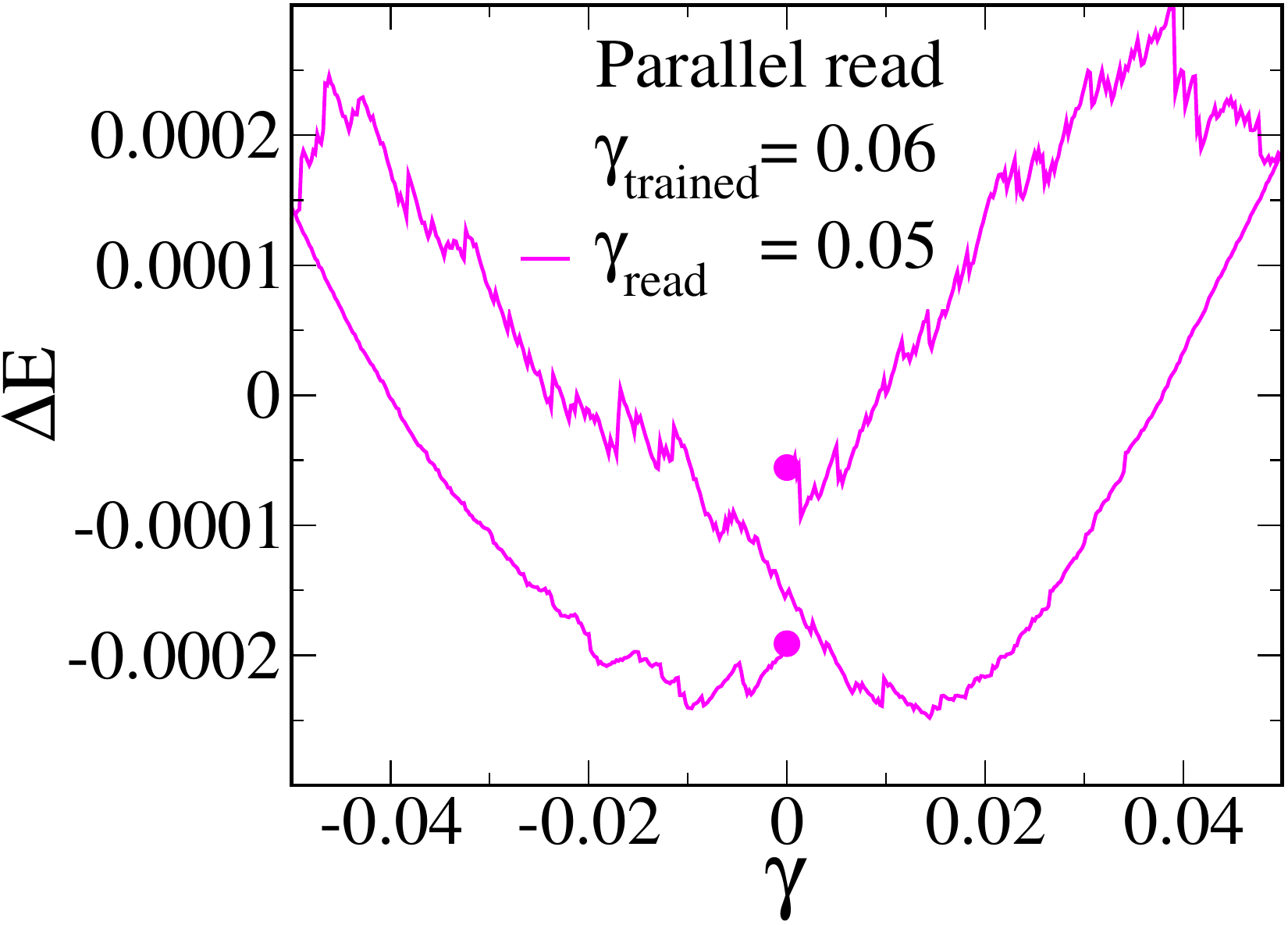}
\includegraphics[scale=0.36]{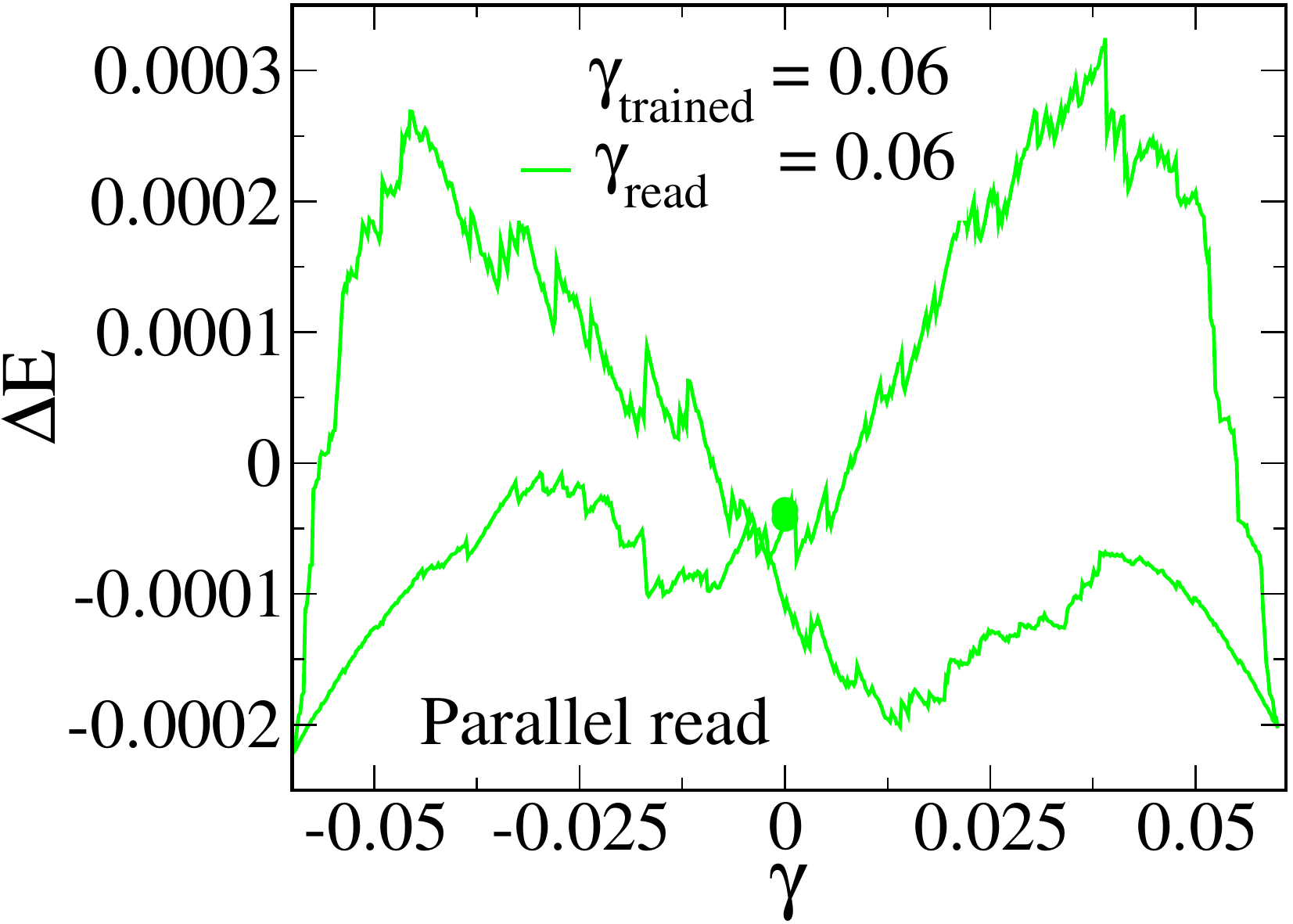}
\includegraphics[scale=0.36]{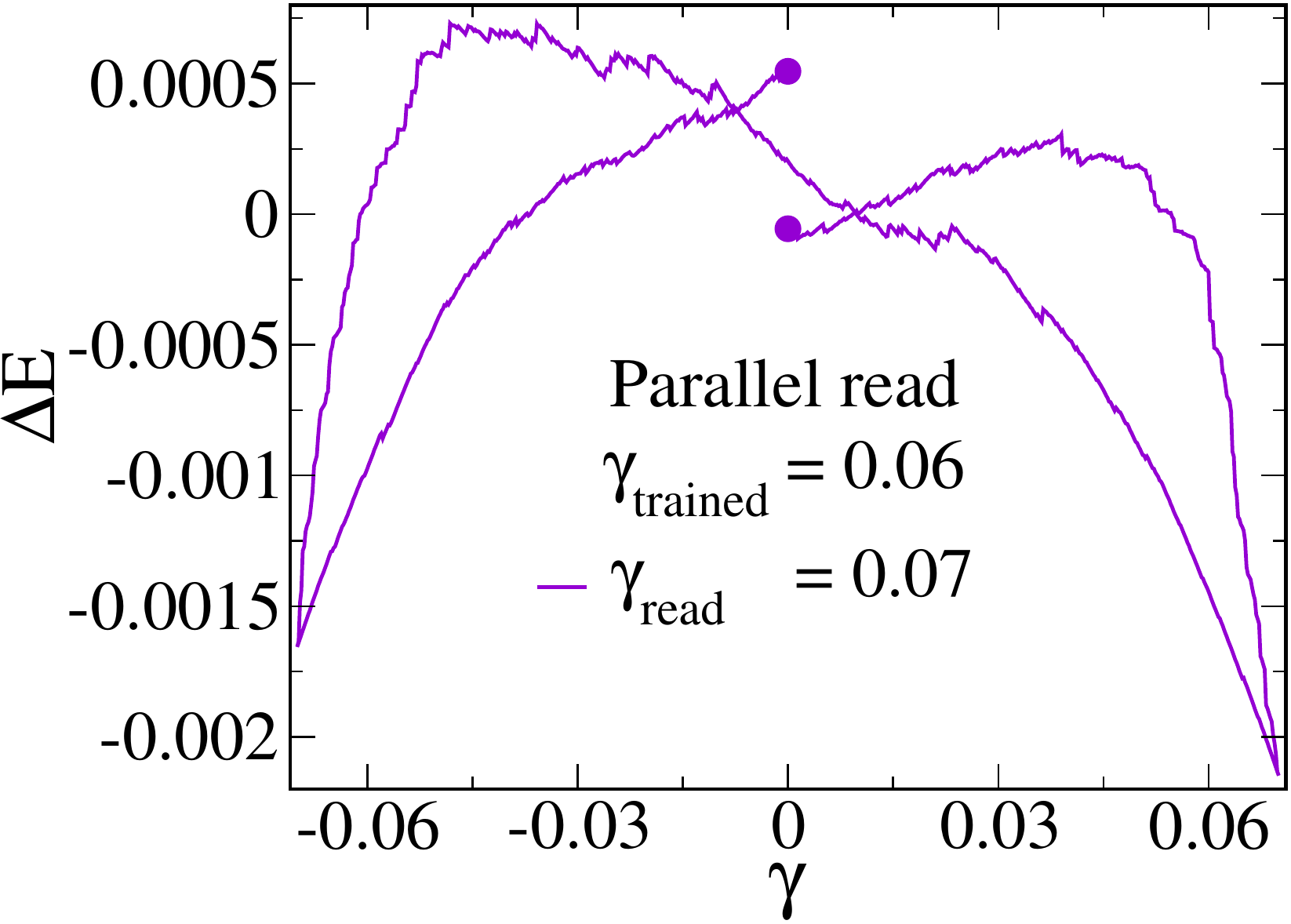}

\caption{Potential energy is plotted as a function of strain $\gamma$ during the  reading cycle. The potential energy curve is fitted to a quadratic function (top panel), and the difference $\Delta E$ obtained by subtracting the quadratic fit from the data is shown (bottom three panels) to clearly display relevant details. The system is trained at $\gamma_{trained}= 0.06$. $\Delta E$ are shown for read cycles for different amplitudes, which are indicated in the legends. While the energy values return to the initial value when  $\gamma_{\text read} =  \gamma_{\text trained}$, they do not do so for other read amplitudes.}
\label{fig-2}
\end{figure} 
The evolution of the system is investigated by measuring energy during the read cycles as a function of strain. When the system is subjected to oscillatory shear deformations, the energy will be proportional to $\gamma^2$ if the system deforms elastically. The observed energies {\it vs.} strain for each read amplitude are fitted to a quadratic function and the difference  $\Delta E$ of the data from the quadratic fits, which highlight relevant details, are studied. The $\Delta E$ curves are shown for three different read amplitudes in Fig. \ref{fig-2}. It is observed that the  $\Delta E$ curves display discontinuous jumps which correspond to plastic rearrangements of particles and correspondingly, transitions between energy minima. When  $\gamma_{read} = \gamma_{trained}$, such jumps in energy are nevertheless organized such that  the energy (and  $\Delta E$) returns to the initial value at the end of a cycle.  When $\gamma_{read} \ne \gamma_{trained}$, however, the sequence of transitions that take place do not lead to the final state being the same as the initial state, which leads to finite signatures in the energy (and the MSD) during the read cycles. 

\paragraph{Position changes during a read cycle: }
\begin{figure}[h!]
\centering
\includegraphics[width = 0.35\textwidth]{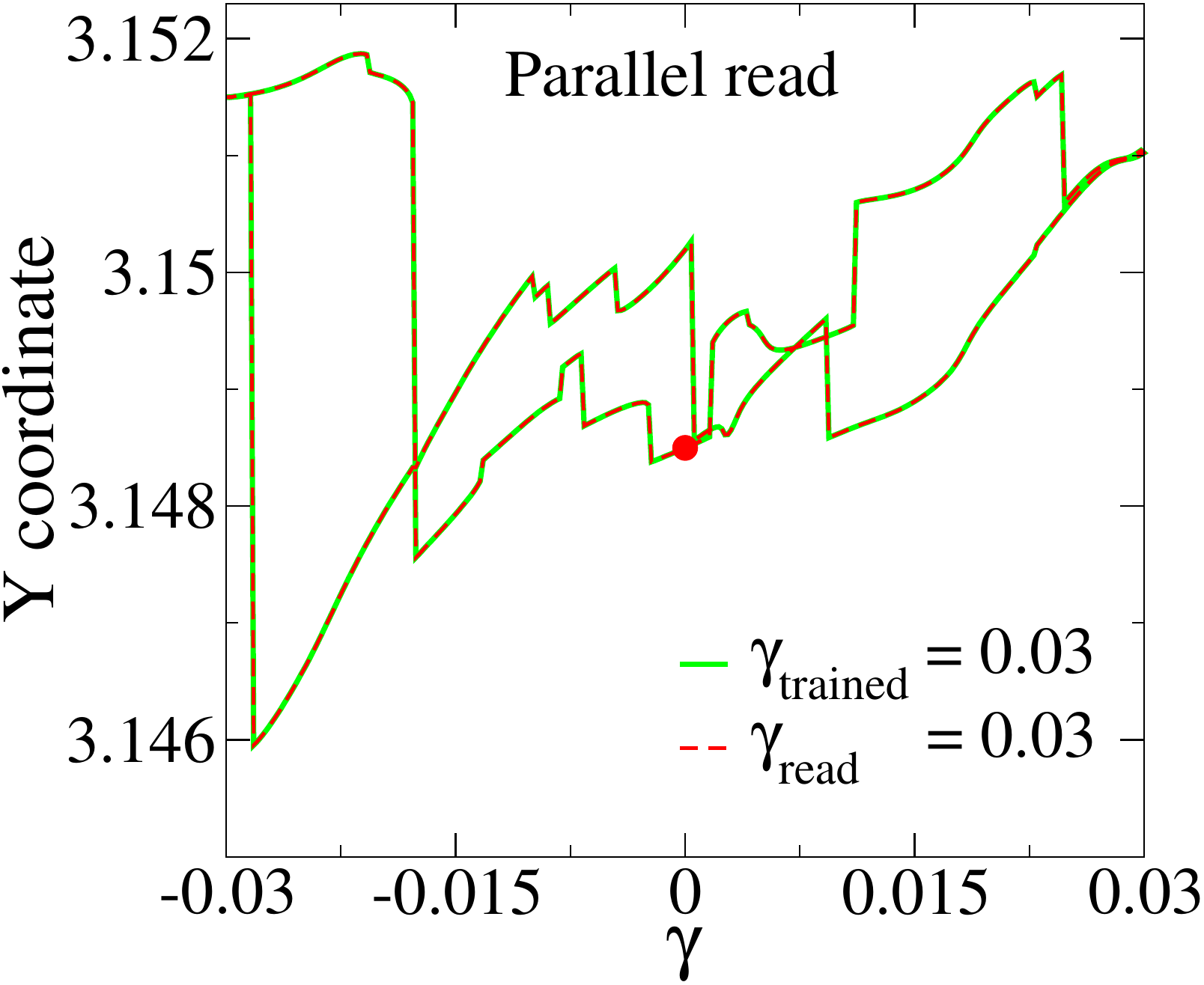}
\includegraphics[width = 0.35\textwidth]{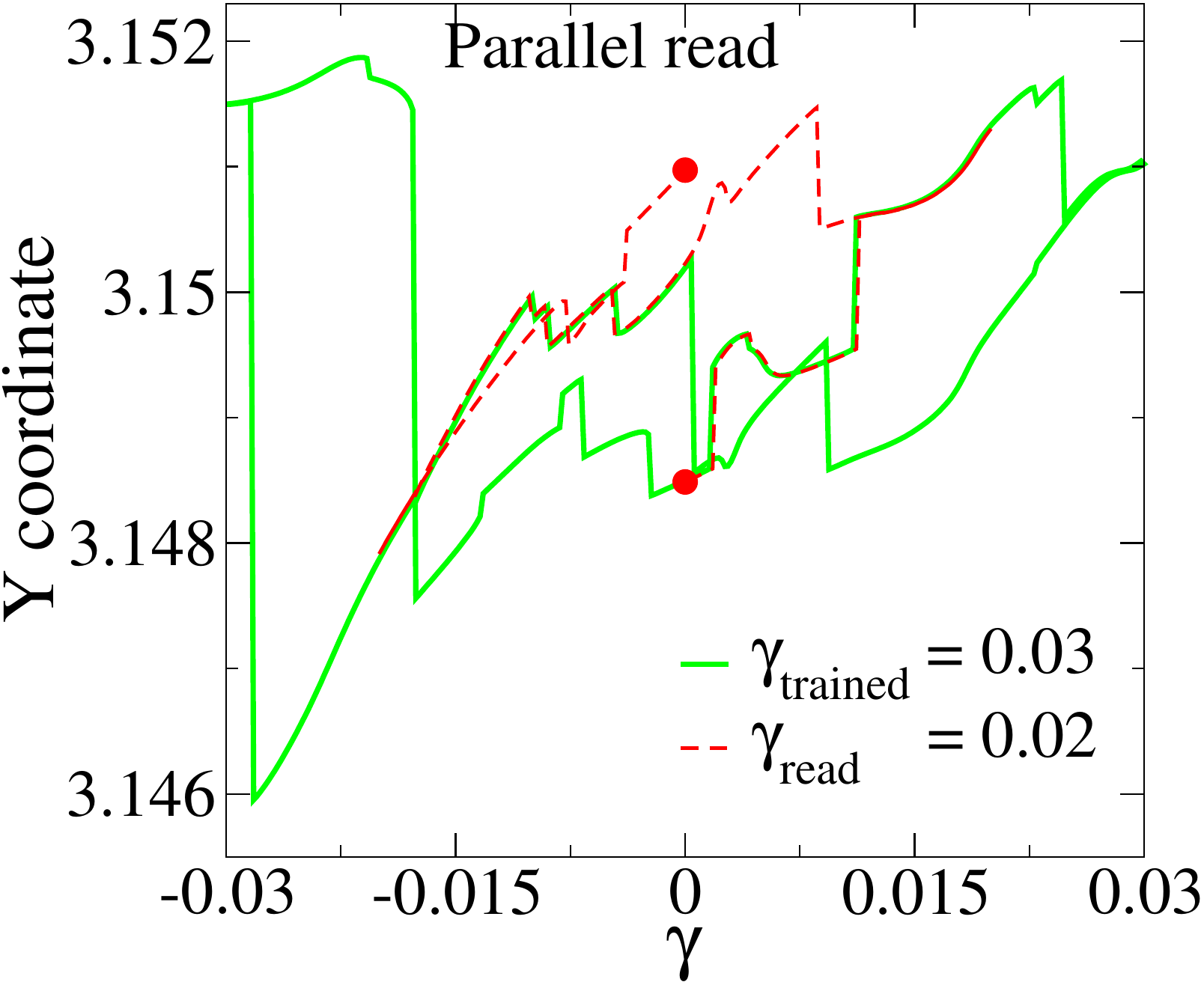}
\includegraphics[width = 0.35\textwidth]{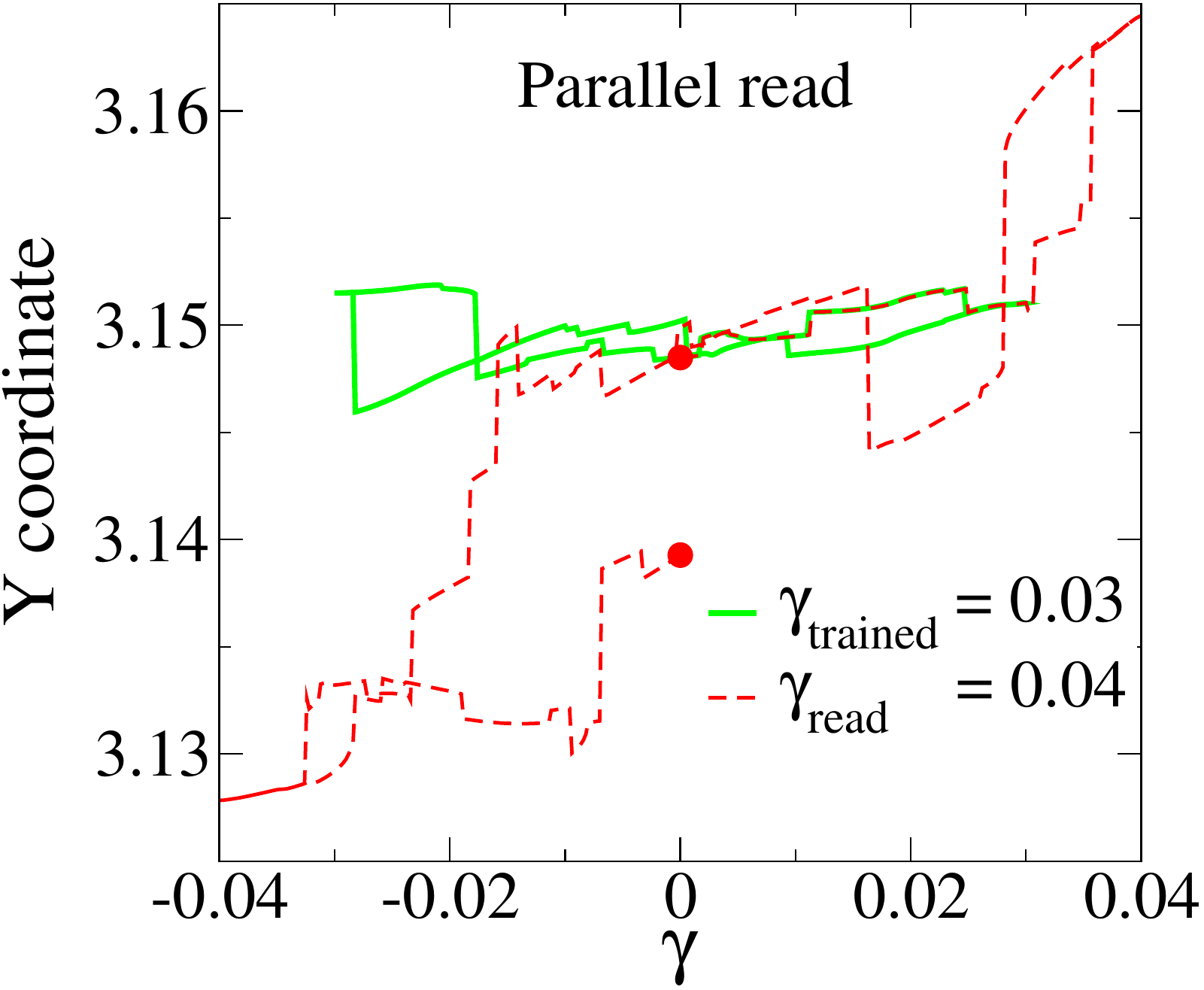}
\caption{The  $Y$ coordinate of a single particle is plotted as a function of strain $\gamma$ for $\gamma_{trained} = 0.03$, $N_{cycles} = 30$. 
Top: The particle follows the same path during the last training cycle (Cycle = 30), and the read cycle at the same amplitude, $\gamma_{read} =  \gamma_{trained} = 0.03$. Middle and bottom panel: When the trained sample is read at different amplitudes $\gamma_{read} =0.02$ and $0.04$ the particle does not retrace the same path.}
\label{fig-3}
\end{figure}
Next, we consider how the position of a particle evolves during the reading cycle, for  $\gamma_{\text trained} = 0.03$ and $N_{cycles} = 30$.  The $Y$ coordinate of a single particle is plotted as a function of strain $\gamma$ for different read amplitudes $\gamma_{\text read}$ in  Fig. \ref{fig-3}. Although the particle position changes discontinuously in all cases (corresponding to jumps between local energy minima), for  $\gamma_{\text read} =  \gamma_{\text trained} = 0.03 $, the particle position during the read cycle clearly retraces the same cycle as the last training cycle. In contrast, for  $\gamma_{\text read} = 0.02, 0.04$ the trajectory of the particle as indicated by its $Y$ coordinate departs strongly from that during the final reading cycle, and the particle does not return to the initial position at the end of the read cycle. 
 
\begin{figure}[htp]
\centering
\includegraphics[width = 0.388\textwidth]{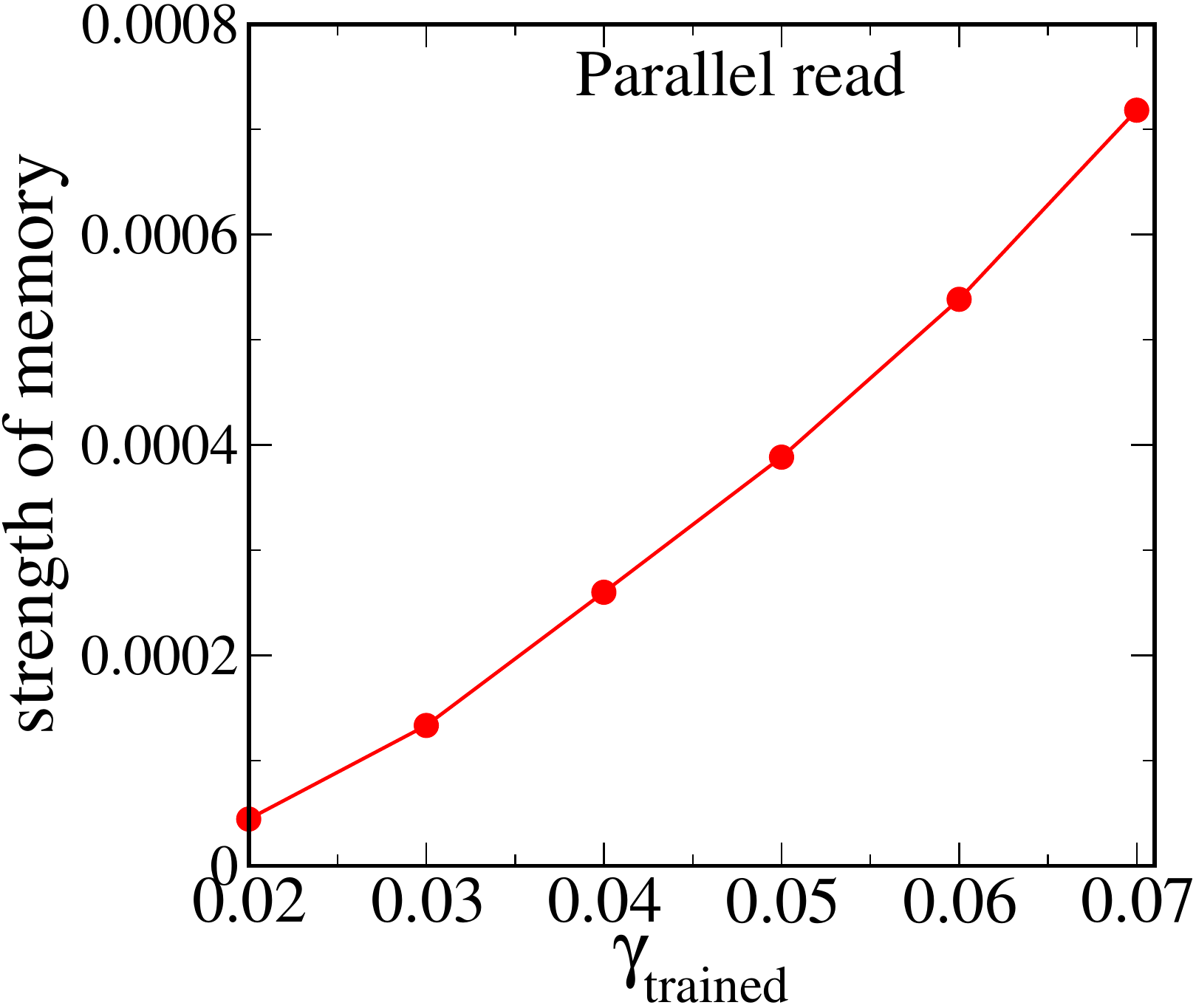}
\caption{The strength of memory is plotted as a function of $\gamma_{\text trained}$. The strength of memory increases with the increase in amplitude of training below yielding amplitude.}
\label{fig-4}
\end{figure}

\paragraph{Strength of the memory:}
We have studied memory effects with various amplitudes below the yielding strain amplitude to understand the dependence of memory behaviour on the amplitudes of training. We define and measure  the {\it strength of the memory} as a function of the training amplitude as follows.  We have observed earlier that there is a non-monotonic increase in the MSD as we increase the amplitude of $\gamma_{\text read}$ for all $\gamma_{\text read}$  which is less than or equal to $\gamma_{\text trained}$. In this regime, the MSD increases initially, but goes through a maximum and  
becomes zero at $\gamma_{\text read}$ = $\gamma_{\text trained}$. 
If the system is partially trained, the MSD at $\gamma_{\text trained}$ may be lower than neighbouring strain values but finite. We thus subtract the MSD at $\gamma_{\text read}$ = $\gamma_{\text trained}$ from the maximum of the MSD below $\gamma_{trained}$, and use it as a measure of the strength of the memory \cite{sreemayi}. The result is presented in Fig. \ref{fig-4}.

\begin{figure}[htp]
\centering
\includegraphics[width = 0.35\textwidth]{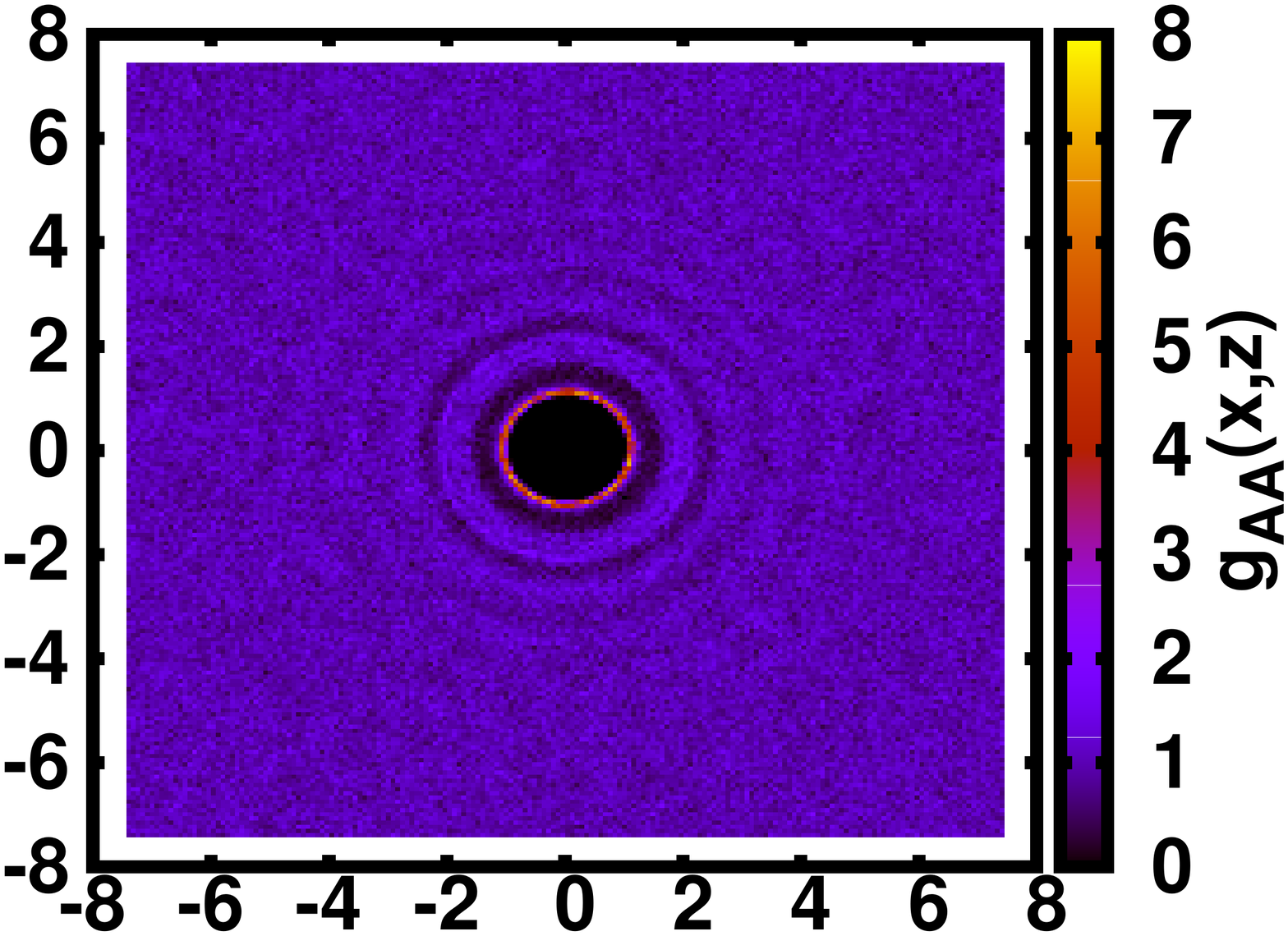}
\includegraphics[width = 0.35\textwidth]{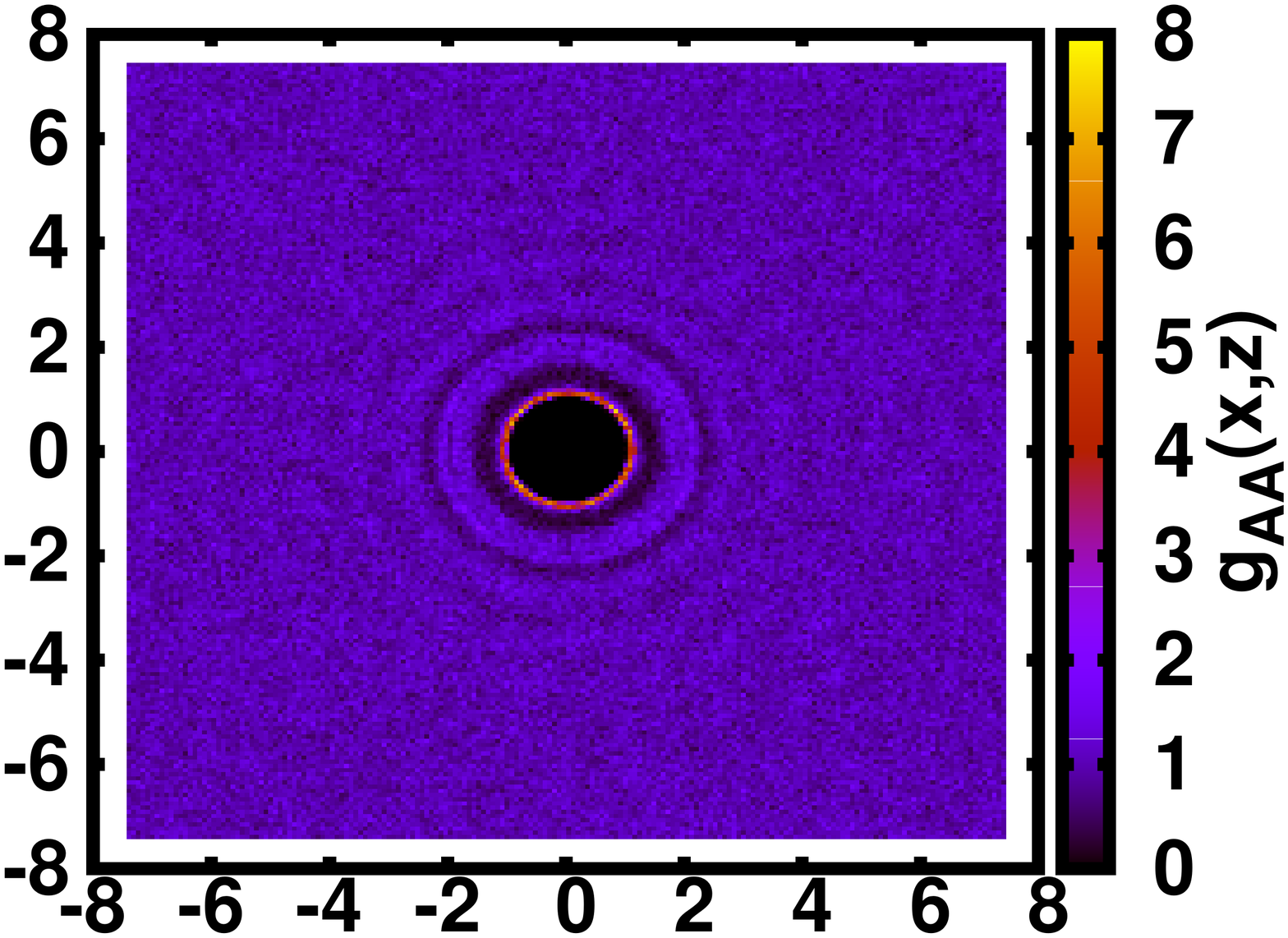}
\caption{Two dimensional pair correlation function, $g_{AA}(x,z)$ for an inherent structure quenched from the liquid (top) and a trained sample with $\gamma_{trained} = 0.06$ (bottom) of the BMLJ system. The data shown here is averaged over $30$ different samples.}
\label{fig-5}
\end{figure}

\paragraph{Structural signatures of memory:} 
In order to assess if the encoding of memory involves clear structural signatures, as in the case studied in \cite{PhysRevE.88.032306}, we compute the two dimensional pair correlation function $g(x,z)$  defined earlier. In Fig:\ref{fig-5}, we show the result for a system trained at $\gamma_{\text trained} = 0.06$, along with the  $g(x,z)$ for an inherent structure quenched from the liquid ({\it i. e.} not subjected to any shear deformation).  We do not see any significant difference between the liquid inherent structure and the trained system, surprisingly, and the correlation function of the trained system does not show any significant anisotropy as seen in \cite{PhysRevE.88.032306}. Although we cannot exclude effects too weak for our numerical estimation to detect, or other subtle effects that require alternate measures, the lack of a clearer structural signature in the case of a model glass is puzzling. 

\paragraph{Application of cyclic shear deformation with a different amplitude to a trained system:}
In the preceding sections, cyclic shear deformation with an amplitude $\gamma_{trained} (< \gamma_c$) is applied to an equilibrated samples repeatedly. 
\begin{figure}[h!]
 \centering
\includegraphics[width = 0.37\textwidth]{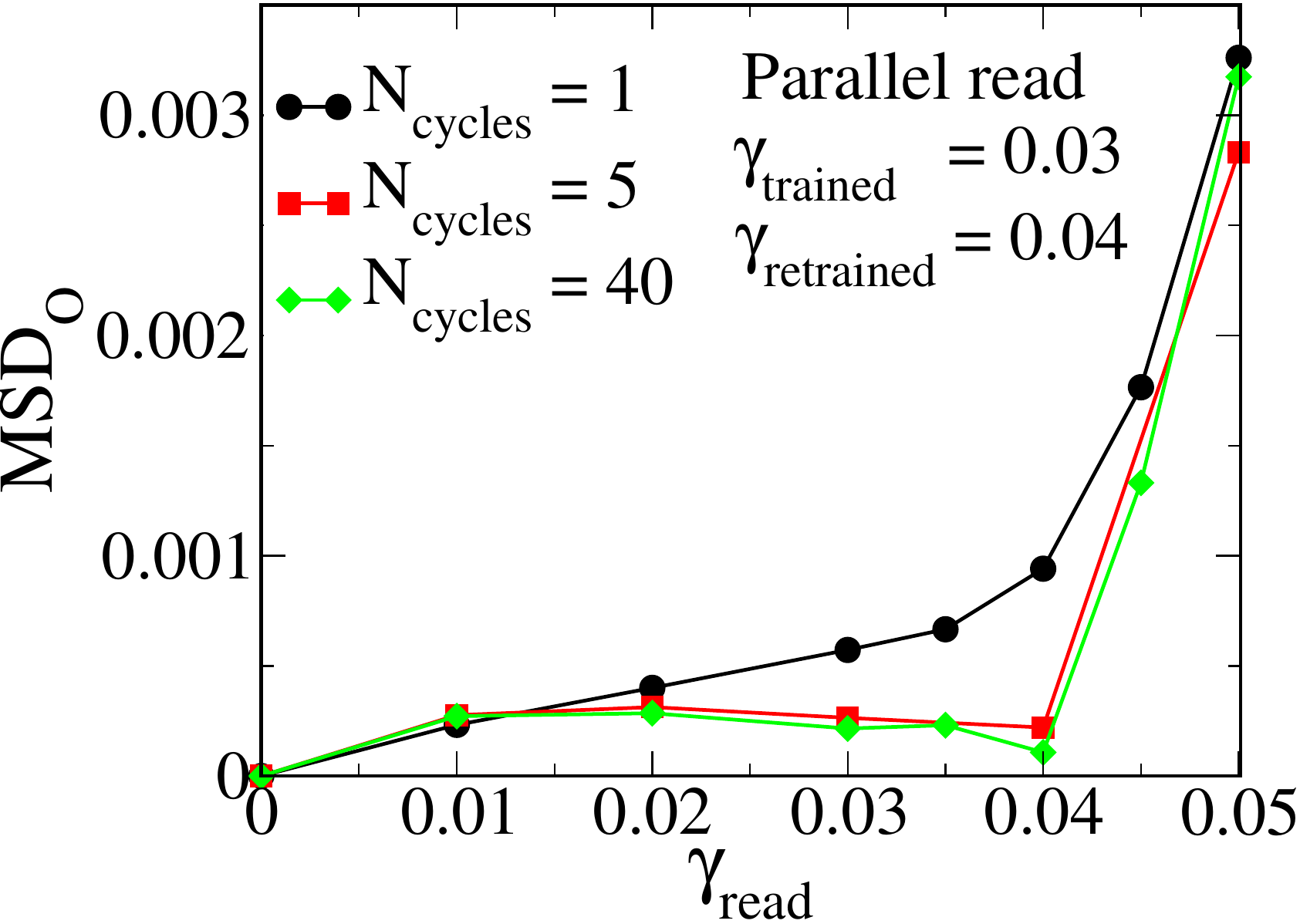}
\includegraphics[width = 0.37\textwidth]{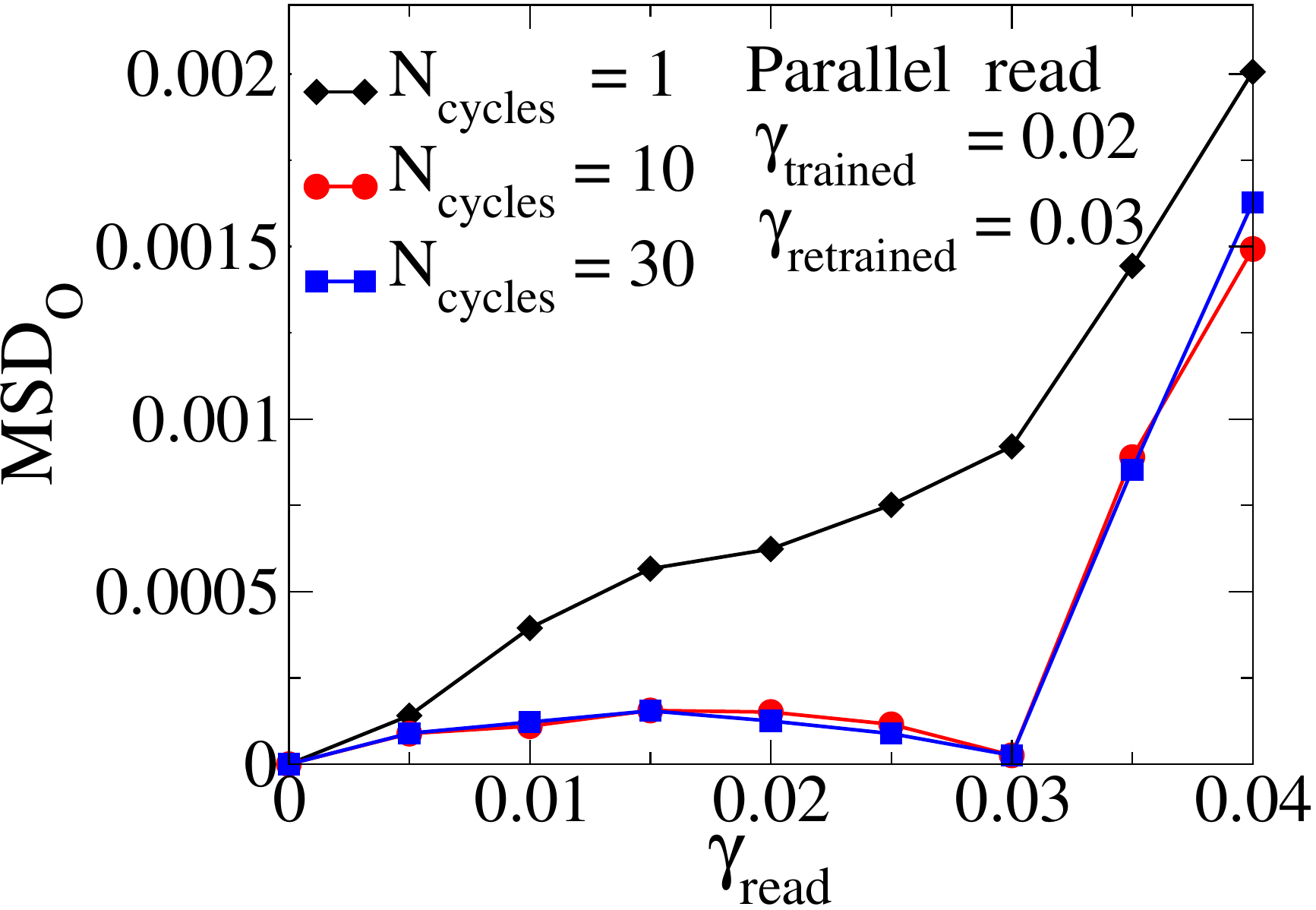}
 \caption{The MSD as a function of $\gamma_{\text read}$ during parallel reading. Top: The system is first trained at $\gamma_{trained}=0.03$ fully ($30$ cycles) and then cycles of shear deformation with amplitude $\gamma_{retrained} = 0.04$ are applied to that trained system. The MSD, even after a single cycle, does not show a memory of the training at $\gamma_{trained}=0.03$. Instead, a change of slope in the MSD is visible at $\gamma=0.04$ (black curve). After a large number of retraining cycles ($40$ cycles) with $\gamma_{retrained}=0.04$ the system shows the usual signature of memory at $\gamma=0.04$, namely a depression/vanishing of the MSD at the retrained amplitude. Bottom: The system is first trained at $\gamma_{trained}=0.02$ fully ($15$ cycles) and then cycles of shear deformation with amplitude $\gamma_{retrained} = 0.03$ are applied to that trained system. The plot does not show a vanishing of the MSD at $\gamma_{read}=0.02$ even after one cycle at the new amplitude (black curve). After a large number of retraining cycles ($30$ cycles) with $\gamma_{retrained}=0.03$, the system displays a vanishing MSD at $\gamma=0.03$.}
\label{fig-6} 
 \centering 
\end{figure}
After a large number of training cycles, the system remembers the amplitude of deformation by which it is trained. We now ask what the effect of applying shear deformation at a second "retaining" amplitude. We ask if such retraining will lead to the system "forgetting" the earlier training, or, in other words, whether the memory will be erased. We consider two cases: (1) The retraining amplitude $\gamma_{retrained}$ is greater than  $\gamma_{trained}$. (2) The retraining amplitude $\gamma_{retrained}$ is smaller than  $\gamma_{trained}$.

\noindent{1. Deformation amplitude is larger than the training amplitude. Erasure of memory:}
We consider a sample which is trained at $\gamma_{trained}$ over a large number of cycles. The trained sample is then deformed cyclically for varying numbers of cycles at  $\gamma_{retrained} > \gamma_{trained}$. We show two such cases in  Fig. \ref{fig-6}, in which we consider a configuration trained at $\gamma_{trained} = 0.03$ trained for $30$ cycles, and $\gamma_{trained} = 0.02$ trained for $15$ cycles. In the first case (Fig. \ref{fig-6}, top panel), we see that even after a single cycle at  $\gamma_{retrained} = .04$, the memory at  $\gamma_{trained} = 0.03$ is erased, in that no signature of trained at that amplitude is present. Further, after a single cycle, a depression of the MSD at $\gamma_{retrained} = .04$ is visible, which evolves with further cycles to a vanishing MSD at $\gamma_{retrained} = .04$. Thus we conclude that retraining a system at a higher amplitude erases the memory, which is consistent with previous observations \cite{PhysRevLett.107.010603,PhysRevE.88.032306} for models of colloidal suspensions. A second example with $\gamma_{trained} = .02$ and $\gamma_{retrained} = .03$ displays the same features.

\begin{figure}[h!]
 \centering
\includegraphics[width = 0.37\textwidth]{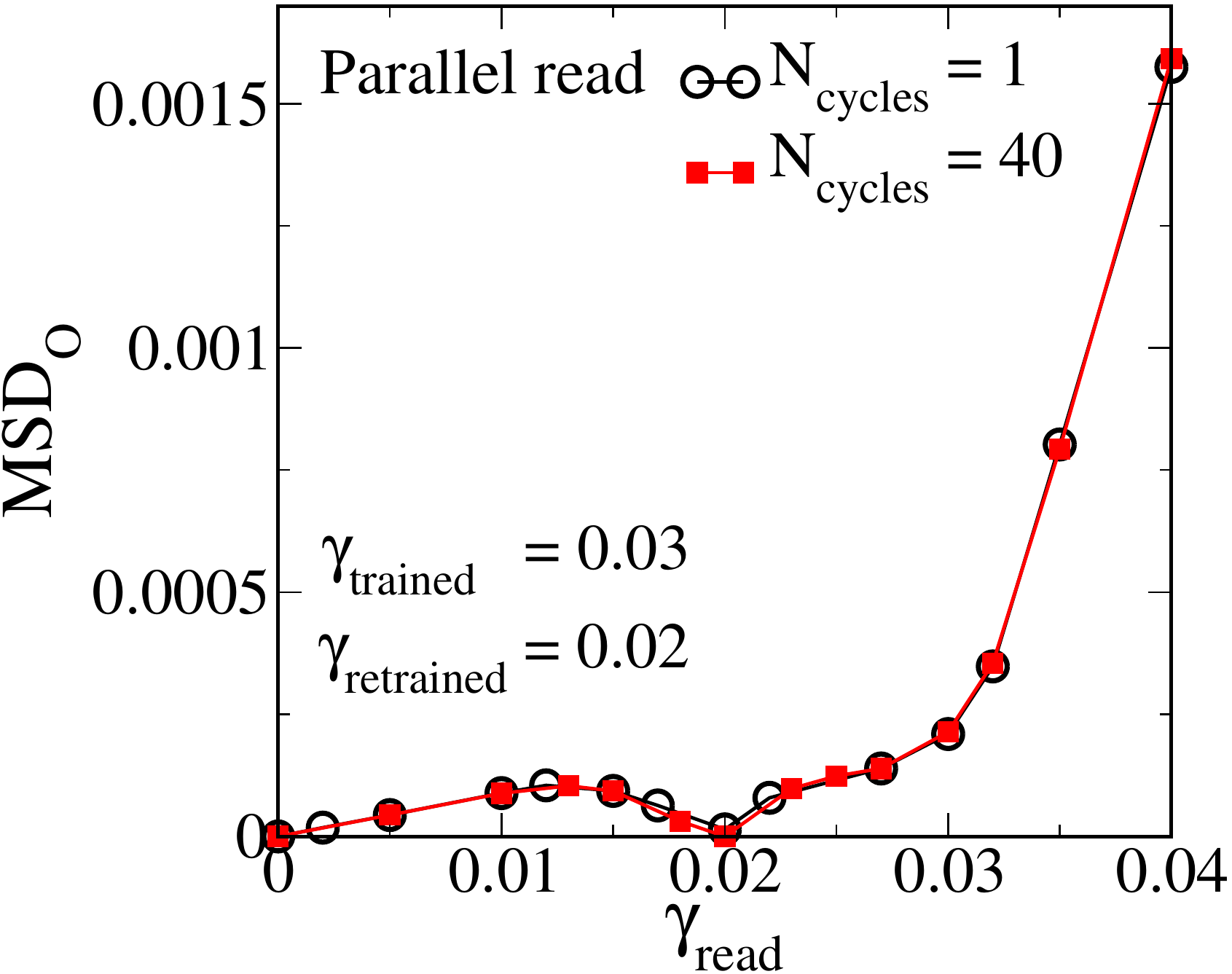}
\caption{The MSD as a function of $\gamma_{\text read}$ during parallel reading. Top: The system is first trained at $\gamma_{trained}=0.03$ fully ($30$ cycles) and then cycles of shear deformation with amplitude $\gamma_{retrained} = 0.02$ are applied to that trained system. The MSD, even after a single cycle, is close to zero at $\gamma_{read} = \gamma_{retrained}$. However, instead of increasing monotonically for larger  $\gamma_{read}$, a depression of the MSD is apparent around $\gamma_{trained}=0.03$ indicating persistence of memory of that amplitude.}
\label{fig-7} 
\end{figure}

\noindent{2. Deformation amplitude is smaller than the training amplitude:} Here we consider once again  a configuration trained at $\gamma_{trained} = 0.03$ trained for $30$ cycles, but apply retraining deformations at $\gamma_{retrained}=0.02$. The results shown in Fig. \ref{fig-7}, indicate that the memory of the new (retraining) amplitude forms even after one cycle, with  a vanishing of the MSD at that amplitude, but the memory of $\gamma_{trained} = 0.03$ is not erased. The MSD at $\gamma_{trained} = 0.03$ is not any longer zero, but is depressed, and shows a change in slope at $\gamma_{read} = \gamma_{trained}$ in a manner that is similar to the signature of multiple memories which we discuss later.

\subsubsection{Sequential read}

\begin{figure}[h!]
\centering
\includegraphics[width = 0.43\textwidth]{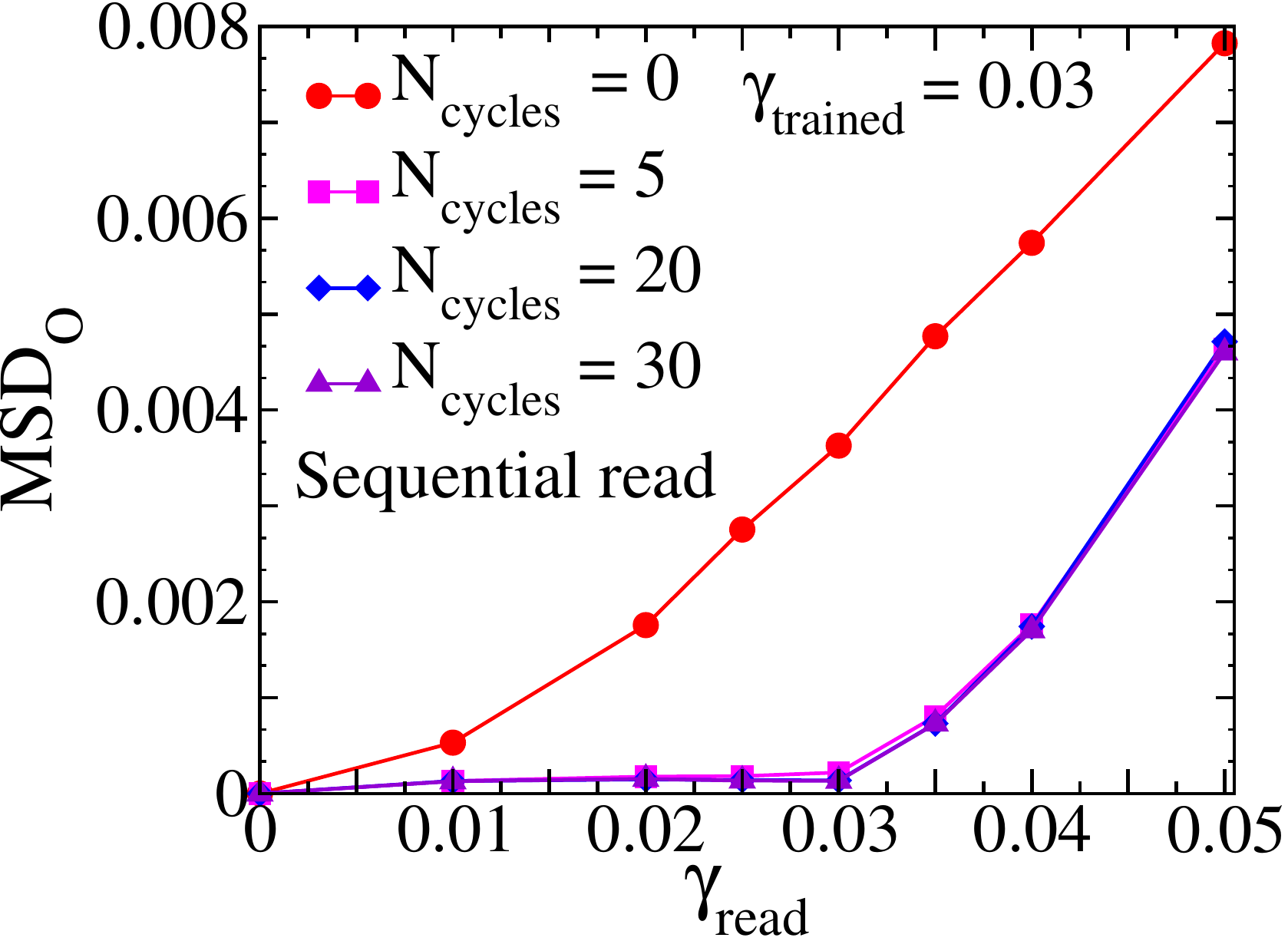}
\includegraphics[width = 0.43\textwidth]{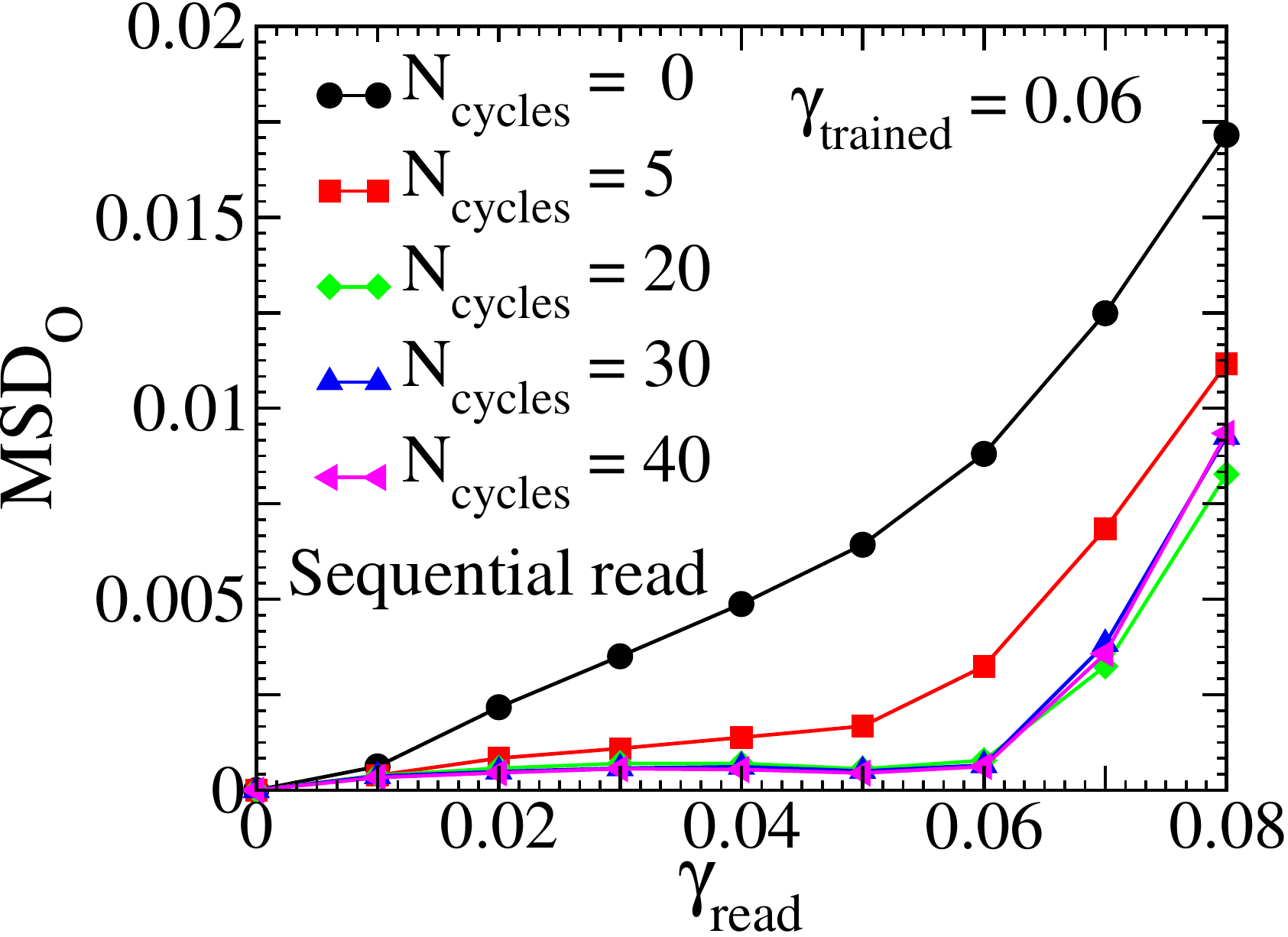}
\caption{The MSD as a function of $\gamma_{\text read}$ for different training cycles with  sequential reading. The MSD is measured with respect to the original configuration. The system is trained at $\gamma=0.03$ (top) $\gamma=0.06$ (bottom). When $\gamma_{\text read}$ is same as $\gamma_{\text trained}$, there is a change in slope of the MSD vs. $\gamma_{\text read}$ curve.}
\label{fig-8} 
 \end{figure}

\begin{figure}[h!]
\centering
\includegraphics[width = 0.40\textwidth]{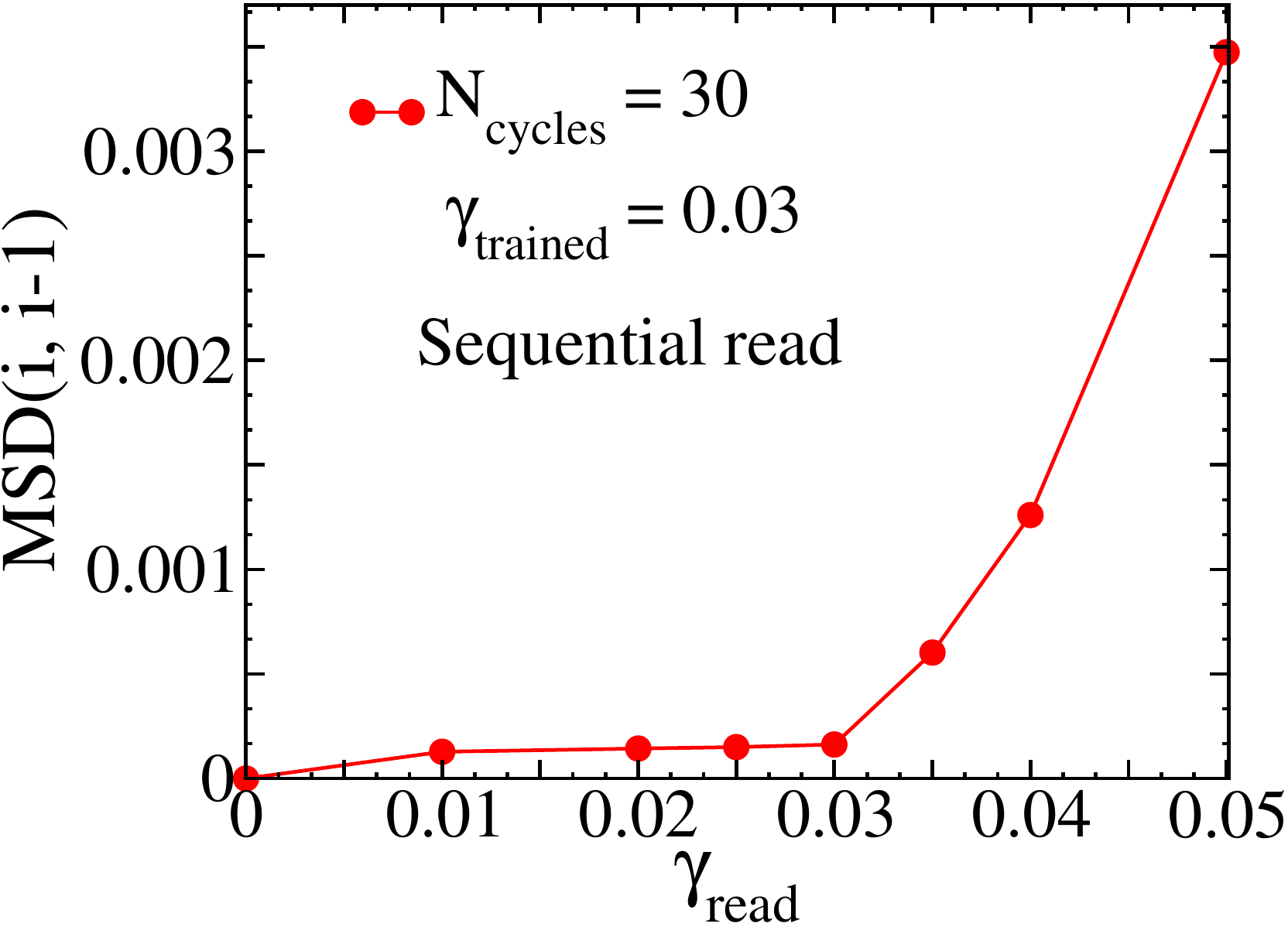}
\includegraphics[width = 0.40\textwidth]{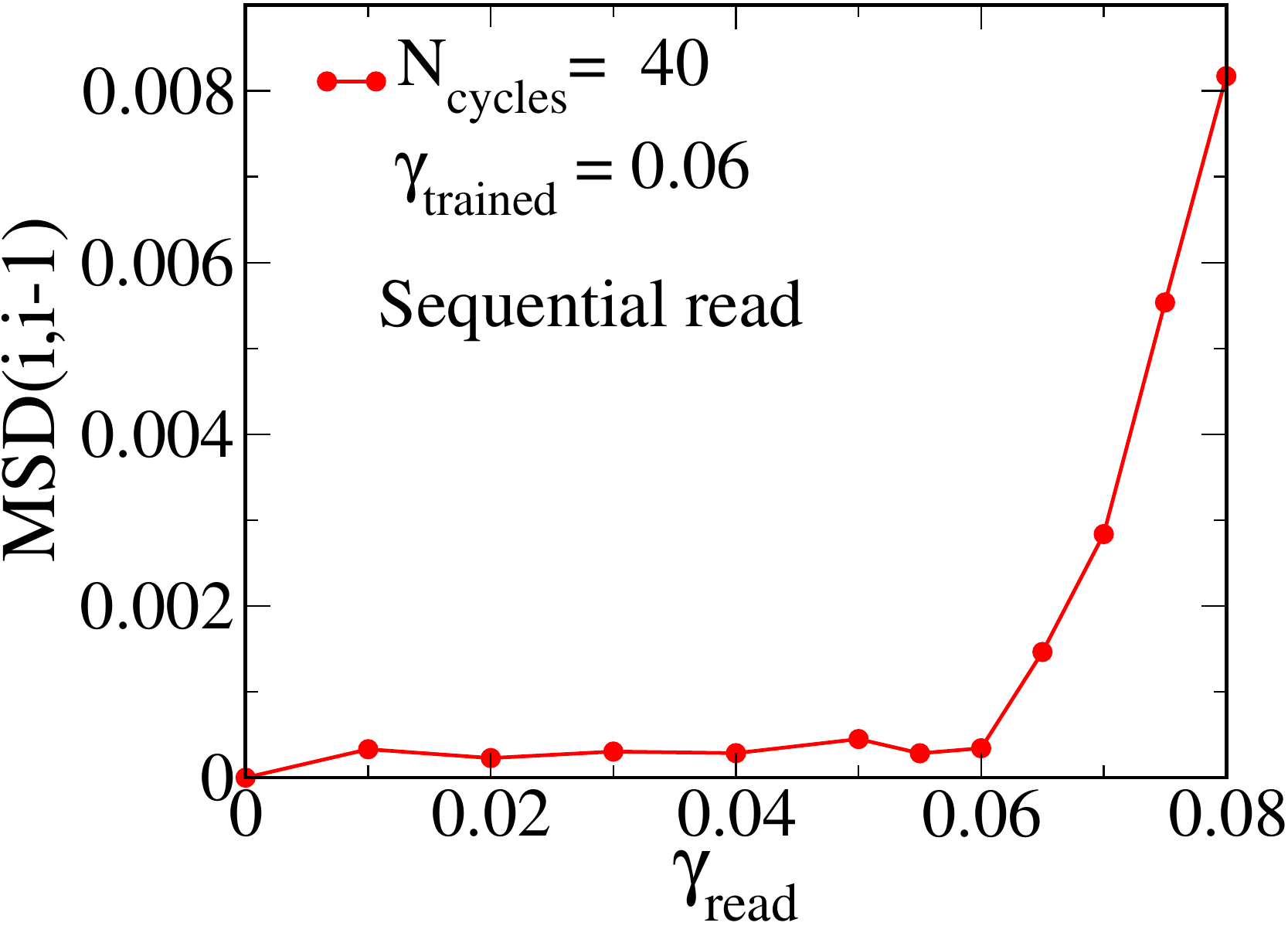}
\caption{The MSD is plotted as a function of $\gamma_{\text read}$ for fully trained system, with sequential reading. The system is trained at $\gamma=0.03$ (top) and $\gamma=0.06$ (bottom). The MSD is measured with respect to the configuration after the previous read cycle.}
\label{fig-9}  
\end{figure}

So far, we have shown results using the parallel read protocol wherein multiple copies of the trained system are subjected to read cycles at different amplitudes. Such a procedure is not available if a measurement is made experimentally, where the same trained system has to be subjected sequentially to read cycles of deformation. We thus consider the analogous sequential read protocol next, wherein after training, the trained configuration is subjected sequentially to a set of read deformations with increasing amplitude. The MSD data, with respect to the starting, trained, configuration are shown in Fig. \ref{fig-8}, for different numbers of training cycles. It is observed that the non-monotonicity of the MSD is greatly diminished and the MSD values at the training amplitude are not strictly zero. Nevertheless, the MSD data reveal a clear memory of the training amplitude, in that the MSD remain small up to the training amplitude and increase rapidly thereafter. 
In Fig. \ref{fig-9}, we show the corresponding results for MSD computed at each amplitude with respect to the configuration at the end of the previous read cycle, for a fully trained system. In this case too, the MSD data reveal a clear memory of the training amplitude, once again with a significantly reduced non-monotonicity of the MSD data. 

\subsubsection{The fraction of active particles: }
\begin{figure}[h!]
\centering
\includegraphics[width = 0.35\textwidth]{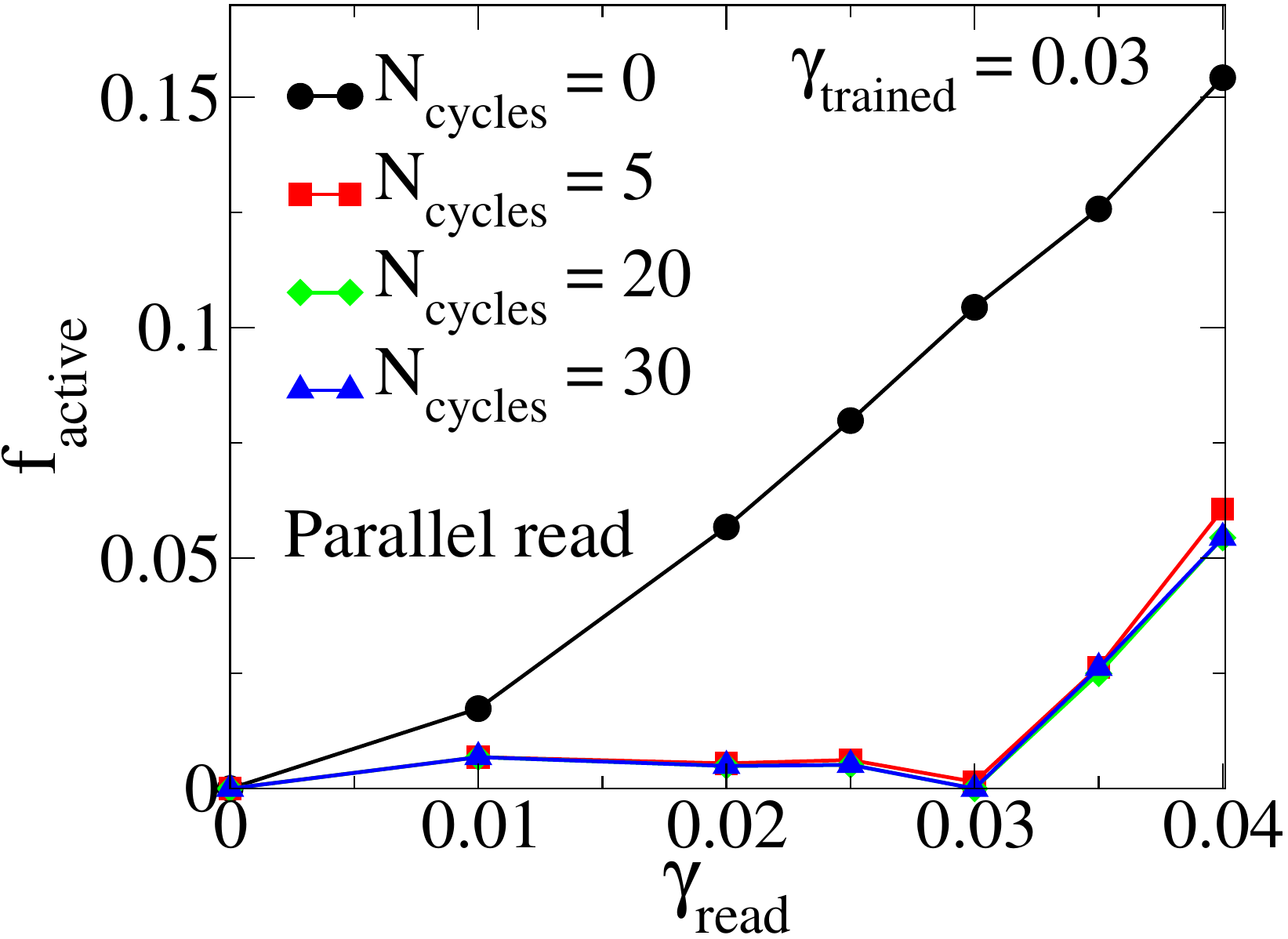}
\includegraphics[width = 0.35\textwidth]{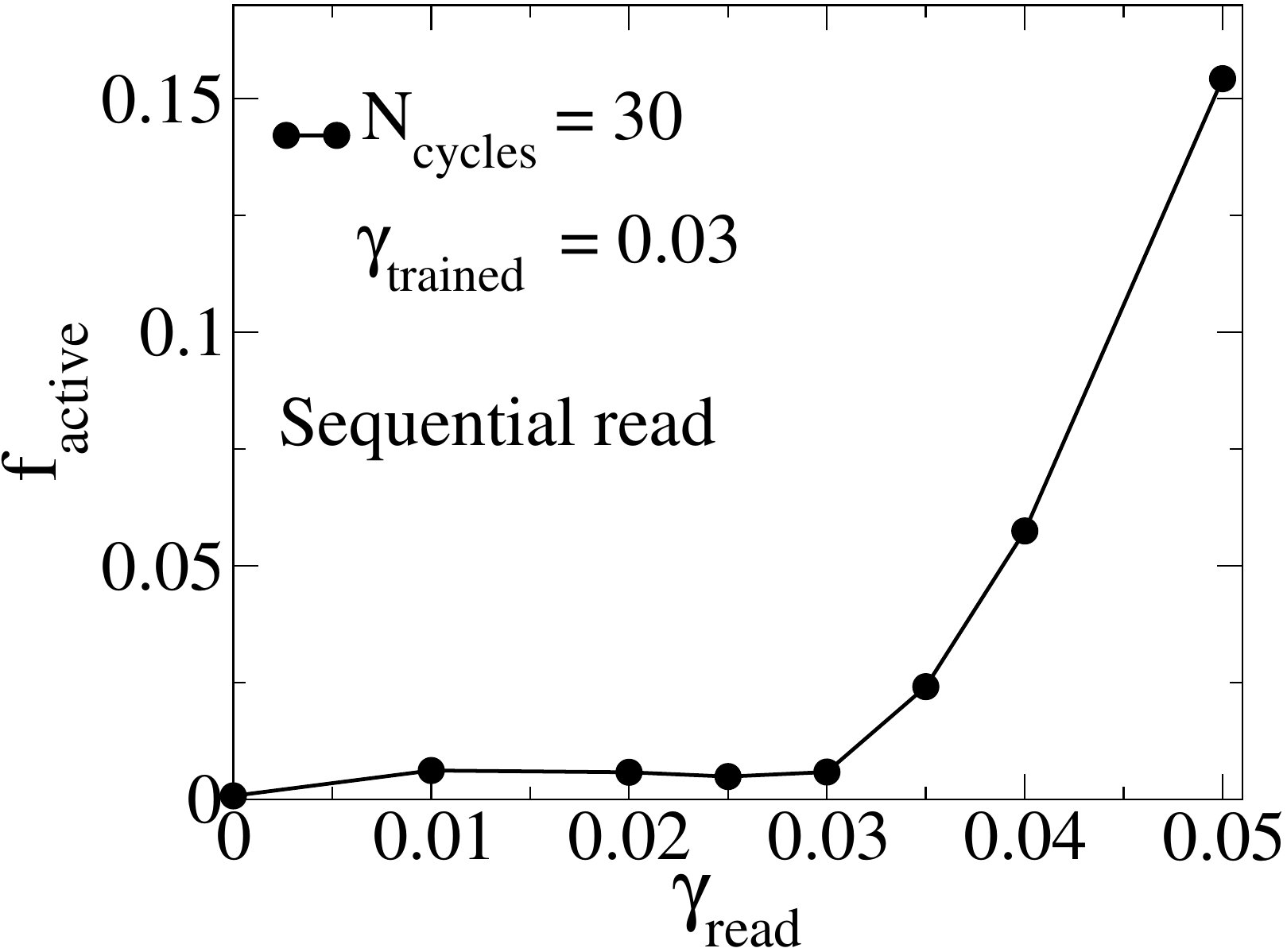}
\caption{The fraction of active particles ($f_{active}$) is plotted as a function of $\gamma_{\text read}$  for a system trained at $\gamma_{trained}=0.03$. Different lines in the top panel correspond to different numbers of training cycles. Top: (Parallel reading) After a large number of training cycles, when the system reaches the steady state, $f_{active}$ becomes zero at $\gamma_{read} = \gamma_{trained}$. Bottom: (Sequential reading) $f_{active}$ increases rapidly as $\gamma_{read}$ crosses $\gamma_{trained}$ for the completely trained system.}
\label{fig-10}  
\end{figure}

In previous related studies, instead of MSD, the fraction of active particles, $f_{active}$, has been considered as the measurement  \cite{PhysRevLett.107.010603,PhysRevE.88.032306}. Here, we perform the analogous measurement, by defining an active particle as one that has moved by a distance greater than $0.1 \sigma_{AA}$ during a read cycle. In Fig.  \ref{fig-10} (top panel), we show the fraction of active particles (with distances measured with respect to the original trained sample), for different numbers of training cycles. After sufficient training, the $f_{active}$ data show a clear signature of memory of the training amplitude, and a non-monotonicity similar to the MSD data. In  Fig.  \ref{fig-10} (bottom panel), we show the corresponding data wherein displacements are measured with respect to the configuration at the end of the previous read cycle, for the fully trained system. Here too, the memory of the training amplitude is clearly revealed. 

\subsubsection{Memory effects in the diffusing state}

\begin{figure}[htp]
\centering
\includegraphics[scale=0.35]{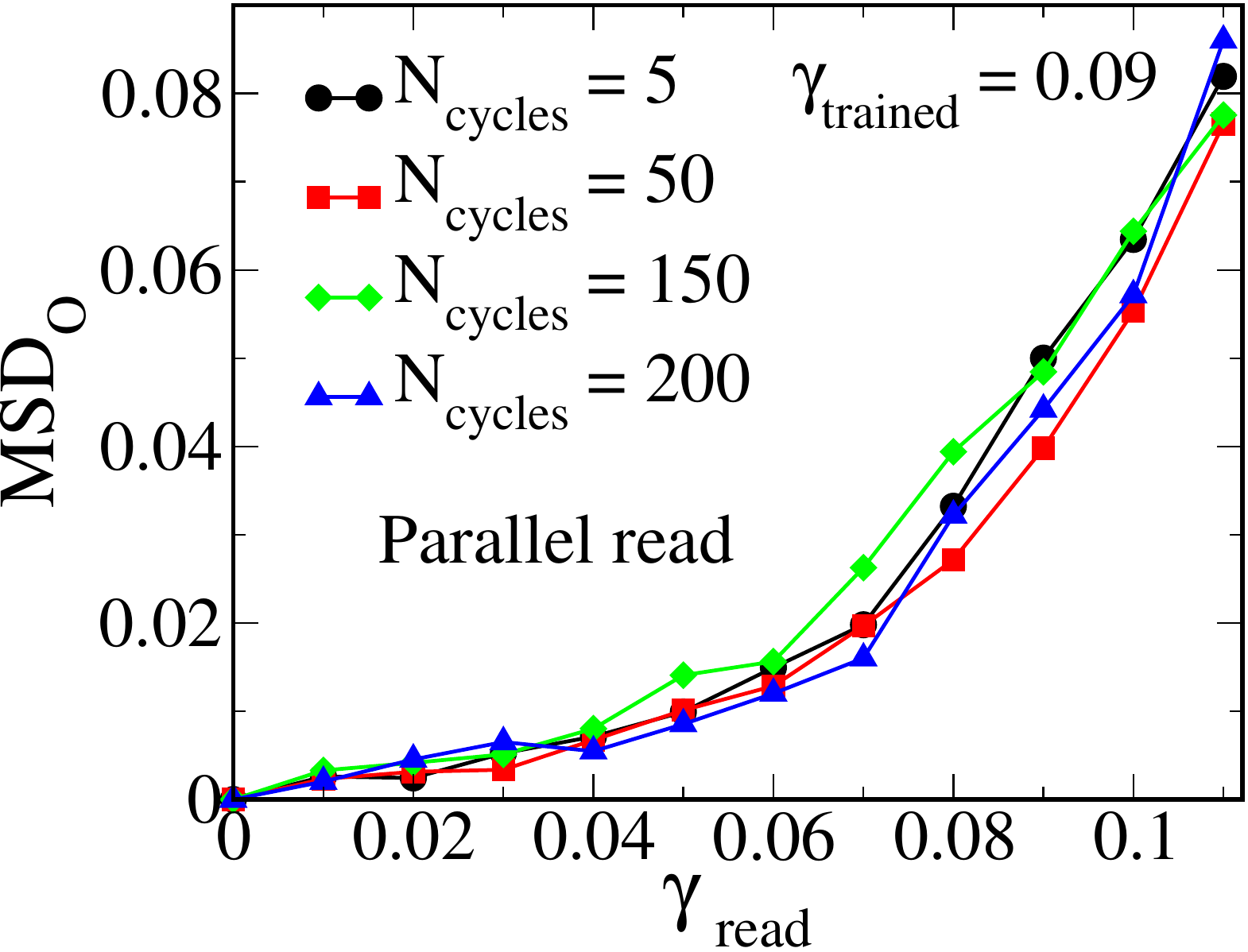}
\includegraphics[scale=0.35]{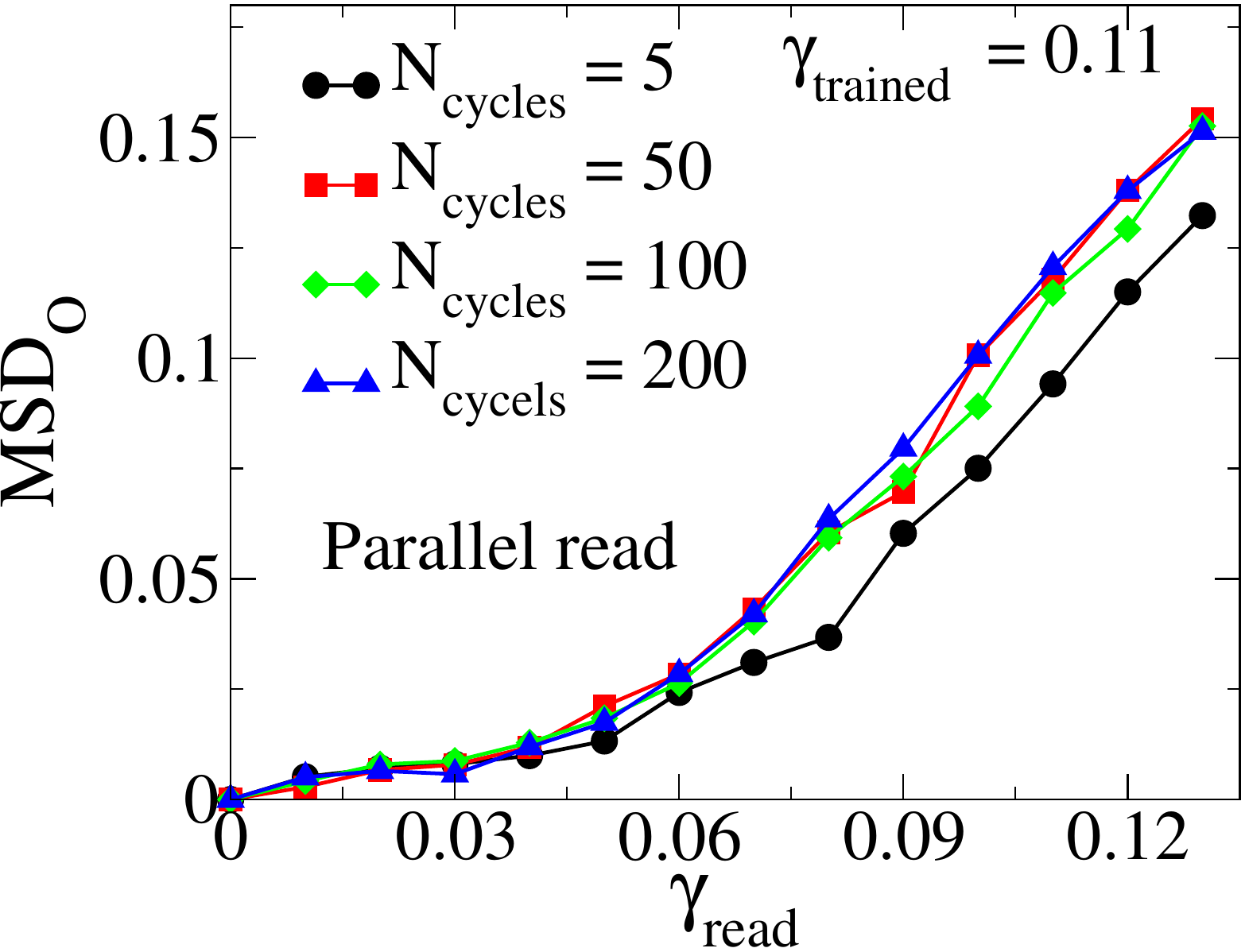}
\caption{The MSD as a function of $\gamma_{\text read}$ for a system which is trained at the amplitude, $\gamma_{\text trained} = 0.09$ (top) $\gamma_{trained} = 0.11$ (bottom) for a different numbers of the training cycles. The MSD increases with increasing $\gamma_{\text read}$, and shows no memory of the training amplitude.}
\label{fig-11}
\end{figure}

We have so far analysed memory effects for the trained system prepared with different amplitudes in the absorbing state,  $\gamma_{trained} < \gamma_c$. 
As already discussed, for  $\gamma_{trained} > \gamma_c$, the system reaches a diffusing state and does not return to the same configuration at the end of successive cycles. Thus, we do not expect that the system will retain any memory of the training amplitude. We test this expectation by performing measurements for two training amplitudes above $\gamma_c$, namely $\gamma_{trained} = 0.09$, and  $\gamma_{trained} = 0.11$. Results of $MSD_0$ shown  in Fig. \ref{fig-11} reveal indeed that there are no signatures of memory of the training amplitude in these cases. 
 
\begin{figure}[htp]
\centering
\includegraphics[scale=0.40]{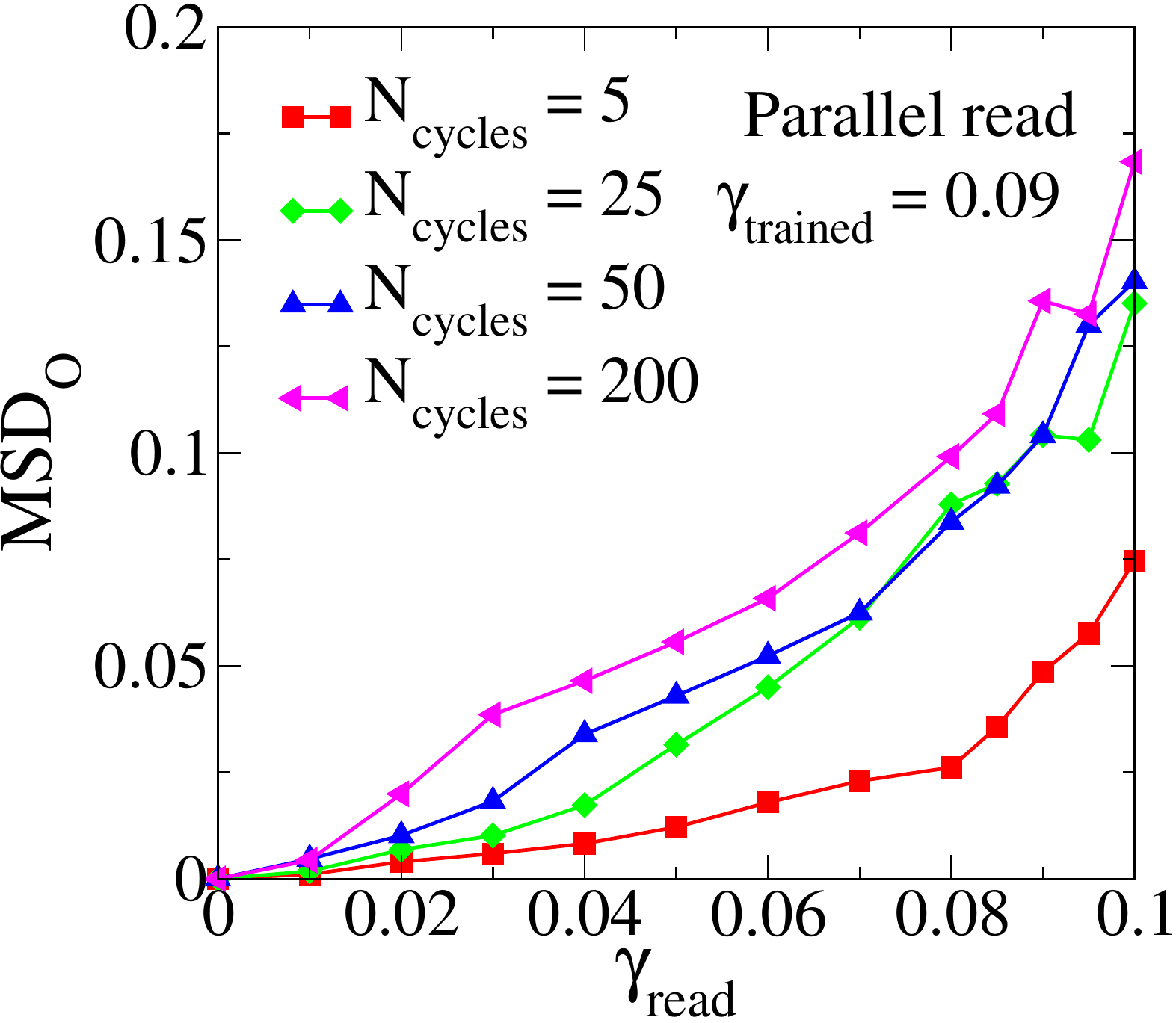}
\includegraphics[scale=0.40]{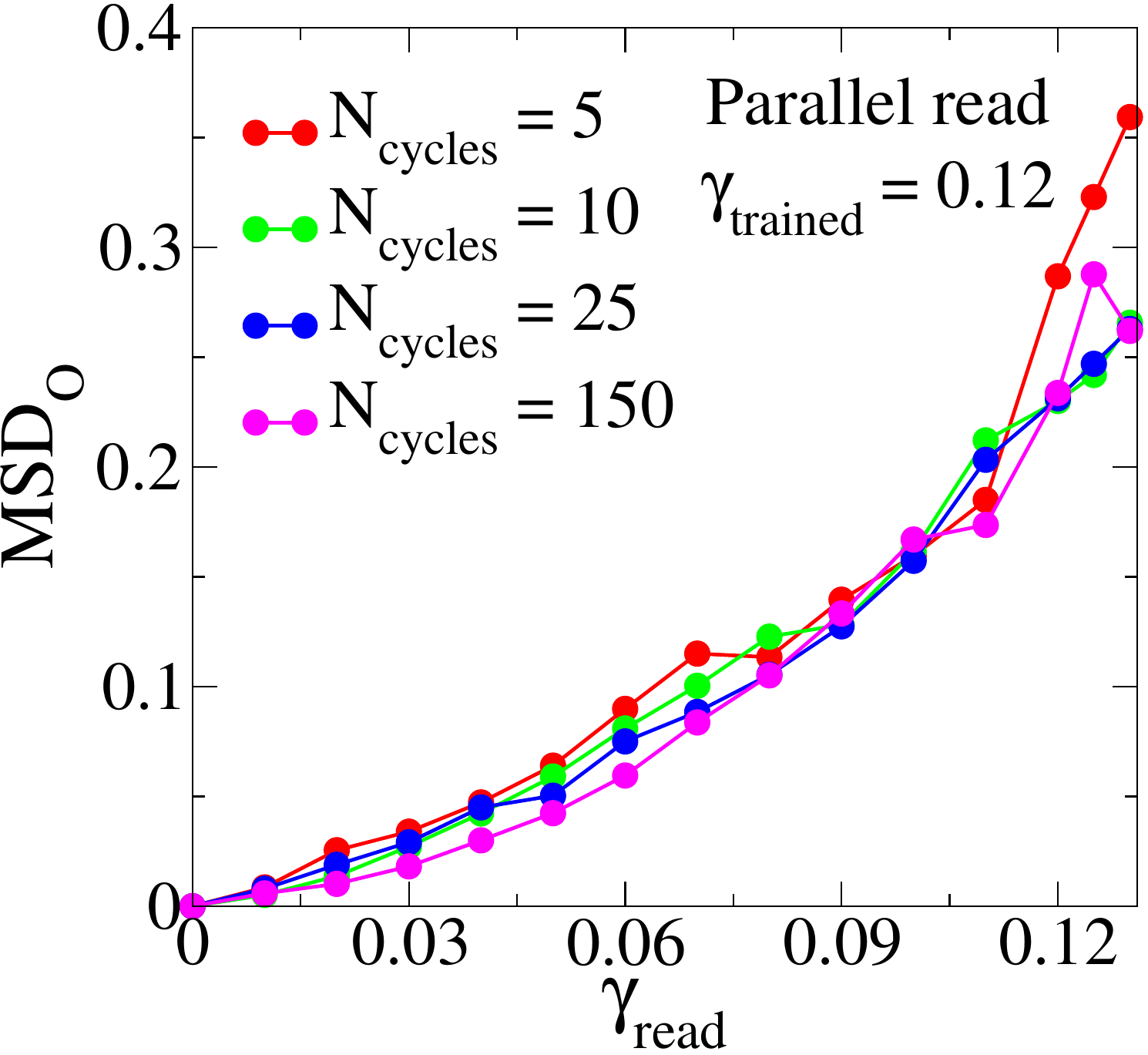}
\caption{The MSD as a function of $\gamma_{\text read}$ for $N = 64000$, $\gamma_{trained} = 0.09$ (top panel) and $\gamma_{trained} = 0.12$ (bottom panel). The different lines correspond to the different numbers of training cycles. No memory effects are observed for these amplitudes, which are larger than the yielding strain.}
\label{fig-12}
\end{figure}

As shown in a recent study\cite{anshul-shearbanding}, however, one may expect shear banding in the diffusing regime. The system size we have used here is too small (4000 particles) for shear banding to be clearly present. Hence, we consider next a larger system of $N = 64000$, and perform the same study.  The results are presented in Fig. \ref{fig-12} for  $\gamma_{trained} = 0.09$, and  $\gamma_{trained} = 0.12$, which do not show any signatures of memory of the training amplitude, confirming the results for the smaller system studied earlier.

\subsection{Multiple memories}

We next consider the case of multiple memories. Here, we train the system by subjecting it to repeated cycles of deformation at two different amplitudes $\gamma_1$ and $\gamma_2$ ($0 \rightarrow \gamma_1 \rightarrow 0 \ \rightarrow -\gamma_1 \rightarrow 0$  $\rightarrow \gamma_2 \rightarrow 0 \ \rightarrow -\gamma_2 \rightarrow 0$) and perform reading as in the case of single memories.  We consider training amplitudes  $\gamma_1=0.02$ and $\gamma_2=0.01$. To assess the dependence of multiple memories on the deformation training amplitudes, we also consider $\gamma_1=0.06$ and $\gamma_2=0.04$. We then consider also the case of encoding three memories as described below.

\subsubsection{Parallel read}

\begin{figure}[h!]
 \centering
\includegraphics[width = 0.42\textwidth]{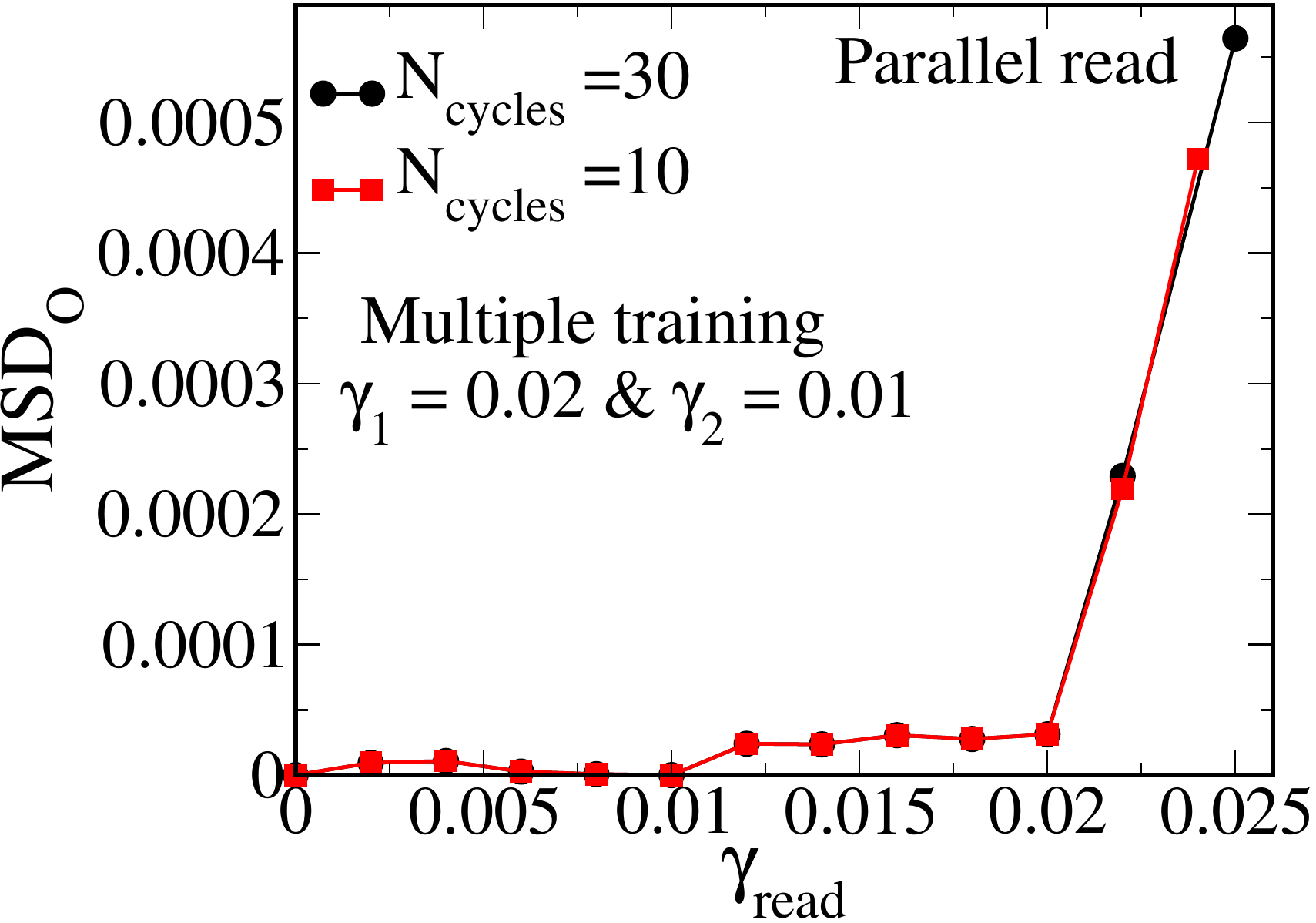}
\includegraphics[width = 0.42\textwidth]{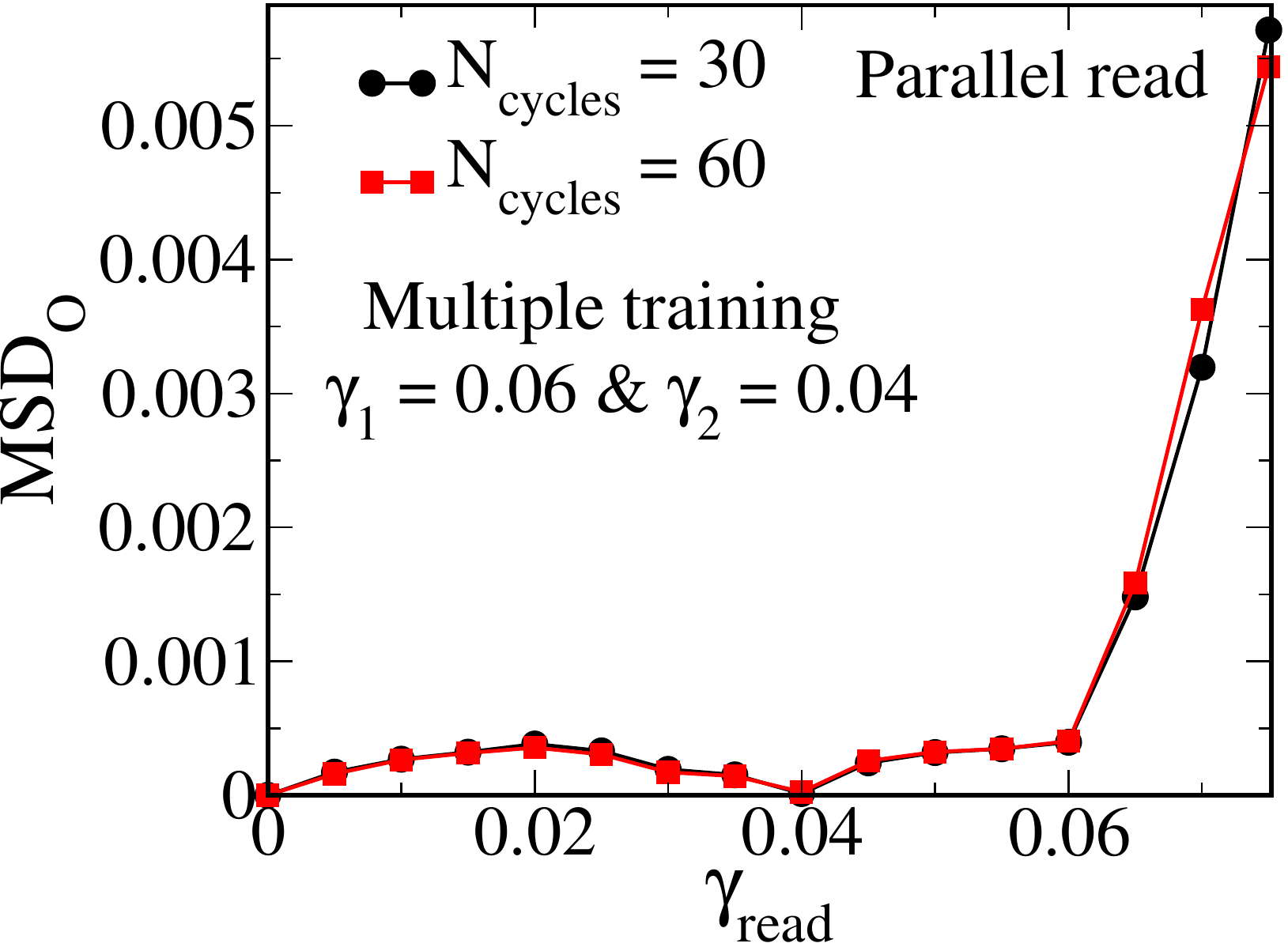}
\caption{The MSD as a function of $\gamma_{\text read}$ for a system which is trained at multiple $\gamma$ (top: $\gamma_1=0.02$ and $\gamma_2=0.01$, bottom: $\gamma_1=0.06$ and $\gamma_2=0.04$). Two kinks are observed at $\gamma_1 = 0.06$ and $\gamma_2= 0.04$ (for other set it is at $\gamma_1 = 0.02$ and $\gamma_2= 0.01$). The different lines correspond to the different numbers of training cycles. Both the memories are present after a large number of training cycles persistently. }
\label{fig-13} 
\end{figure}

In  Fig. \ref{fig-13}, we show results using parallel read for the two sets of training amplitudes, for two different numbers of training cycles. As seen clearly, the data reveal signatures of memory of both amplitudes, although they are different for the two amplitudes. For the smaller amplitude, the MSD goes to zero, whereas it remains finite at the larger amplitude. At the larger amplitude, however, a sharp change in the MSD values is seen, which serves as a clear signature of memory of that training amplitude. As previously discussed \cite{MemFiocco}, both these memories are persistent, and do not diminish in strength with increased number of training cycles. 

\begin{figure}[h!]
 \centering
\includegraphics[width = 0.42\textwidth]{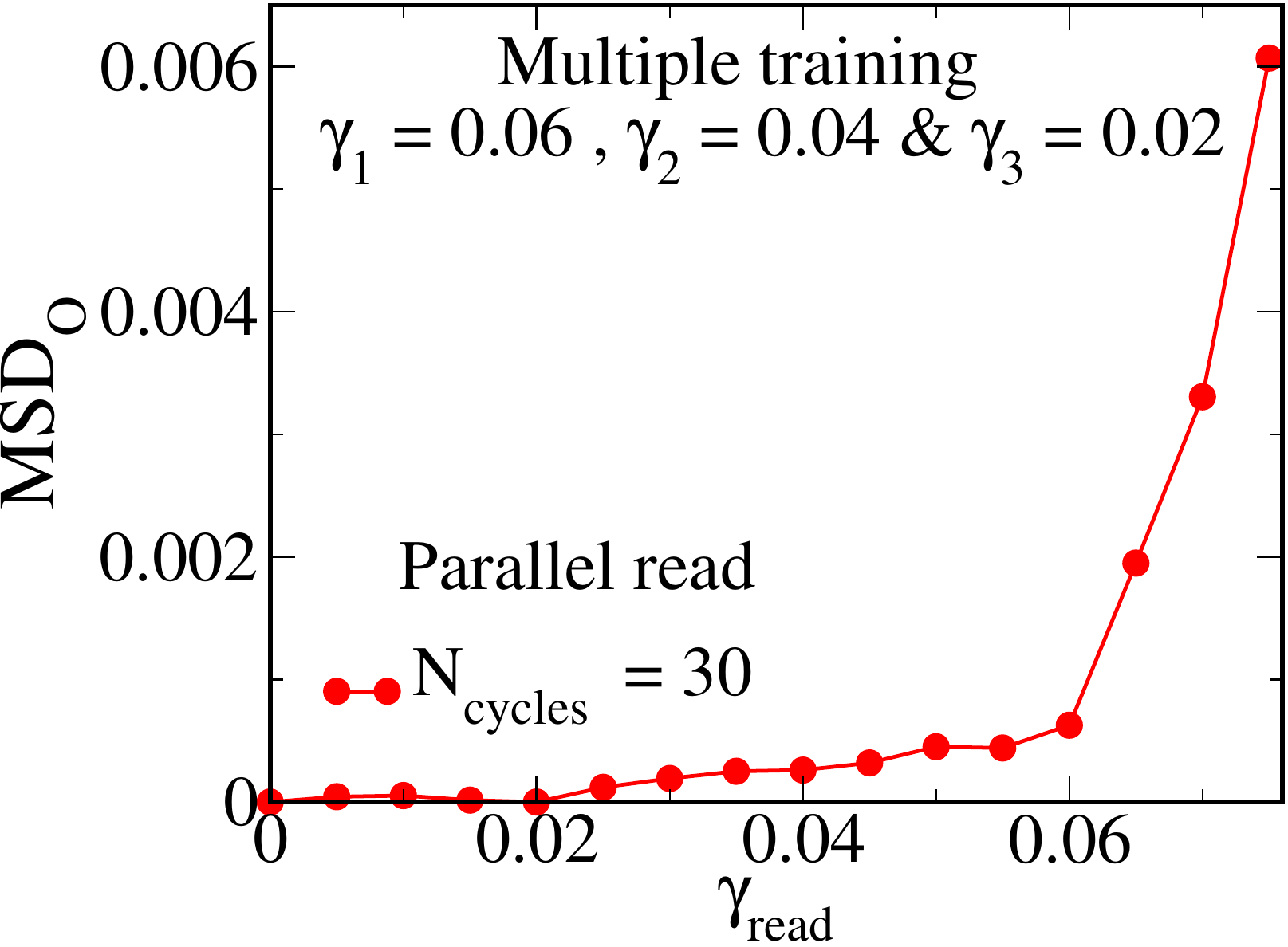}
\caption{\small{The MSD as a function of $\gamma_{\text read}$ for training with three different amplitudes. The system is trained at $\gamma_1=0.06$, $\gamma_2 =0.04$, $\gamma_3 = 0.02$. The memory is observed at $\gamma=0.06$ and $\gamma =0.02$} }
\label{fig-14} 
\end{figure}

\begin{figure}[htp]
\centering
\includegraphics[width = 0.40\textwidth]{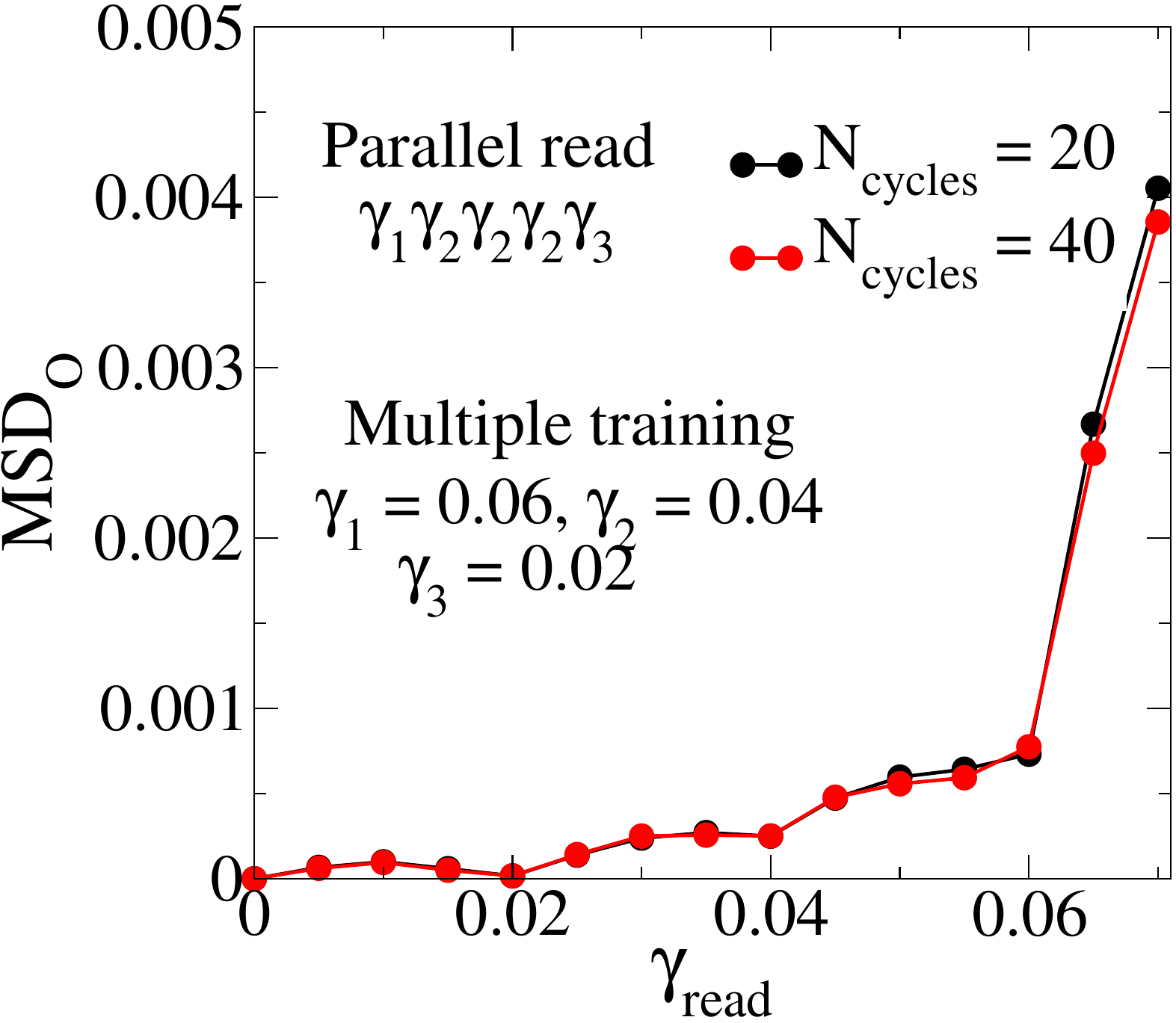}
\caption{The MSD as a function of $\gamma_{read}$ during parallel reading. The system is trained for training cycles where a single cycle has three different shear deformation amplitude (sub)cycle repeated according to the pattern: $\gamma_1 \gamma_2 \gamma_2 \gamma_2 \gamma_3$. Here $\gamma_1 =0.06$ $\gamma_2 =0.04$ and $\gamma_3 =0.02$. After a large number of training cycles, signatures of all three memories are clearly seen and these memories are persistent.}
\label{fig-15}
\end{figure}

We next consider whether a larger number of memories can be encoded. To this end, we train configurations with three different amplitudes ($0 \rightarrow \gamma_1 \rightarrow 0 \ \rightarrow -\gamma_1 \rightarrow 0$  $\rightarrow \gamma_2 \rightarrow 0 \ \rightarrow -\gamma_2 \rightarrow 0$ $\rightarrow \gamma_3 \rightarrow 0 \ \rightarrow -\gamma_3 \rightarrow 0$.), with $\gamma_1=0.06$, $\gamma_2=0.04$  and $\gamma_3=0.02$. When subjected to read cycles, we find (as shown in  Fig. \ref{fig-14}) signatures of memory only at the smallest and largest of the amplitudes, namely $\gamma_3=0.02$ and $\gamma_1=0.06$. In order to assess the role of training protocol, we consider a different sequence of training deformations, with a repetition of the pattern $\gamma_1 \gamma_2 \gamma_2 \gamma_2 \gamma_3$. 
In this case, as shown in Fig. \ref{fig-15}, all three training amplitudes have corresponding dips in the MSD revealing that all these memories are encoded in the trained system. 

\subsubsection{Sequential reading}

\begin{figure}[h!]
 \centering
 \includegraphics[width = 0.40\textwidth]{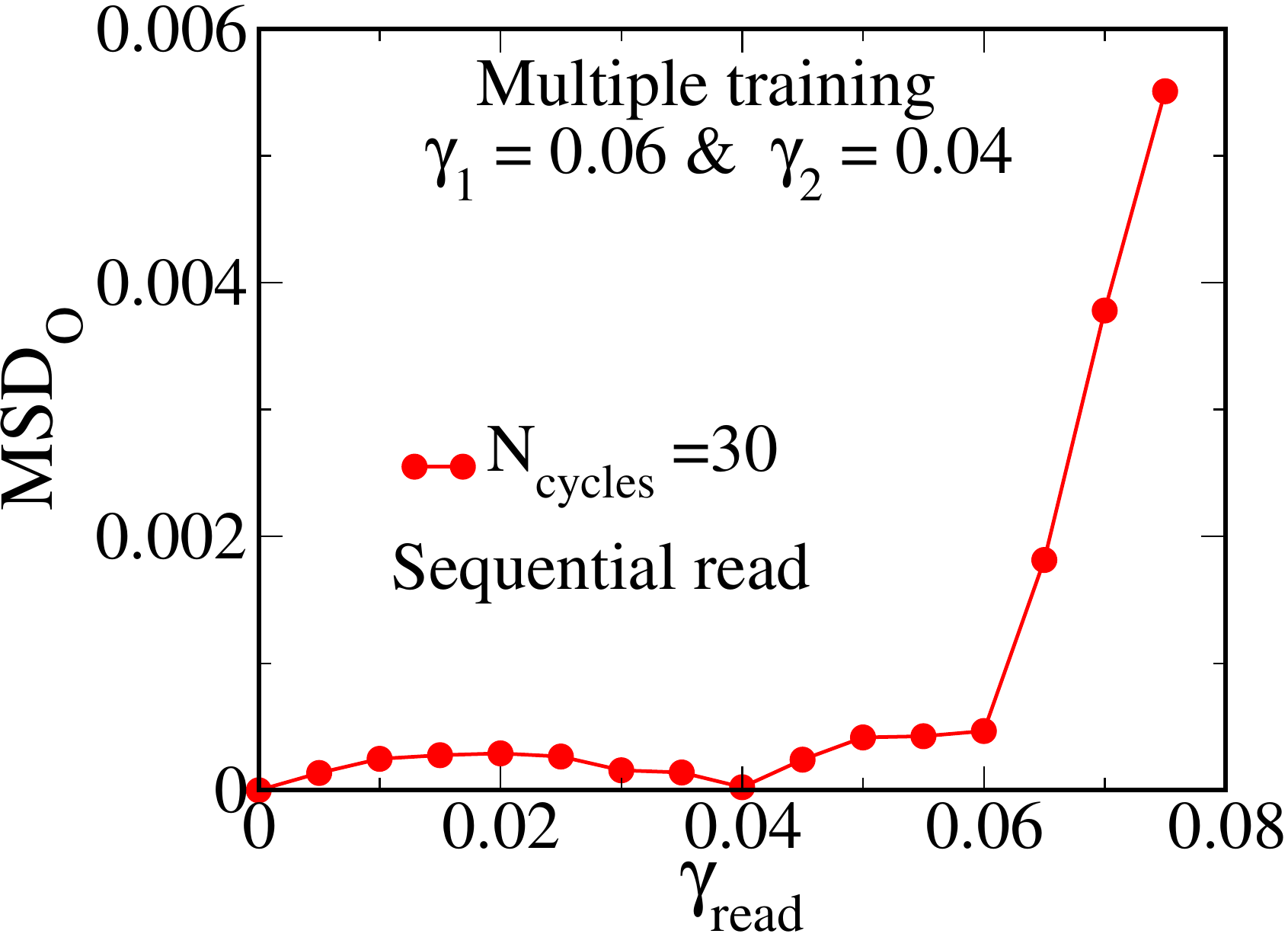}
\includegraphics[width = 0.40\textwidth]{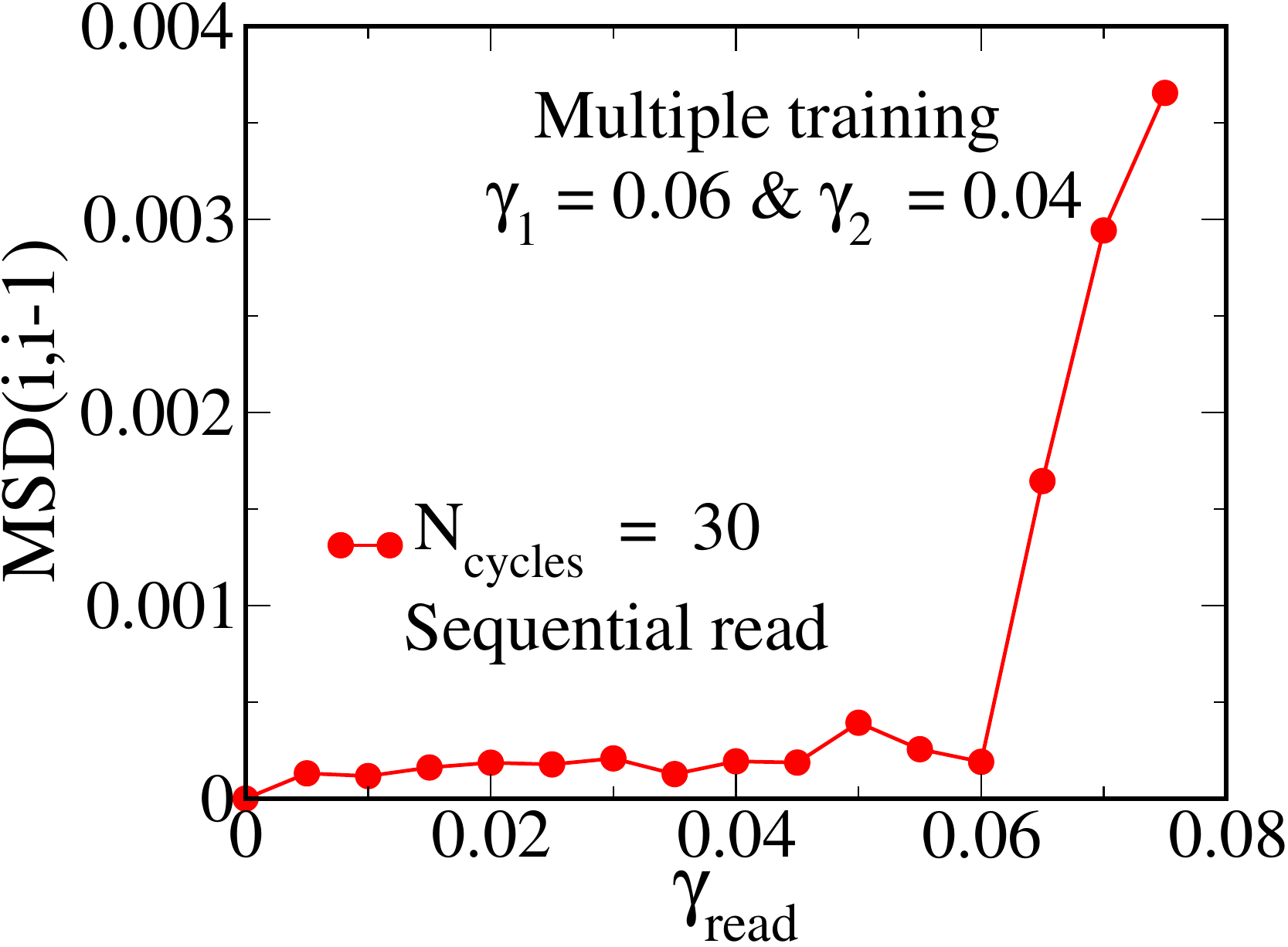}
 \caption{The MSD is plotted as a function of $\gamma_{\text read}$ (sequential) for a system trained at multiple amplitudes. Top: The MSD is measured with respect to the original configurations. Bottom: The MSD is measured with respect to the final configuration of the previous read cycle.}
\label{fig-16} 
 \end{figure}

We next employ sequential reading as done before for single memories for the case of multiple memories, with two different training amplitudes with $\gamma_1=0.06$, $\gamma_2=0.04$. As shown in Fig.\ref{fig-16}, when the MSD is measured with respect to the trained configuration, sequential read generates data which capture the encoding of multiple memories as clearly as the parallel read, but when the MSD is measured with respect to the final configuration of the previous read cycle, a less distinct signature is seen at the smaller of the training amplitudes, $\gamma_2=0.04$. While there may be variations of the procedure used here that will generate a clear signature of multiple memories even in this case, we do not pursue this investigation further in that direction. 

\subsubsection{Application of cyclic shear deformation with different amplitudes to a trained (at multiple amplitudes) system}

Similar to the case of single memory, we wish to investigate the effect of applying cyclic deformation at a new amplitude in a multiply trained system. The system is trained at two different amplitudes $\gamma_1 = 0.06$ and $\gamma_2 = 0.04$, and is then subjected to a single cycle of shear deformation with $\gamma_3$ repeatedly.  
We have three cases (1). $\gamma_3$ is smaller than both $\gamma_1$ and $\gamma_2$. (2). $\gamma_3$ is less than $\gamma_1$ but greater than $\gamma_2$ and (3). $\gamma_3$ is larger than both $\gamma_2$ and $\gamma_1$. We consider these cases in turn. 

\paragraph{Retraining amplitude is smaller than both the training amplitudes:}

\begin{figure}[h!]
\centering
\includegraphics[width = 0.40\textwidth]{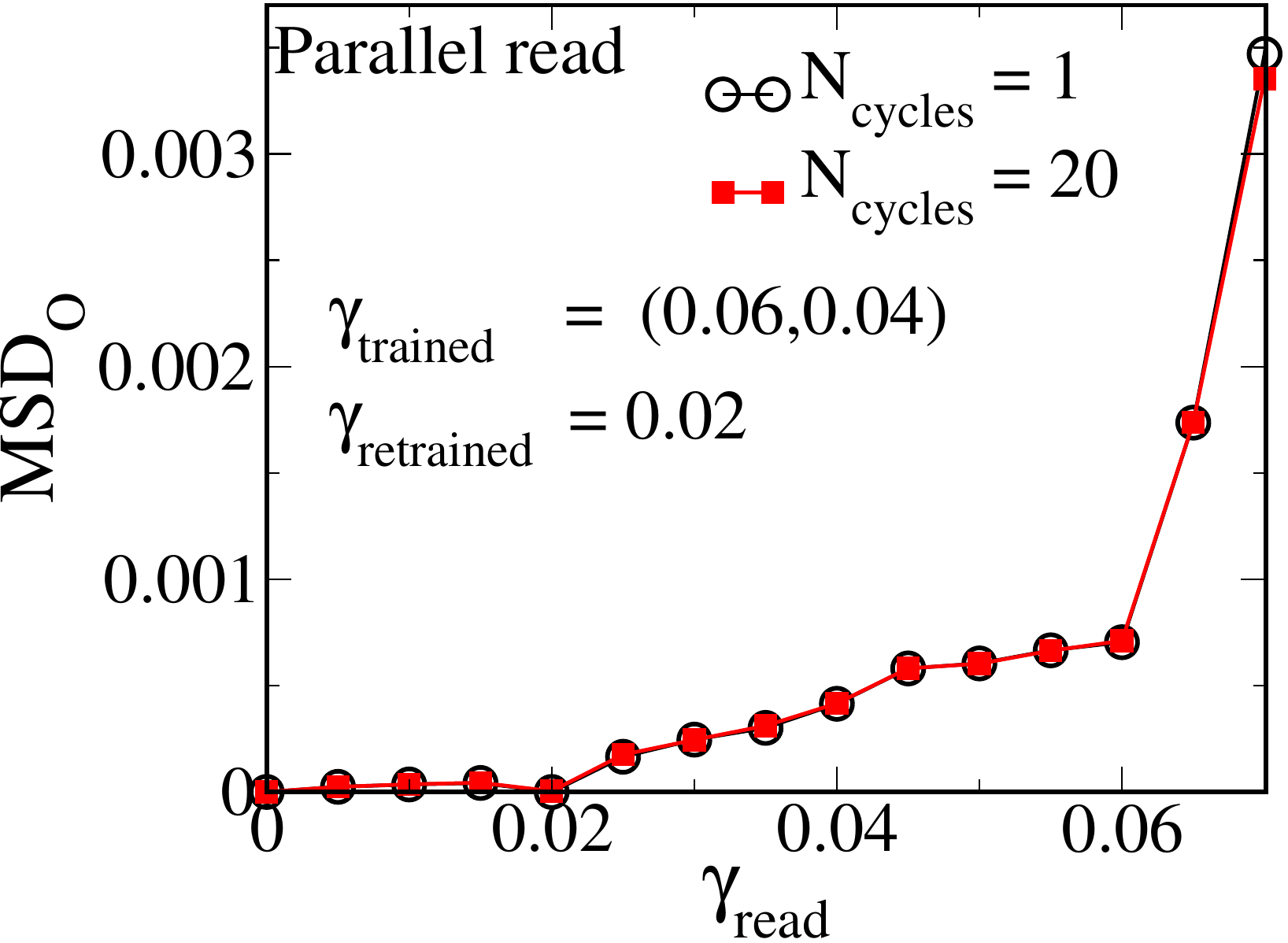}
\caption{The MSD is plotted as a function of $\gamma_{\text read}$  for different numbers of retraining cycles. A single cycle of shear deformation with amplitude $\gamma_3=0.02$ is applied to the system trained at $\gamma_1 = 0.06$ and $\gamma_2=0.04$. The new configuration has memory at $\gamma_3=0.02$ and $\gamma_1=0.06$ but no distinct memory of $\gamma_2 =0.04$. Memory signatures at $\gamma_1=0.06$ remains robustly even after a large number of retraining cycles at $\gamma_3=0.02$.}
\label{fig-17}
\end{figure}

We consider configurations trained at $\gamma_1 = 0.06$ and $\gamma_2 = 0.04$. We then apply cyclic shear deformation with $\gamma_3 =0.02$. The results are shown in Fig.\ref{fig-17}. Kinks in the MSD curves  at $\gamma_1= 0.06$ (largest $\gamma$) and  $\gamma_3=0.02$ indicate that the memory of these amplitudes is encoded, and remain even after a large number of cycles at $\gamma_3 =0.02$. No clear signature is visible at  $\gamma_2 = 0.04$. As in the case of triple memories, it may be possible that this signature will remain if a different training protocol is used, but we do not investigate it further. 

\begin{figure}[h!]
\centering
\includegraphics[width = 0.40\textwidth]{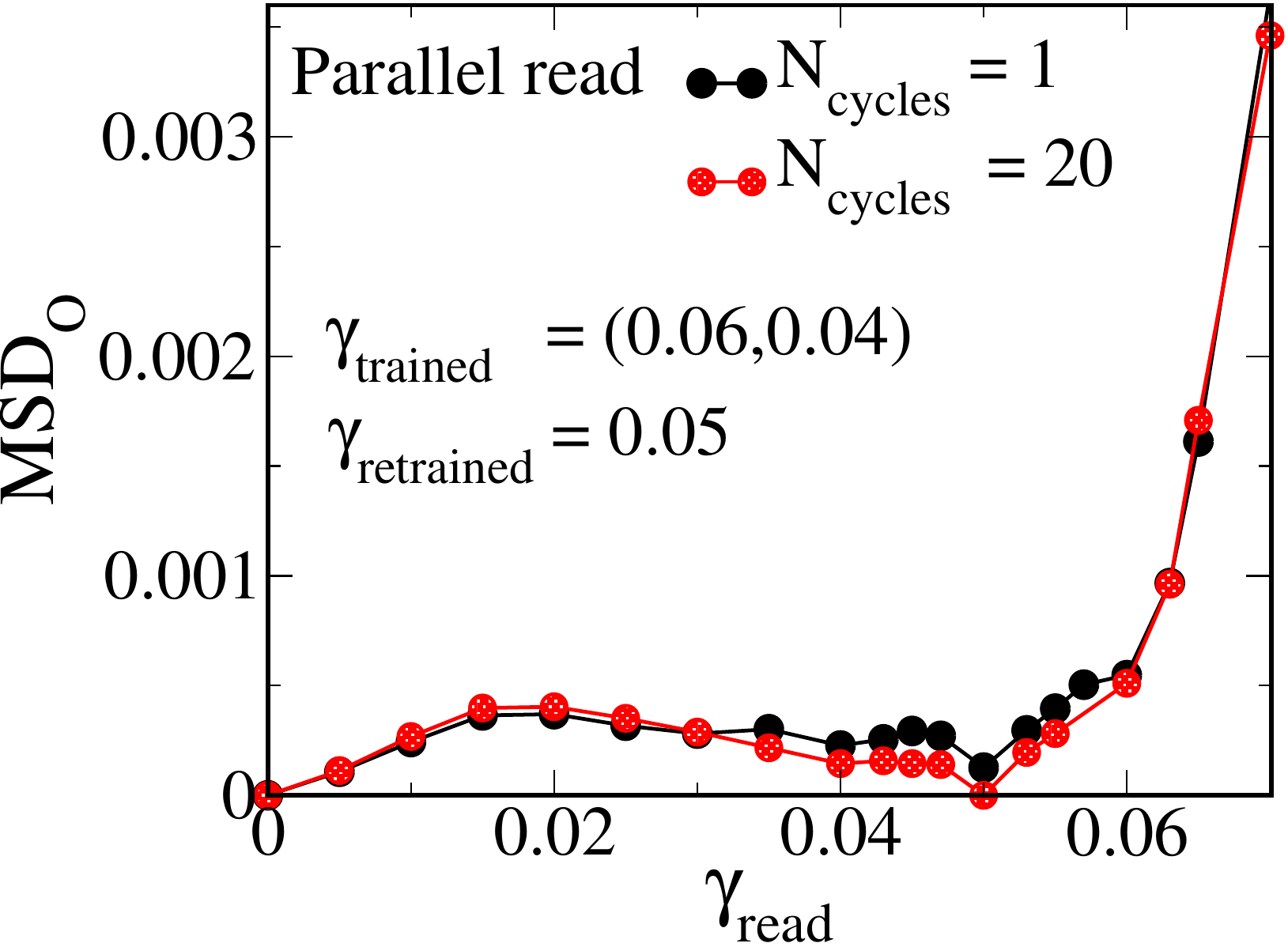}
\caption{The MSD is plotted as a function of $\gamma_{\text read}$. A single cycle shear deformation with amplitude $\gamma_3=0.05$ is applied to the system which is trained at two different amplitudes, $\gamma_1 = 0.06$ and $\gamma_2 =0.04$. The new system does not have a clear signature of  memory at $\gamma_2=0.04$ but has features revealing memory at  $\gamma_1=0.06$ and at $\gamma_3=0.05$, which remain after a large number of retraining cycles at  $\gamma_3=0.05$. }
\label{fig-18}
 \centering
\end{figure}

\paragraph{Deformation amplitude is smaller than one of the training amplitudes but larger than the other:}
We consider configurations trained at $\gamma_1 = 0.06$ and $\gamma_2 = 0.04$. We then apply cyclic shear deformation with $\gamma_3 =0.05$. The results are shown in Fig.\ref{fig-18}. Even after a single deformation at  $\gamma_3 =0.05$, the memory at $\gamma_2 = 0.04$ is erased, while the memory at  $\gamma_1 = 0.06$ is weak but present. In addition, a strong signature of memory at $\gamma_3 =0.05$ appears after a single cycle. 

\begin{figure}[h!]
 \centering
\includegraphics[width = 0.40\textwidth]{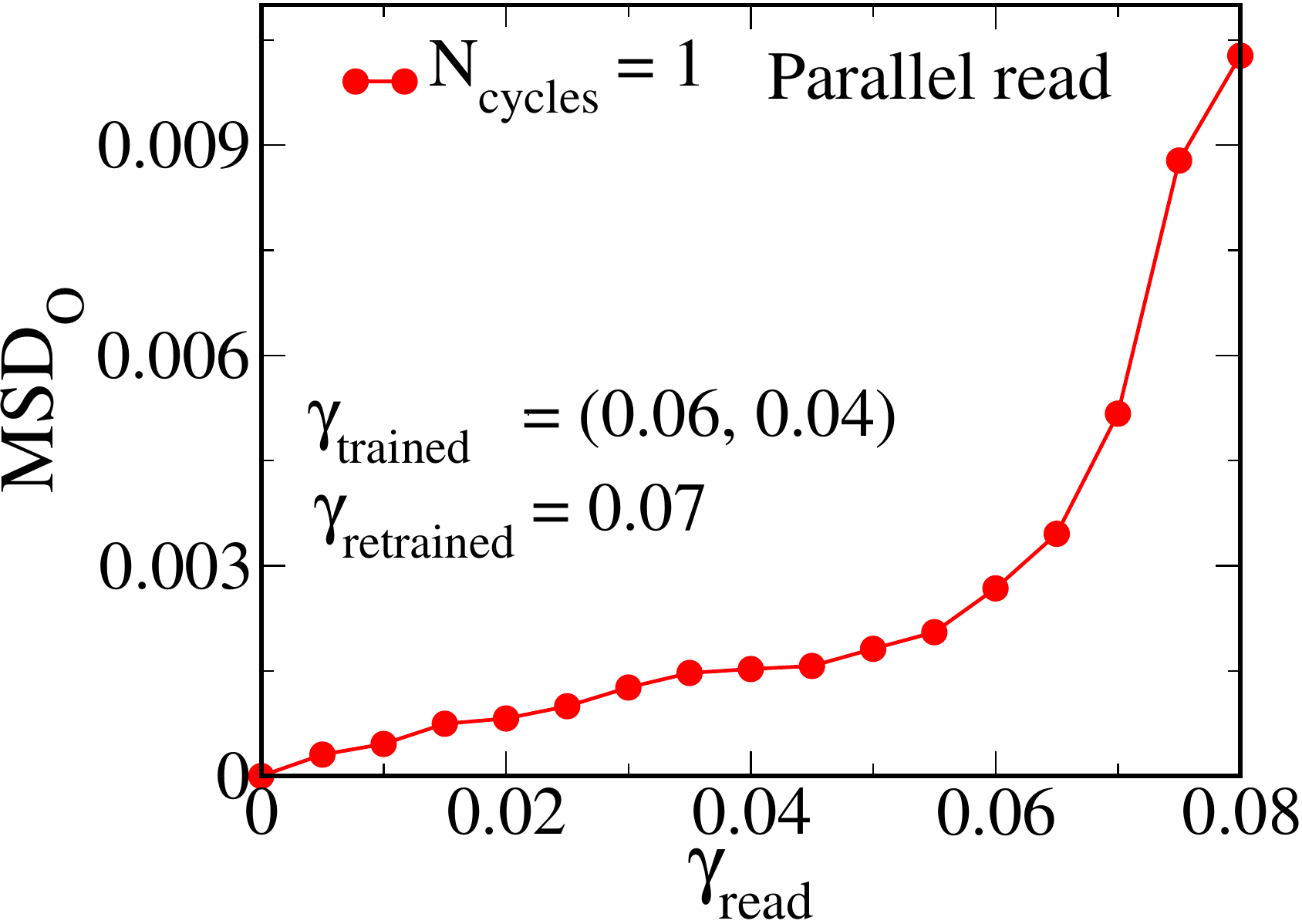}
 \caption{The MSD is plotted as a function of $\gamma_{\text read}$. A single cycle of shear deformation with amplitude $\gamma=0.07$ is applied to the system which is trained at multiple $\gamma$, $\gamma_1 = 0.06$ and $\gamma_2=0.04$, which erases memories of both these amplitudes.}
\label{fig-19} 
 \centering
\end{figure}

\paragraph{Deformation amplitude is larger than both the training amplitudes:}
We consider configurations trained at $\gamma_1 = 0.06$ and $\gamma_2 = 0.04$. We then apply cyclic shear deformation with $\gamma_3 =0.07$. As shown in Fig. \ref{fig-19}, a single cycle of shear deformation with $\gamma_3=0.07$ erases both the memories. This is consistent with the observation for the case of single memory that deformation by a larger amplitude erases stored memories. 

\section{Results: Soft Sphere binary mixture system}  \label{secIV}
The differences in memory effects in the BMLJ glass former described above and in \cite{MemFiocco} and the model considered in \cite{PhysRevLett.107.010603,PhysRevE.88.032306} have previously been rationalised in terms of the observation that trained configurations of BMLJ that reach an absorbing state nevertheless traverse a non-trivial energy landscape during a read cycle, which involve multiple transitions between energy minima. Thus, even though the configurations reach the same configuration at the end of a cycle, their trajectories during the cycle are non-trivial. This is in contrast with the model of a sheared colloidal suspension studied  in \cite{PhysRevLett.107.010603,PhysRevE.88.032306} wherein, upon reaching an absorbing state, sheared configurations return to the same configuration at the end of each shear cycle. More importantly, the particles merely undergo affine deformations during the cycle and do not interact with other particles. They thus experience a trival, flat, landscape during the read cycles. It is thus interesting to ask if there are cases that deviate from both these scenarios. Such a case is presented in the study of sheared soft spheres. The soft sphere system at densities below the jamming point, under AQS deformation, traverses a trivial landscape, in the sense that the energie of the inherent structures always remain zero. However, this system displays different regimes in applied strain, for which the nature of the absorbing states are different. For small applied strain, the system reaches absorbing states which are similar to those of the model studied in \cite{PhysRevLett.107.010603,PhysRevE.88.032306} wherein during a shear cycle, particles do not interact with other particles and show smooth affine displacements. These are termed {\it point reversible} states  \cite{PhysRevESchreck}. At higher amplitudes, a new regime is encountered wherein particles return to their original positions at the end of a shear cycle, but during the cycle, they may collide or interact with other particles that they come into contact with. These states are referred to as {\it loop reversible} states. It is thus interesting to consider the nature of memory in point reversible and loop reversible states. For the volume fraction $\phi = 0.61$, the transition from point to loop reversible states occurs around  $\gamma_{c} = 0.07$. Accordingly, we consider $\gamma_{\text trained}=0.03$ (which belongs to the point reversible regime) and $\gamma_{\text trained} = 0.12$ (which belongs to the loop reversible range) to study the nature of memory effects in the case of training at a single amplitude. We will subsequently consider multiple memories, which are described later. 

\subsection{Single memory}

\subsubsection{Parallel reading} 

\begin{figure}[h!]
 \centering
 \includegraphics[width = 0.43\textwidth]{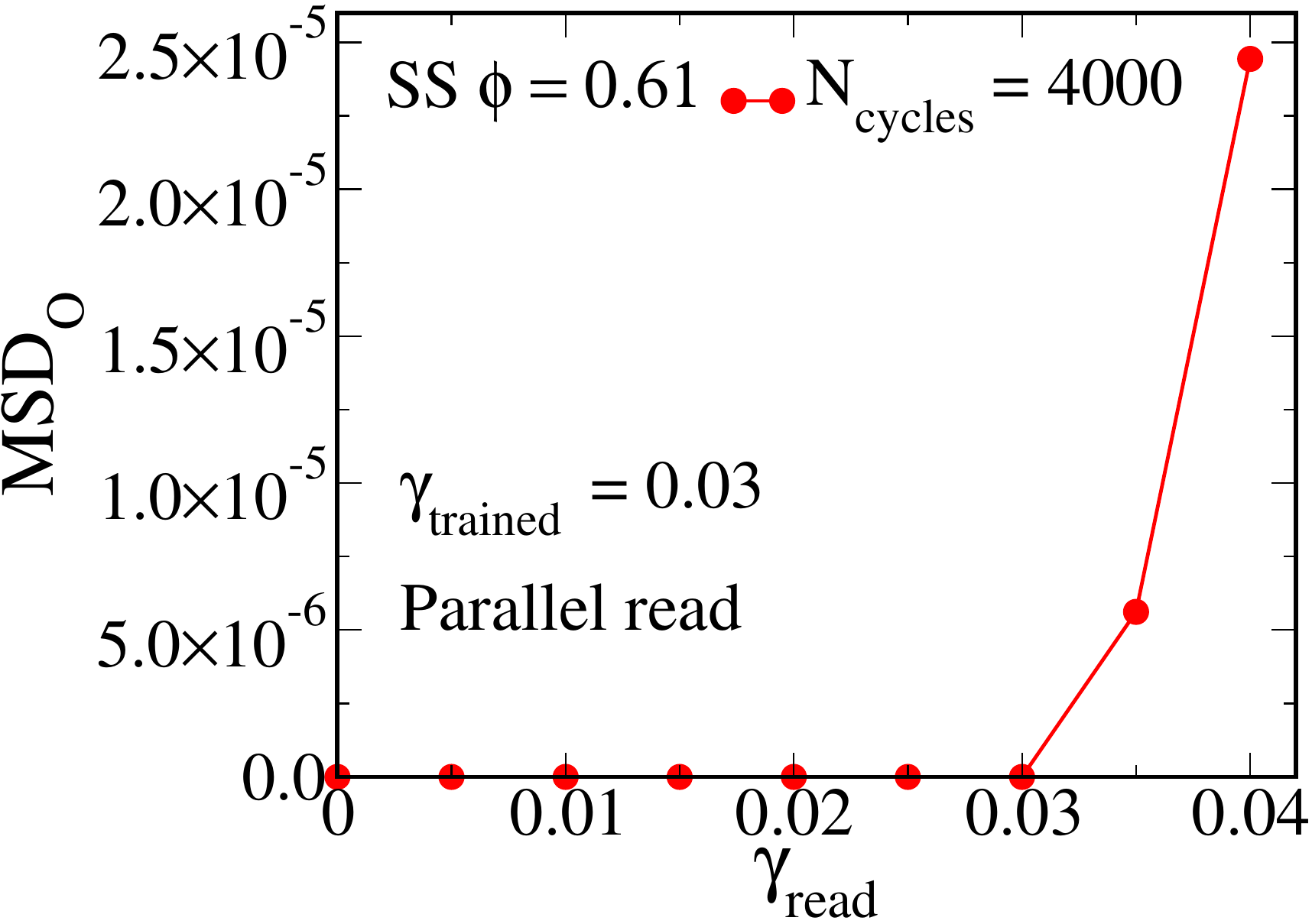}
\includegraphics[width = 0.43\textwidth]{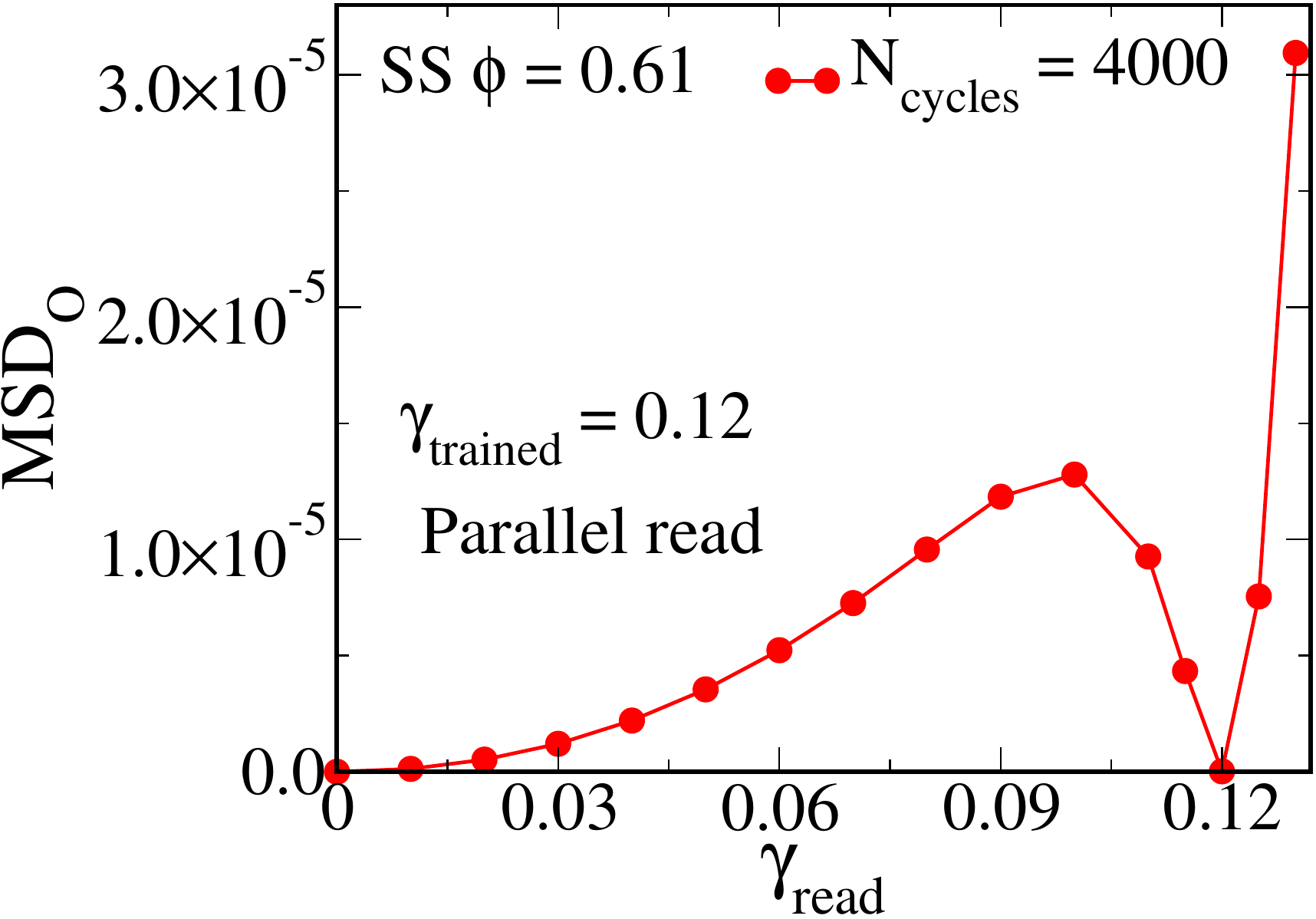}
 \caption{The MSD as a function of $\gamma_{\text read}$ for single training. Top: The system is trained at $\gamma_{trained}=0.03$. At this amplitude, system is point reversible. We find that the MSD is zero for all $\gamma_{\text read}$ $\leq$ $\gamma_{\text trained}$. Then there is a change of slope. Bottom: The system is trained at $\gamma_{trained}=0.12$. At this amplitude, the system is loop reversible. We find that the MSD is zero only when $\gamma_{\text read}$ = $\gamma_{\text trained}$}
\label{fig-20} 
 \end{figure}

We start with an equilibrated system and then train it for a sufficient number of cycles with a single amplitude. The number of training cycles needed for the soft sphere system to reach the absorbing state is much larger than for the BMLJ configurations. The trained system is subjected to read cycles using parallel reading. We have a system which is trained at $\gamma=0.03$, (where the system is point reversible) and another system which is trained at $\gamma=0.12$ (where the system is loop reversible). The results are shown in Fig. \ref{fig-20}. We observe that when the training amplitude belongs to the point reversible regime, the MSD is zero for all the amplitudes below the training amplitude,  similarly to the model in \cite{PhysRevLett.107.010603,PhysRevE.88.032306}. When the amplitude is in the loop reversible range, the memory behaviour is similar to the BMLJ system discussed earlier, in that the MSD is finite both above and below the training amplitude.

\subsubsection{Sequential reading}
\begin{figure}[h!]
 \centering
 \includegraphics[width = 0.42\textwidth]{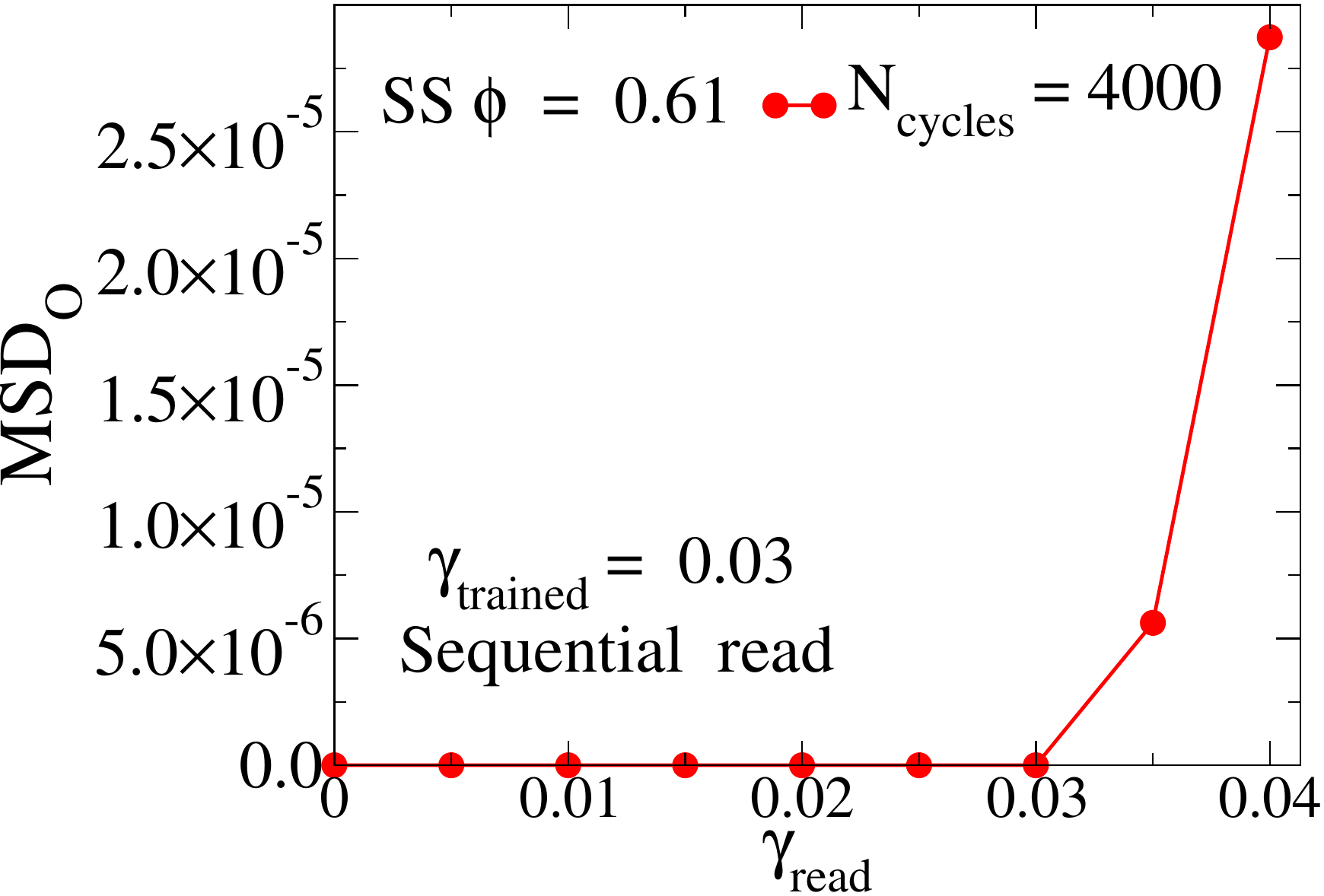}
\includegraphics[width = 0.42\textwidth]{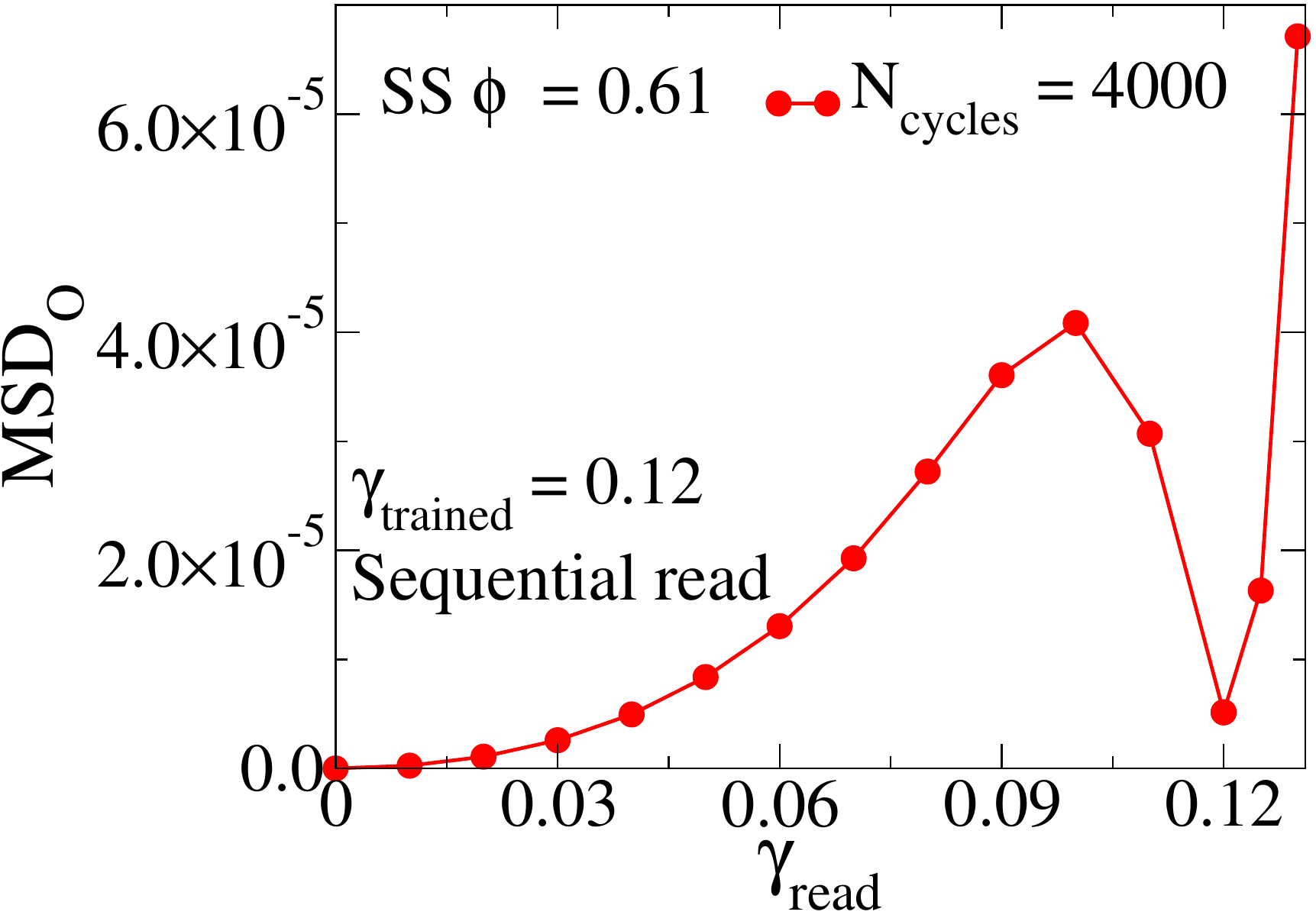}
\caption{The MSD as a function of $\gamma_{\text read}$  during sequential reading. Top: The system is trained at $\gamma=0.03$. At this amplitude, the system is point reversible. We find that the MSD is zero for all $\gamma_{\text read}$ $\leq$ $\gamma_{\text trained}$, and increases rapidly thereafter. Bottom: The system is trained at $\gamma=0.12$. At this amplitude, the system is loop reversible. We observe that the MSD decreases sharply around and is very small at $\gamma_{\text read}$ = $\gamma_{\text trained}$  This memory behaviour is different from parallel reading, since the system is not fully reversible at $\gamma_{\text read}$ = $\gamma_{\text trained}$, although it displays a very clear memory signature.}
\label{fig-21} 
 \end{figure}

We also perform sequential reading for the cases considered above, and present the results in Fig. \ref{fig-21}. We observe that when the amplitude is in the point reversible range, the memory behaviour is the same as for parallel reading. The MSD is zero at all $\gamma_{read}$ below $\gamma_{\text trained}$ and it increases with increasing read amplitude above. This behaviour is expected since the application of  $\gamma_{\text read}$ ($< \gamma_{\text trained}$) does not change the system in any way. When the training amplitude is in the loop reversible range, the system is not reversible at $\gamma_{\text read} = \gamma_{\text trained}$ (MSD is not zero, although very small) but a sharp change in MSD occurs across $\gamma_{\text trained}$ and the resulting MSD behaves very similarly to the parallel read case. 

\begin{figure}[htp]
\centering
\includegraphics[scale=0.35]{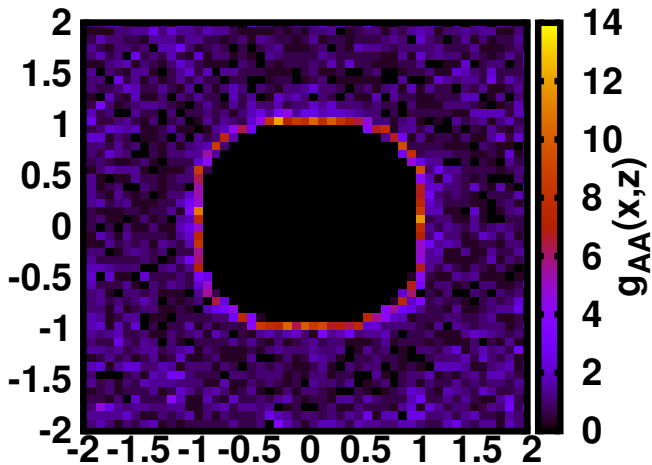}
\caption{Two dimensional pair correlation function ($g_{AA}(x,z)$) for the trained system in the shear plane $xz$. The system is trained at $\gamma_{trained} = 0.23$. The data are averaged over $40$ different samples.}
\label{fig-22}
\end{figure}

\paragraph{Structural signature of memory:} We study the nature of structural change due to training by computing the two dimensional pair correlation function $g(x,z)$ for a training amplitude in the point reversible regime. We choose $\phi = 0.54$ (in order to have better clarity) at which density the system is point reversible at $\gamma= 0.23$. The results are presented in Fig. \ref{fig-22} indicating that $g(x,z)$ for the trained system shows significant anisotropy and significantly different from $g(x,z)$ that of the fluid (not shown) which is isotropic. This behaviour is analogous to the observations in \cite{PhysRevE.88.032306} although the observed anisotropy is different owing to the differences in the shearing protocol. Other partial pair correlation functions display similar behaviour (not shown). 

\subsection{Memory effects in the diffusing states}

\begin{figure}[htp]
\centering
\includegraphics[scale=0.38]{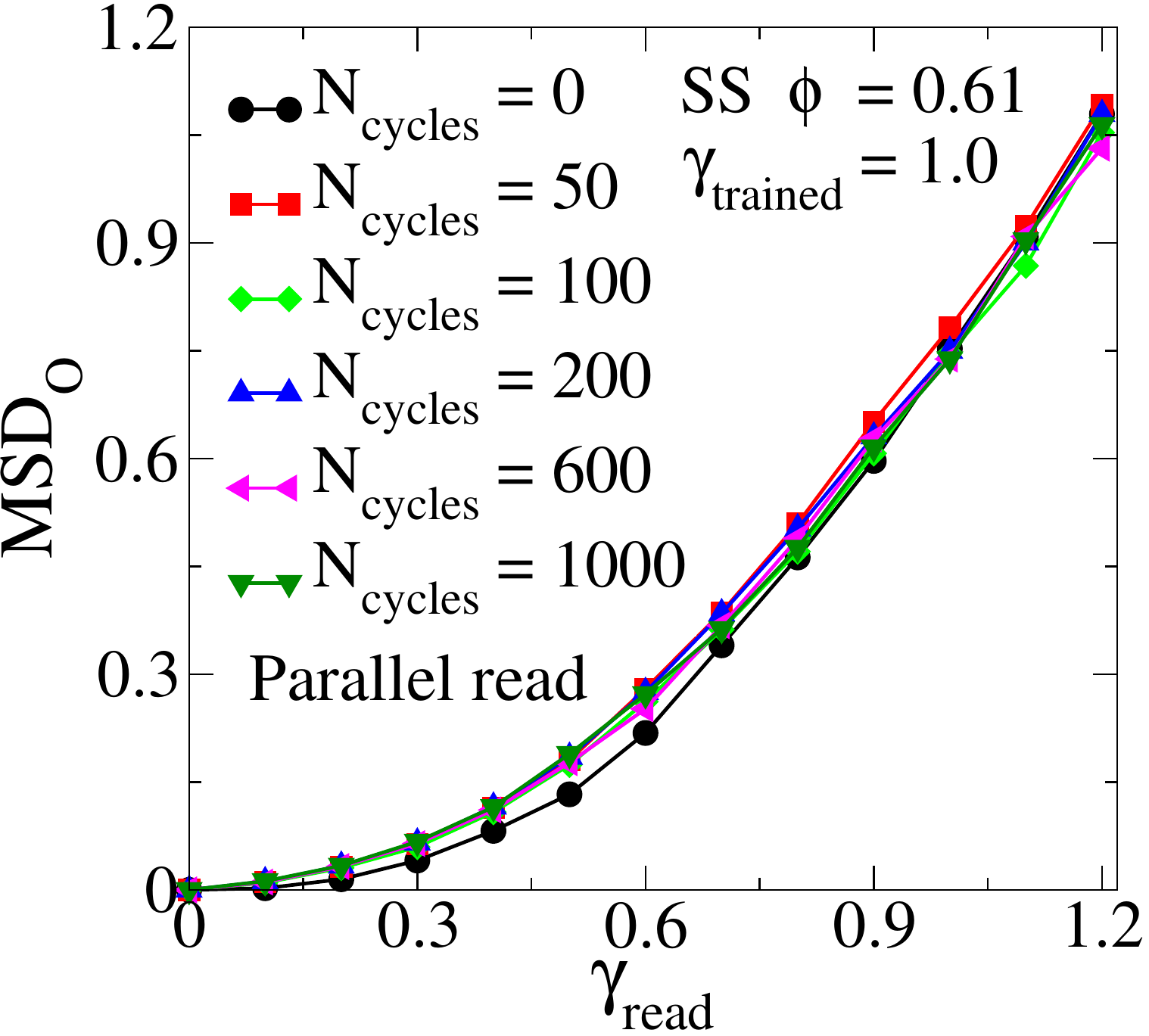}
\includegraphics[scale=0.38]{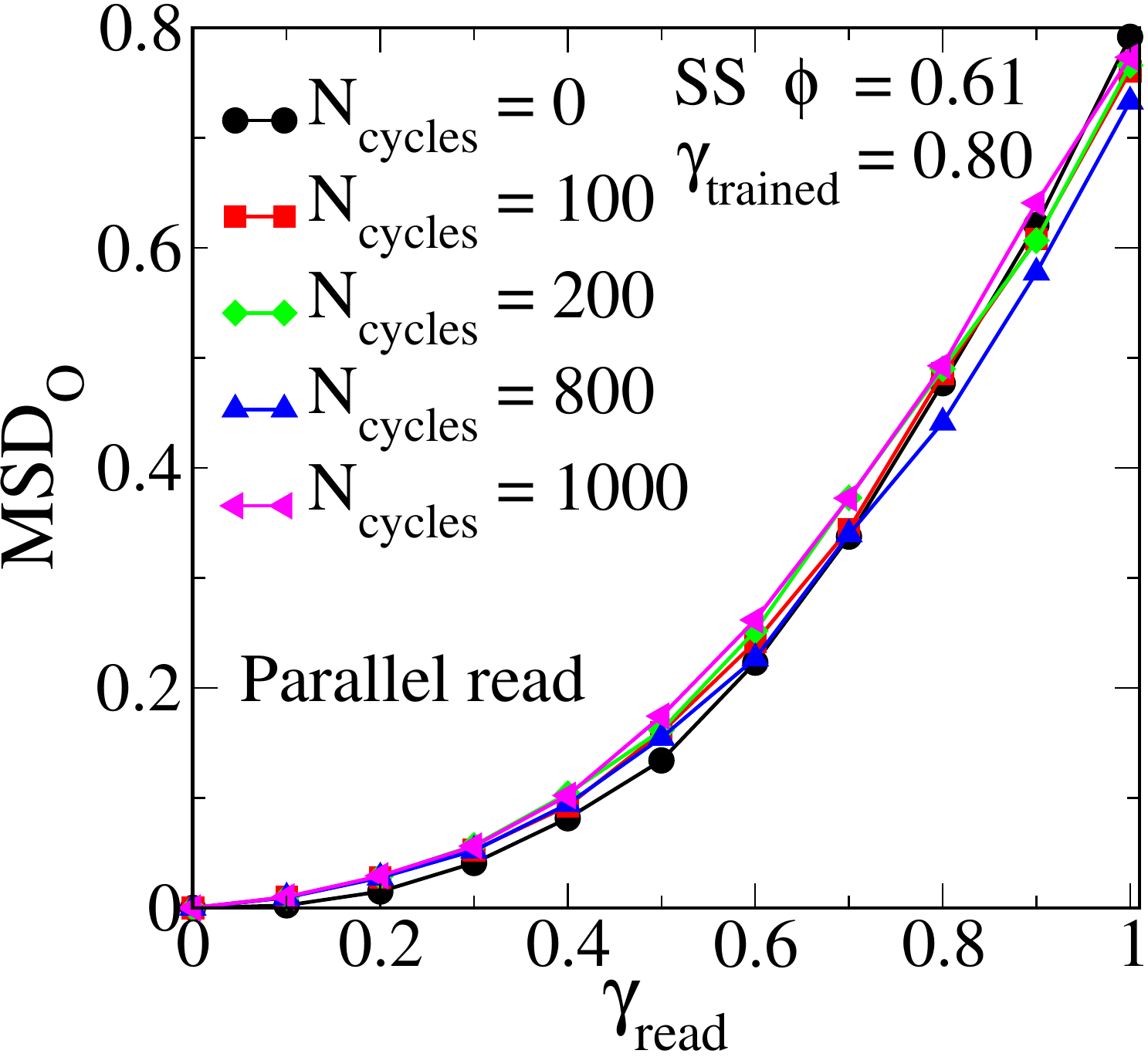}
\caption{The MSD is plotted as a function of $\gamma_{\text read}$ for configurations trained at $\gamma = 1.0$ (top) $\gamma = 0.8$ (bottom). Both these amplitudes belong to diffusive regime. The different lines correspond to different numbers of training cycles. We do not observe any memory signatures in these cases.}
\label{fig-23}
\end{figure}

At very large amplitude of shearing the soft sphere system shows diffusive behaviour, like the BMLJ. The MSD increases with  increasing  accumulated strain linearly.
We consider whether any memory effects are present in this regime, by analysing configurations trained at  $\gamma _{trained}= 1.0$ and  $\gamma_{trained} = 0.8$, at packing fraction $\phi = 0.61$. The results are presented in Fig. \ref{fig-23}. As before, the system is trained over a large number of training cycles and after training, the system is read using parallel reading. It is observed that the MSD increases smoothly as $\gamma_{\text read}$ increases and there are no signatures of memory of the training amplitudes. 

\subsection{Multiple memories}

We consider three cases in studying multiple memories in the soft sphere system: (i) Both the training amplitudes are below $\gamma_c$, (ii) Both training amplitudes are above $\gamma_c$, and (iii) One training amplitude is below $\gamma_c$, and the other is above $\gamma_c$. Here,  $\gamma_c$ refers to the stain at which a transition is observed from the point reversible to loop reversible states. We consider each of these cases employing both parallel and sequential read protocols.

\begin{figure}[h!]
 \centering
 \includegraphics[width = 0.40\textwidth]{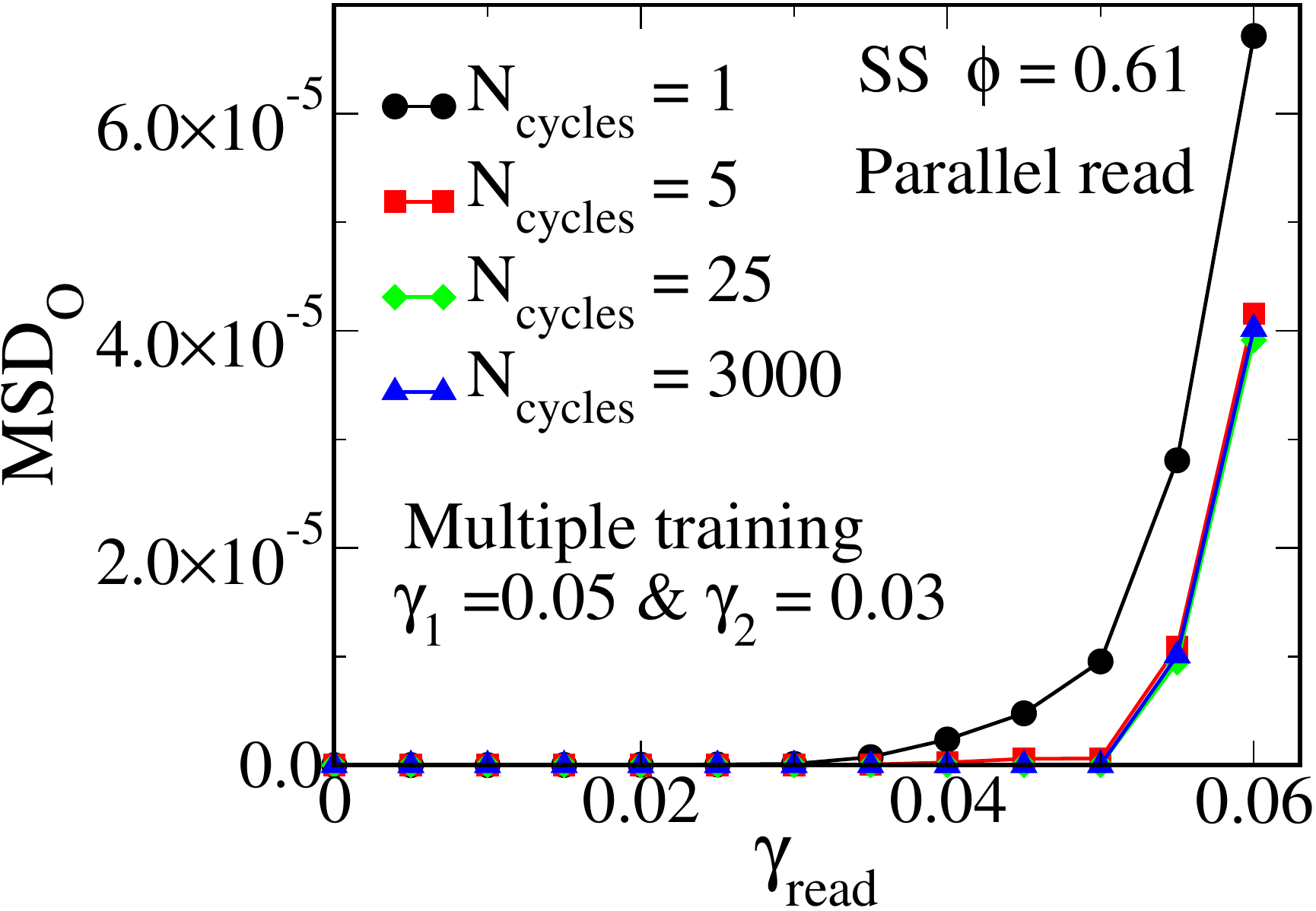}
\includegraphics[width = 0.40\textwidth]{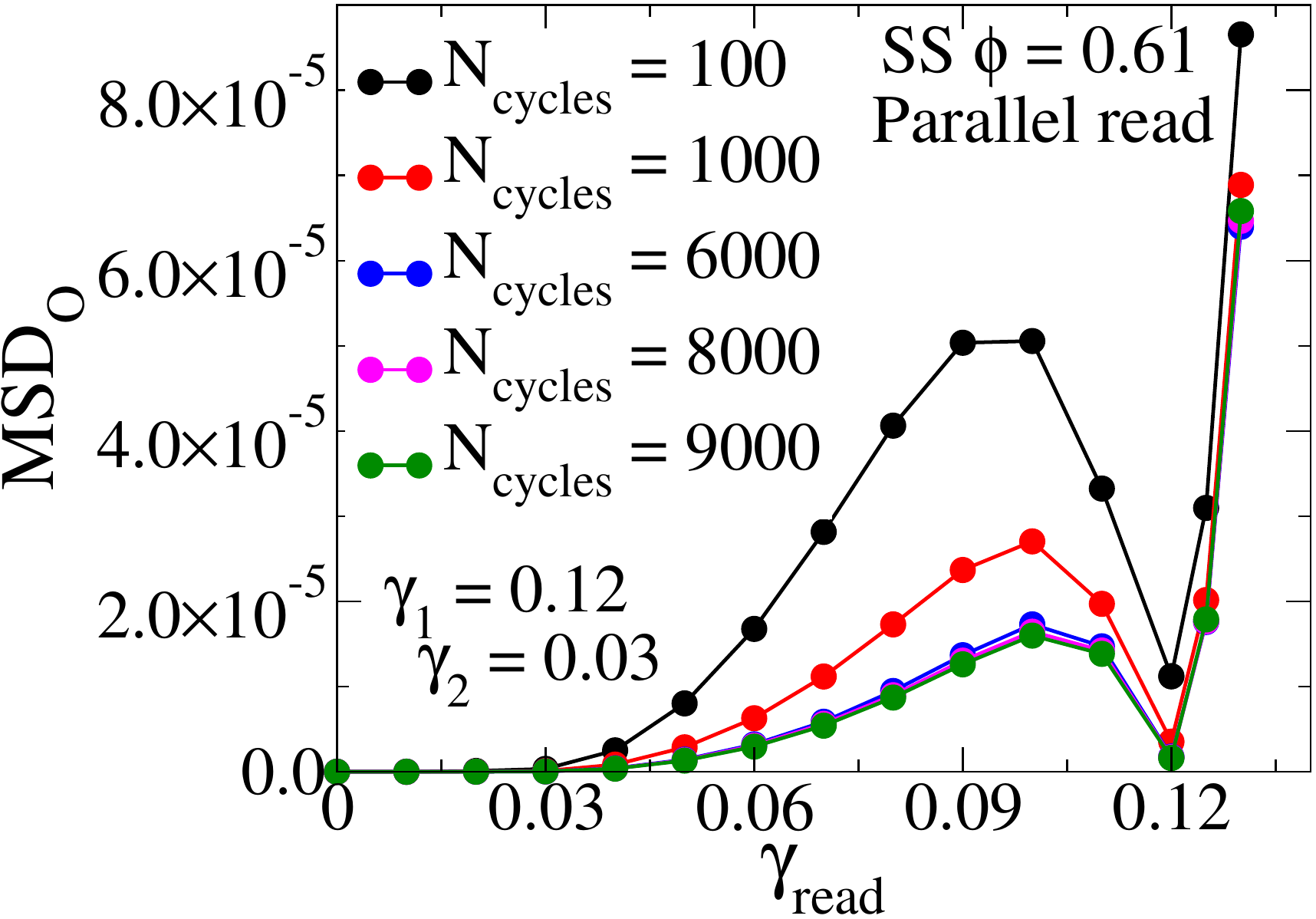}
\includegraphics[width = 0.40\textwidth]{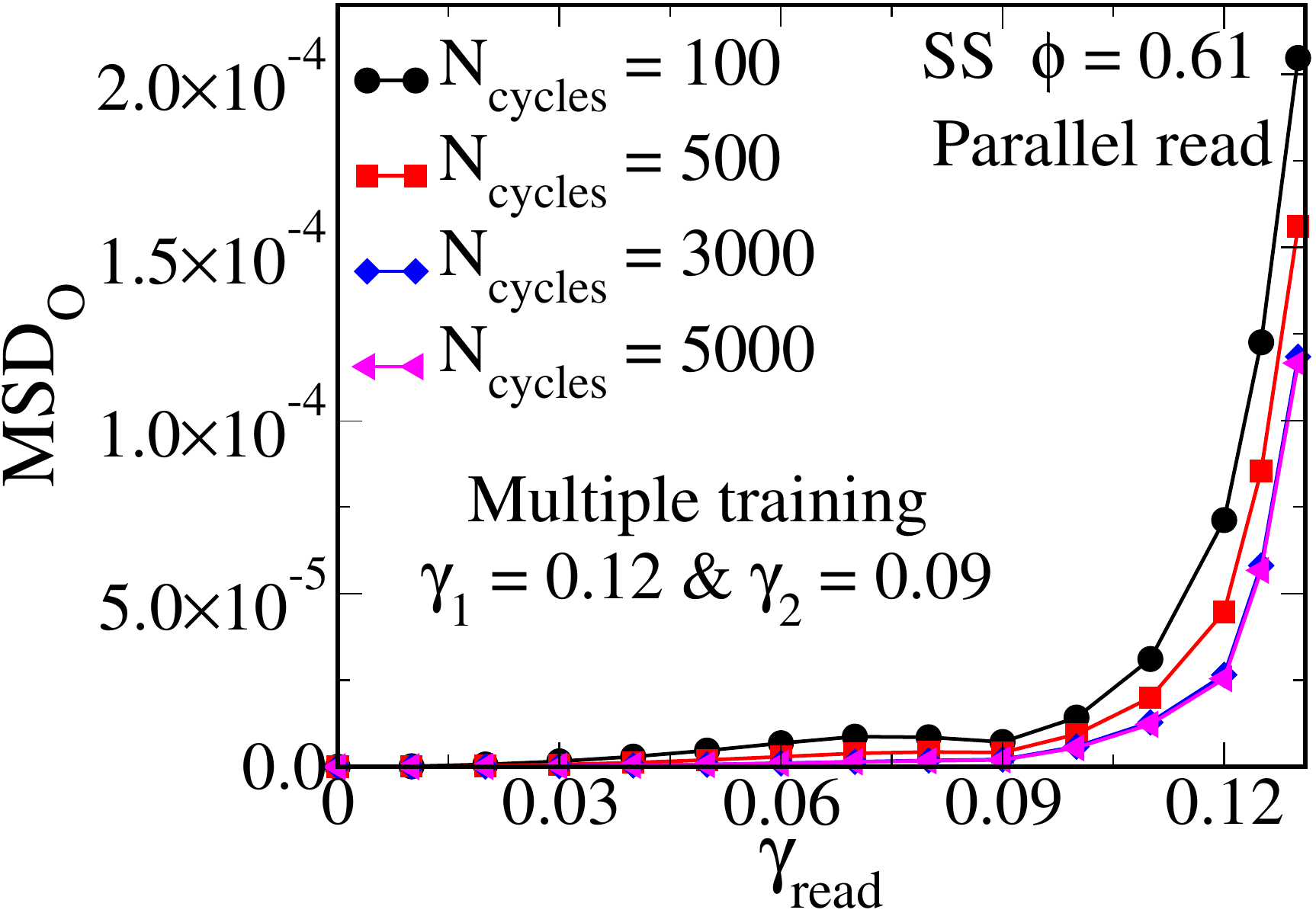}
 \caption{The MSD as a a function of $\gamma_{\text read}$ using the parallel read protocol. Different lines correspond to different numbers of training cycles. Top panel: The system is trained $\gamma_1=0.05$ and $\gamma_2=0.03$. At these amplitudes, the system is point reversible. After one training cycle, the MSD is zero below $\gamma_2=0.03$ and finite above, but with more training cycles, the MSD becomes zero for all $\gamma_{\text read} < \gamma_1=0.05$. Thus, only the memory of the largest amplitude remains.  Middle panel: The system is trained $\gamma_1=0.12$ and $\gamma_2=0.03$. At the amplitude $\gamma=0.12$, the system is in loop reversible at $\gamma=0.03$, the system is point reversible. The MSD remains zero below $\gamma_{read}=\gamma_2$ and exhibits a minimum value approaching zero at  $\gamma_{read}=\gamma_1$, each being a clear signature of memory in the respective regimes.  Bottom panel: The system is trained $\gamma_1=0.12$ and $\gamma_2=0.09$. At both these amplitudes, the system is loop reversible. While the MSD is close to zero below $\gamma_{read}=\gamma_2$, no clear signature of memory is present near $\gamma_{read}=\gamma_1$. The presence of multiple memories in this case cannot be concluded from these observations.  }
\label{fig-24} 
\end{figure}

\begin{figure}[htp]
\centering
\includegraphics[width = 0.41\textwidth]{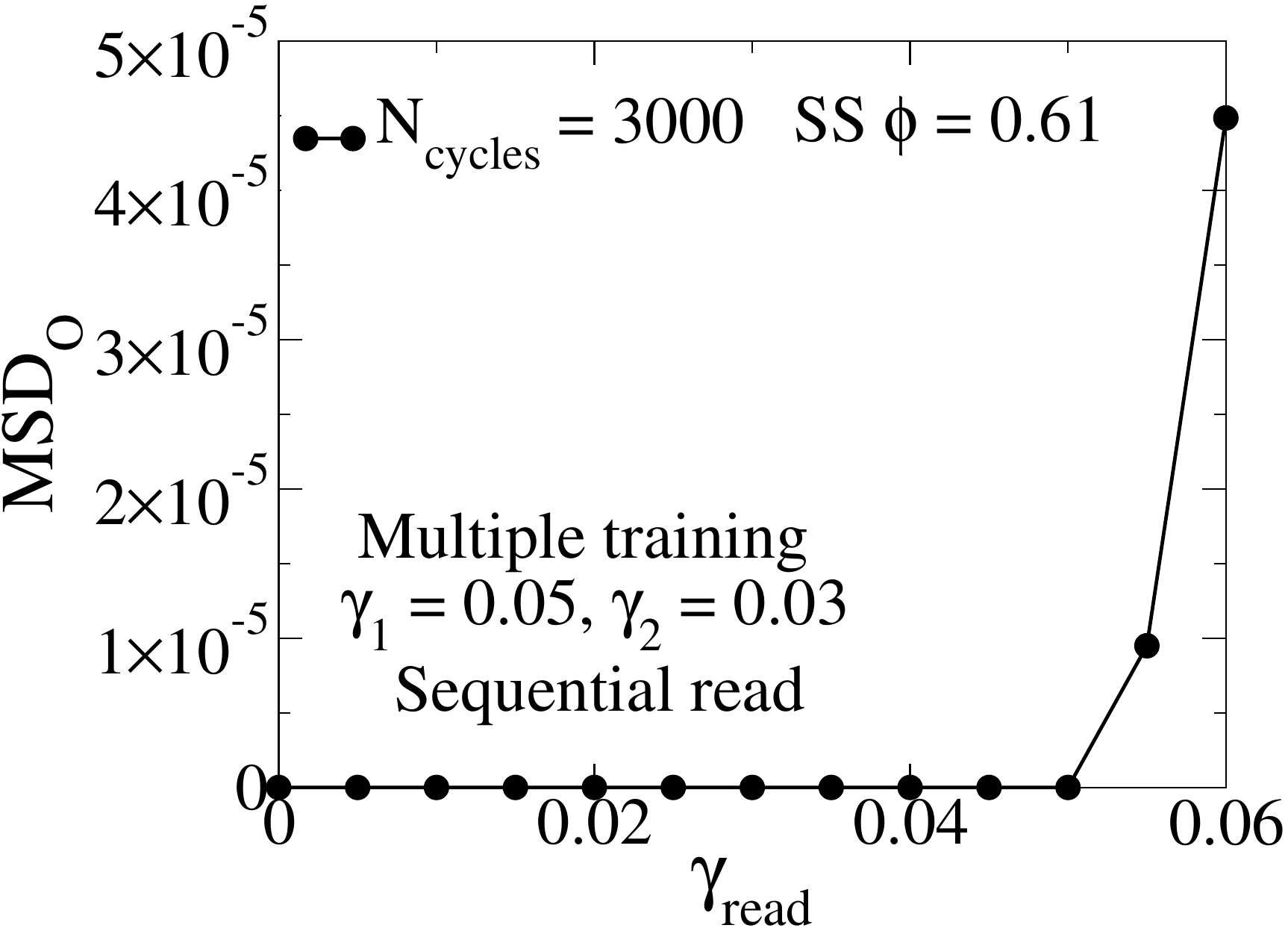}
\includegraphics[width = 0.41\textwidth]{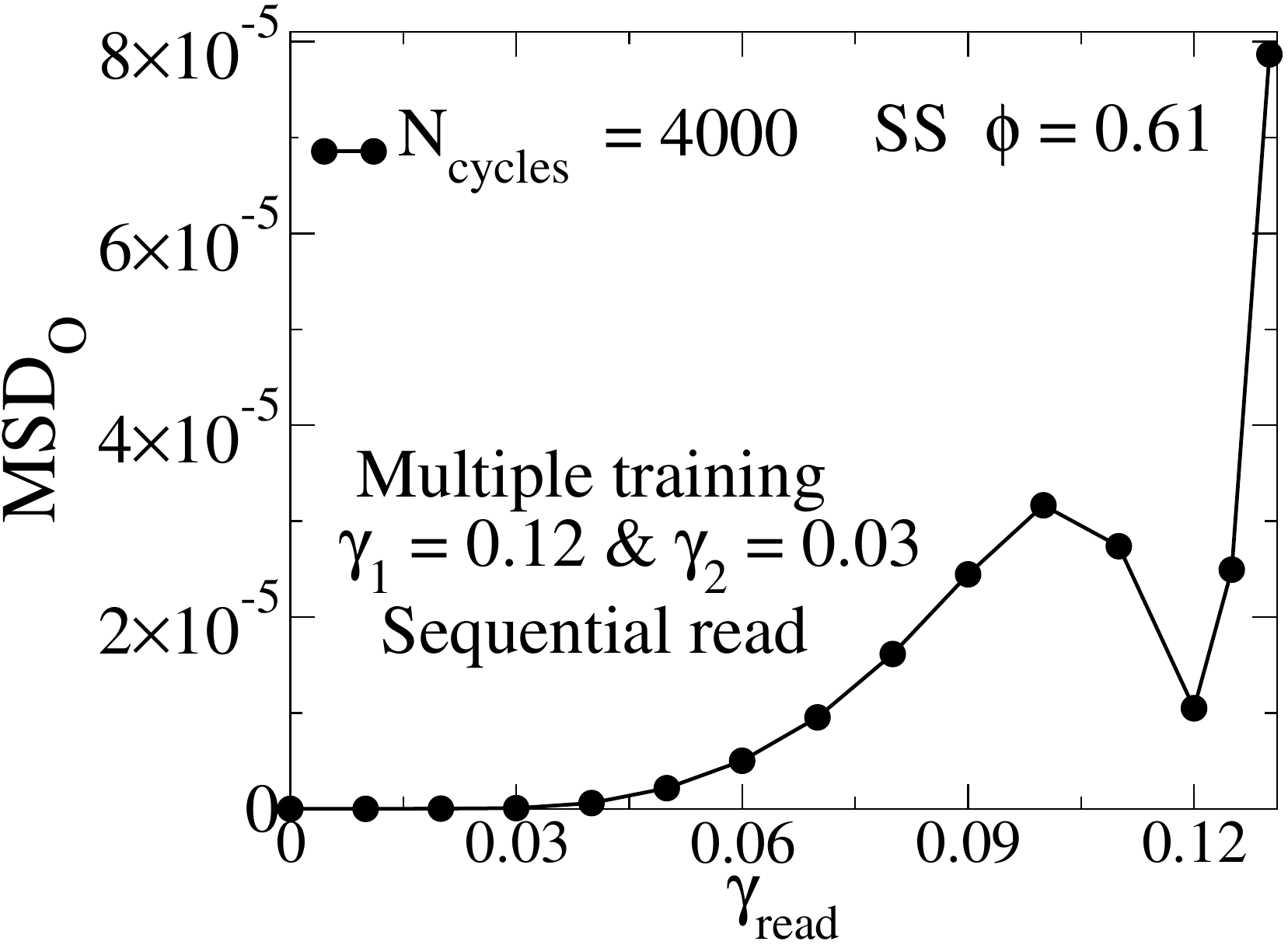}
\includegraphics[width = 0.41\textwidth]{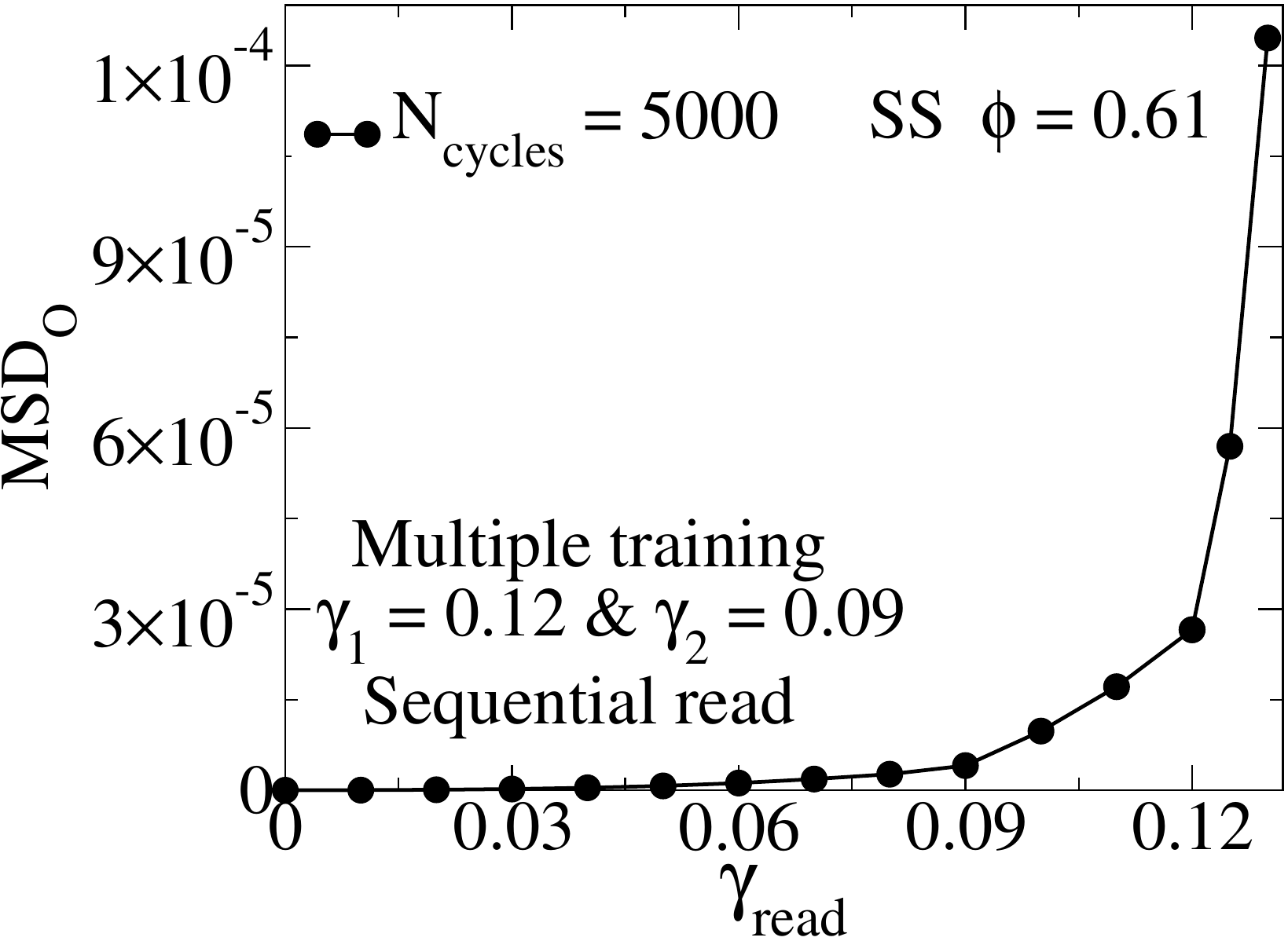}
\caption{The MSD as a a function of $\gamma_{\text read}$ using the sequential read protocol. Different lines correspond to different numbers of training cycles. Top panel: The system is trained $\gamma_1=0.05$ and $\gamma_2=0.03$. At these amplitudes, the system is point reversible. After one training cycle, the MSD is zero below $\gamma_2=0.03$ and finite above, but with more training cycles, the MSD becomes zero for all $\gamma_{\text read} < \gamma_1=0.05$. Thus, only the memory of the largest amplitude remains. Middle panel: The system is trained $\gamma_1=0.12$ and $\gamma_2=0.03$. At the amplitude $\gamma=0.12$, the system is in loop reversible at $\gamma=0.03$, the system is point reversible. The MSD remains zero below $\gamma_{read}=\gamma_2$ and exhibits a minimum value at  $\gamma_{read}=\gamma_1$, each being a clear signature of memory in the respective regimes.  Bottom panel: The system is trained $\gamma_1=0.12$ and $\gamma_2=0.09$. At both these amplitudes, the system is loop reversible. While the MSD is close to zero below $\gamma_{read}=\gamma_2$, no clear signature of memory is present near $\gamma_{read}=\gamma_1$. The presence of multiple memories in this case cannot be concluded from these observations. }
\label{fig-25}
\end{figure}

\subsubsection{Parallel reading} We perform parallel reading to the trained samples. The results are shown in Fig. \ref{fig-24}. We first consider configurations  trained at amplitudes $\gamma_1 =0.05$ and $\gamma_2 =0.03$. The system is point reversible at both the amplitudes. We train the system for $3000$ cycles. After training, the system is read using the parallel reading protocol. Below the higher training amplitude, the MSD is zero for all amplitudes. This implies that the multiple memories are transient when both the amplitudes are in the point reversible range, consistently with previous observations. 
We next consider training amplitudes $\gamma_1= 0.12$ (at which the system is loop reversible) and $\gamma_2=0.03$  (at which the system is point reversible). After training for $4000$ cycles the system is read using the parallel reading protocol. We observe that below $\gamma_{read}=\gamma_2$, the MSD remains zero, and at $\gamma_{read}=\gamma_1$ the MSD exhibits a minimum, approaching zero for large enough training cycles. Thus, both these memories are retained with expected signatures. Finally we consider training amplitudes  $\gamma_1 =0.12$ and $\gamma_2 =0.09$. The system is loop reversible at both the amplitudes. We train the system for $4000$ cycles. After training, the system is read using the parallel reading protocol. We observe that the MSD approaches zero for all amplitudes below  $\gamma_{read}=\gamma_1$ but no distinct signature of memory is found at  $\gamma_{read}=\gamma_1$. These observations are both surprising, since neither conforms to the expected memory behaviour in analogy with the BMLJ glass. The case of multiple memories in the loop reversible regime thus require further investigation. 

\subsubsection{Sequential reading}

We perform  sequential reading for each of the cases considered above, and show the results in Fig. \ref{fig-25}. We observe that the memory behaviour is the same as in the case of the  parallel reading protocol. 

\section{Summary and Conclusions} \label{secV}

We have performed numerical investigations of memory effects in two model systems, the Kob-Andersen binary mixture (BMLJ) and a soft sphere mixture. In the former case, our results extend and elaborate on the results and observations from earlier work \cite{MemFiocco}, in particular in the form of considering different protocols for reading the encoded memories. The latter case offers an interesting extension of previous studies, in that it offers an example displaying features that are distinct from the earlier studied cases of a model glass and a model of colloidal suspensions  \cite{PhysRevLett.107.010603,PhysRevE.88.032306}, exhibiting features found in both these earlier examples. In the loop reversible regime of this model, the memory effects seen are, to a large extent, similar to the case of the model glass, even while the system samples a trivial energy landscape. In considering structural signatures of memory, we find that the model glass studied does not reveal the features seen previously for the model of colloidal suspensions, and thus prompting further investigations on the manner in which the memory is encoded in this system. On the other hand, the soft sphere system in the point reversible regime does exhibit the expected structural signatures. These results taken together offer a detailed characterisation of memory effects in the studied model systems, some aspects of which require further investigations to more clearly delineate. In addition to such investigations, a careful study of well chosen, simple, model systems may be a fruitful direction to fully comprehend memory effects in the type of driven systems studied here. Analysis in \cite{JPCMFiocco} and \cite{Mungan2018} may offer useful starting points. 

\acknowledgements 
The work presented here and plans for future investigations have benefitted from discussions with P. Leishangthem, and with Sidney Nagel, Natham Keim, Joseph Paulsen, Muhittin Mungan, Thomas Witten and Ajay Sood, during the program "Memory Formation in Matter" at the KITP, UCSB.  This research was supported in part by the National Science Foundation under Grant No. NSF PHY17-48958. We gratefully acknowledge TUE-CMS and SSL, JNCASR, Bengaluru for computational resources and support. S. S. gratefully acknowledges support through the J. C. Bose fellowship, SERB, DST, India.

\bibliography{biblio_memory}

\end{document}